\documentclass[reprint,
superscriptaddress,
 amsmath,amssymb,
 aps,
prb,
]{revtex4-1}
\usepackage{hyperref}
\usepackage{graphicx}
\usepackage{dcolumn}
\usepackage{bm}

\usepackage{cleveref}
\usepackage{color}
\usepackage{comment}
\usepackage{float}
\newcommand{\ve}[1]{\boldsymbol{#1}} 

\usepackage{booktabs}
\usepackage{multirow}
\usepackage{makecell}
\usepackage{color, colortbl}

\definecolor{Gray}{gray}{0.95}
\newcolumntype{g}{>{\columncolor{Gray}}c}

\preprint{APS/123-QED}
\begin{document}
\title{Tomographic identification of all molecular orbitals in a wide binding energy range}

\author{Anja Haags}
\email{a.haags@fz-juelich.de}
\affiliation{Peter Gr\"unberg Institut (PGI-3), Forschungszentrum J\"ulich, 52425 J\"ulich, Germany}
\affiliation{J\"ulich Aachen Research Alliance (JARA), Fundamentals of Future Information Technology, 52425 J\"ulich, Germany}
\affiliation{Experimentalphysik IV A, RWTH Aachen University; Aachen, Germany}

\author{Dominik Brandstetter}
\affiliation{Institute of Physics, NAWI Graz, University of Graz, 8010 Graz, Austria}%

\author{Xiaosheng Yang}
\affiliation{Peter Gr\"unberg Institut (PGI-3), Forschungszentrum J\"ulich, 52425 J\"ulich, Germany}
\affiliation{J\"ulich Aachen Research Alliance (JARA), Fundamentals of Future Information Technology, 52425 J\"ulich, Germany}
\affiliation{Experimentalphysik IV A, RWTH Aachen University; Aachen, Germany}

\author{Larissa Egger}
\affiliation{Institute of Physics, NAWI Graz, University of Graz, 8010 Graz, Austria}%

\author{Hans Kirschner}
\affiliation{Physikalisch-Technische Bundesanstalt (PTB), 10587 Berlin, Germany.}

\author{Alexander Gottwald}
\author{Mathias Richter}
\affiliation{Physikalisch-Technische Bundesanstalt (PTB), 10587 Berlin, Germany.}

\author{Georg Koller}
\affiliation{Institute of Physics, NAWI Graz, University of Graz, 8010 Graz, Austria}%

\author{Fran\c{c}ois C. Bocquet}
\author{Christian Wagner}
\affiliation{Peter Gr\"unberg Institut (PGI-3), Forschungszentrum J\"ulich, 52425 J\"ulich, Germany}
\affiliation{J\"ulich Aachen Research Alliance (JARA), Fundamentals of Future Information Technology, 52425 J\"ulich, Germany}

\author{Michael G. Ramsey}
\affiliation{Institute of Physics, NAWI Graz, University of Graz, 8010 Graz, Austria}%

\author{Serguei Soubatch}
\affiliation{Peter Gr\"unberg Institut (PGI-3), Forschungszentrum J\"ulich, 52425 J\"ulich, Germany}
\affiliation{J\"ulich Aachen Research Alliance (JARA), Fundamentals of Future Information Technology, 52425 J\"ulich, Germany}

\author{Peter Puschnig}%
 \email{peter.puschnig@uni-graz.at}
\affiliation{Institute of Physics, NAWI Graz, University of Graz, 8010 Graz, Austria}%

\author{F. Stefan Tautz}
\email{s.tautz@fz-juelich.de}
\affiliation{Peter Gr\"unberg Institut (PGI-3), Forschungszentrum J\"ulich, 52425 J\"ulich, Germany}
\affiliation{J\"ulich Aachen Research Alliance (JARA), Fundamentals of Future Information Technology, 52425 J\"ulich, Germany}
\affiliation{Experimentalphysik IV A, RWTH Aachen University; Aachen, Germany}
\date{\today}

\begin{abstract}
In the past decade, photoemission orbital tomography (POT) has evolved into a powerful tool to investigate the electronic structure of organic molecules adsorbed on surfaces. Here we show that POT allows for the comprehensive experimental identification of all molecular orbitals in a substantial binding energy range, in the present case  more than 10 eV. Making use of the angular distribution of photoelectrons as a function of binding energy, we exemplify this by extracting orbital-resolved partial densities of states (pDOS) for 15 $\pi$ and 23 $\sigma$ orbitals from the experimental photoemission intensities of the prototypical organic molecule bisanthene (C$_{28}$H$_{14}$) on a Cu(110) surface.
In their entirety, these experimentally measured orbital-resolved pDOS for an essentially complete set of orbitals serve as a stringent benchmark for electronic structure methods, which we illustrate by performing density functional theory (DFT) calculations employing four frequently-used exchange-correlation functionals. By computing  the respective molecular-orbital-projected densities of states of the bisanthene/Cu(110) interface, a one-to-one comparison with experimental data for an unprecedented number of 38 orbital energies becomes possible. The quantitative analysis of our data reveals that the range-separated hybrid functional HSE performs best for the investigated organic/metal interface. At a more fundamental level, the remarkable agreement between the experimental and the Kohn-Sham orbital energies over a binding energy range larger than 10\,eV suggests that---perhaps unexpectedly---Kohn-Sham orbitals approximate Dyson orbitals, which would rigorously account for the electron extraction process in photoemission spectroscopy but are notoriously difficult to compute, in a much better way than previously thought.

\end{abstract}

\maketitle

\section{\label{sec:intro}Introduction}

Interfaces between organic molecules and metals play a central role in functional devices in nanoscience and nanotechnology.\cite{Ishii1999,Kahn2003,Ueno2008,Braun2009,MoleculeMetalInterface,Willenbockel2014,Liu2017,Ferri2017} When a molecule adsorbs on a metallic surface, several processes determine the resulting electronic structure of the interface. First, the molecule gets physically attracted by van der Waals interactions.\cite{Berland2015,Hermann2017}  Then, the ionization potential (IP) and electron affinity (EA) levels of the molecule are renormalized because of the proximity to the metal or, in a dense layer, also because of neighboring molecules, both as a result of polarization effects.\cite{Neaton2006,Garcia-Lastra2009,Thygesen2009,Soubatch2009,Puschnig2012} When the molecule approaches the substrate further, the molecular orbitals start to overlap with the metallic states, and thus hybridized states arise at the interface.\cite{Yamane2007,Ziroff2010,Berkebile2011,Wiessner2013,Ules2014,Yang2022} If the resulting level alignment permits, the native electronic states may be further reconfigured and charge may be transferred, resulting in the population of formerly unoccupied molecular states or the depopulation of formerly occupied ones. Many examples of such charge transfers have been reported.\cite{Duhm2008,Rangger2009,Puschnig2009a,Ziroff2010,Heimel2013,Hofmann2015,Schonauer2016,Hollerer2017}

\begin{figure}[bt]
\begin{center}	\includegraphics[width=\columnwidth]{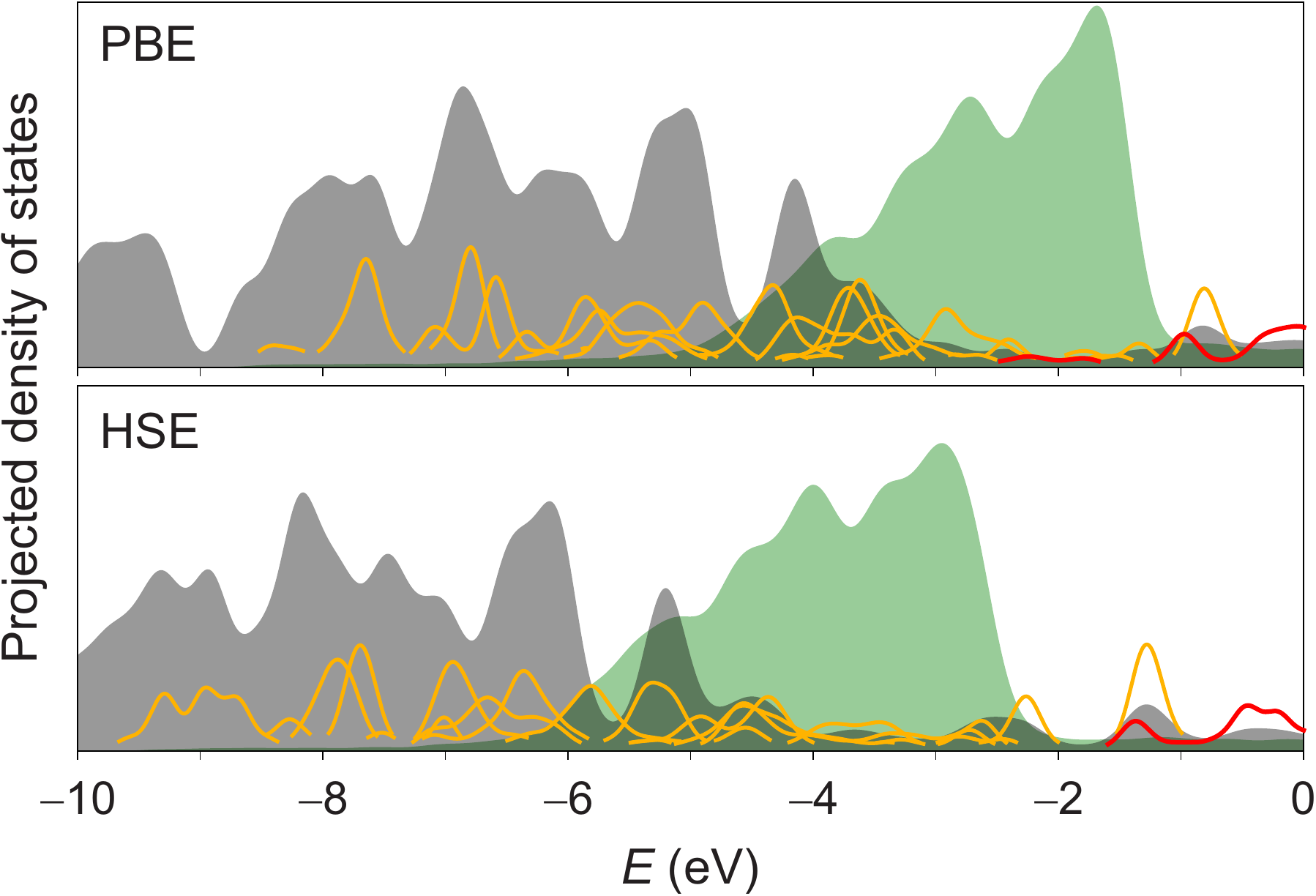}
\end{center}
	\caption{\label{fig:pDOS_theory}  Projected density of states of bisanthene/Cu(110) computed using two different exchange-correlation functionals (see main text). The green and gray shaded areas depict the density of states projected on copper and carbon atoms, respectively. The molecular-orbital-projected density of states (MOPDOS) of the former lowest unoccupied molecular orbital of bisanthene, populated due to charge donation from the substrate, is shown in red. The MOPDOS of all other molecular $\pi$ orbitals are shown by orange curves. }
\end{figure}

All these mechanisms must be taken into account if a valid theoretical description of organic/metal interfaces is to be attempted. The most commonly applied framework for this purpose is density functional theory (DFT). It provides the ground-state electron density and the ground state total energy as the primary observables, which are constructed from the single-particle Kohn-Sham orbitals and energies. It is common practice to relate these Kohn-Sham orbital energies to single-electron excitation energies, which are accessible from experimental techniques such as photoemission or scanning tunneling spectroscopy. Yet, this lacks a rigorous theoretical justification, except for the  highest occupied Kohn-Sham orbital which can be identified with the ionization potential of the system \cite{Seidl1996,Kronik2012}. For lower lying Kohn-Sham orbitals no such equivalent connection is currently known. Furthermore, the equivalence between the highest occupied Kohn-Sham orbital and the ionization potential holds, strictly speaking, only for the \emph{exact} exchange-correlation functional.

In practical DFT calculations, however, one has to resort to approximations for the exchange-correlation functional, and moreover is usually interested in the electronic density of states (DOS) well beyond the Fermi energy. This is illustrated in Fig.~\ref{fig:pDOS_theory} which displays DFT results for the prototypical organic/metal interface bisanthene (C$_{28}$H$_{14}$) on Cu(110) obtained with the PBE \cite{Perdew1996} and HSE \cite{Heyd2004,Heyd2006} exchange-correlation functionals, representatives of the $2^\textrm{nd}$ and $4^\textrm{th}$ rung of the Jacob's ladder, respectively.  It is revealing to decompose the total DOS into contributions from the substrate (green), the molecular subsystem (gray), and ultimately, from individual molecular orbitals (yellow and red lines).  Here, the yellow lines correspond to the molecular-orbital-projected density of states (MOPDOS) for all 14 occupied $\pi$ orbitals of isolated bisanthene. Thus, each curve represents the offspring of a molecular Kohn-Sham level in the coupled system. The figure shows that there are a few molecular orbitals close to the Fermi energy, \emph{i.e.}, above the Cu $d$~band. It is common practice to analyze the energy level alignment of these frontier orbitals to search, \emph{e.g.}, for a possible charge transfer between the molecule and the metal surface.  In the present case of bisanthene/Cu(110), the calculations do indeed suggest a (partial) charge transfer into the lowest unoccupied molecular orbital (LUMO, red curve), however, as the comparison between the PBE and GGA results  demonstrates, to a different extent. This situation calls for an experimental method that is capable of providing such orbital-resolved densities of states as benchmarks for theory.

Fig.~\ref{fig:pDOS_theory} also conveys another important message: most of the molecular density of states is found in a broad binding energy interval \emph{below} the $d$ band. In this energy range, the molecular density of states appears to be even more sensitive to the different exchange-correlation potentials than it does above the $d$ band. Thus, it would be desirable to use this rich information on orbital binding energies for benchmarking density functional approaches and also methods that go beyond them, \emph{e.g.}, the $GW$ method in the framework of many-body perturbation theory.\cite{Blase2011,Faber2014,Draxl2014,Marom2017,Golze2019} Before this can be done, however, two questions need to be answered in the affirmative. First, can Kohn-Sham orbitals far below the frontier orbitals be interpreted as Dyson orbitals,\cite{Truhlar2019,Krylov2020} and thus be assumed to carry information about electron extraction, even if no rigorous theoretical proof for this conjecture is known?\cite{Stowasser1999,Chong2002} And second, can such an orbital-resolved DOS actually be measured in this energy range? In the present work we demonstrate the latter and thereby experimentally confirm the former.

The most direct way to measure the occupied electronic structure over a wide range of binding energies is ultraviolet photoemission spectroscopy.\cite{Kahn2003,Dori2006,Hwang2007,Ueno2008,Heimel2008,Korzdorfer2009} Due to photoionization cross section effects, \emph{i.e.}, the dependence of the measured spectra on the polarization of the light and the electron emission angle, a one-to-one comparison of such photoemission spectra with a calculated density of states can be inconclusive and sometimes even misleading. However, by also measuring the angular distribution of the photoemission intensity, \emph{i.e.}, using angle-resolved photoemission spectroscopy (ARPES), this complexity can be turned into an advantage. In the last decade it has been shown that for many molecular films adsorbed on (metallic) surfaces, the angular distribution of the photoelectrons contains orbital fingerprints. This approach to reveal the orbital patterns and their corresponding binding energies is known as photoemission orbital tomography (POT). In the simplest case, if the final state of the emitted electron is approximated by a plane wave, the angular distribution can be understood in terms of the squared modulus of the Fourier transform of the initial-state Dyson orbital, \emph{i.e.}, as its momentum-space image.\cite{Puschnig2009a,Dauth2014,Woodruff2016,Puschnig2017} 

The concept of POT has been used to deconvolve measured energy distribution curves into the contributions of individual molecular (Kohn-Sham) orbitals, thereby providing an \emph{experimentally determined} orbital-resolved partial density of states, referred to as pDOS in the following.\cite{Dauth2011,Puschnig2011,Puschnig2017,Zamborlini2017,Yang2019,Kliuiev2019,Brandstetter2021,Saettele2021} In principle, this quantity is the exact experimental counterpart of the theoretically calculated MOPDOS (orange curves in Fig.~\ref{fig:pDOS_theory}). So far, however, the available experimental data and deconvolution analyses have mostly been limited to a few molecular orbitals, \emph{e.g.}, the frontier $\pi$ orbitals in a comparatively small energy window below the Fermi energy. Here we go far beyond the current state of the art and measure and unravel the orbital energy level alignment of bisanthene/Cu(110) over an energy range of more than 10\,eV below the Fermi energy ($E_\mathrm{F}$), whereby we even include $\sigma$ orbitals in the orbital tomography approach \cite{Haags2022}.  

\begin{figure}[bt]
\begin{center}
   \includegraphics[width=\columnwidth]{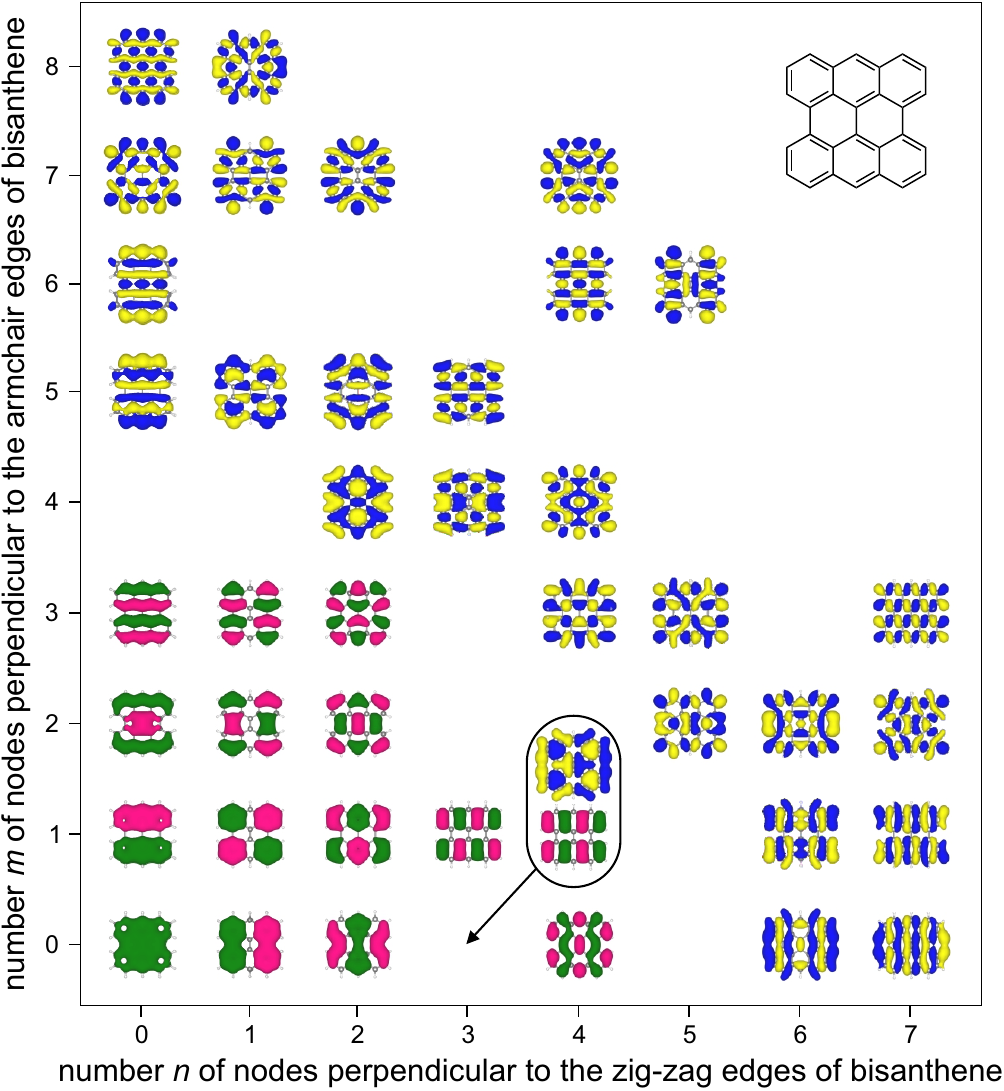}
\end{center}
	\caption{\label{fig:orbitals}  Real-space representations of orbitals of the isolated bisanthene molecule calculated using the PBE functional. The phase of $\pi$ and $\sigma$ orbitals is depicted using magenta/green and blue/yellow colors, respectively. The orbitals are positioned according to their number of (approximate) nodal surfaces perpendicular to the zig-zag ($n$) and armchair ($m$) edges of the molecule, which defines the orbital nomenclature used in this work. Note that the orbitals $\pi_{(3,0)}$ and $\sigma_{(3,0)}$ are depicted in the inset, while their position in the graph is indicated with the arrow. }
\end{figure}

\begin{figure}[bt]
\begin{center}	\includegraphics[width=\columnwidth]{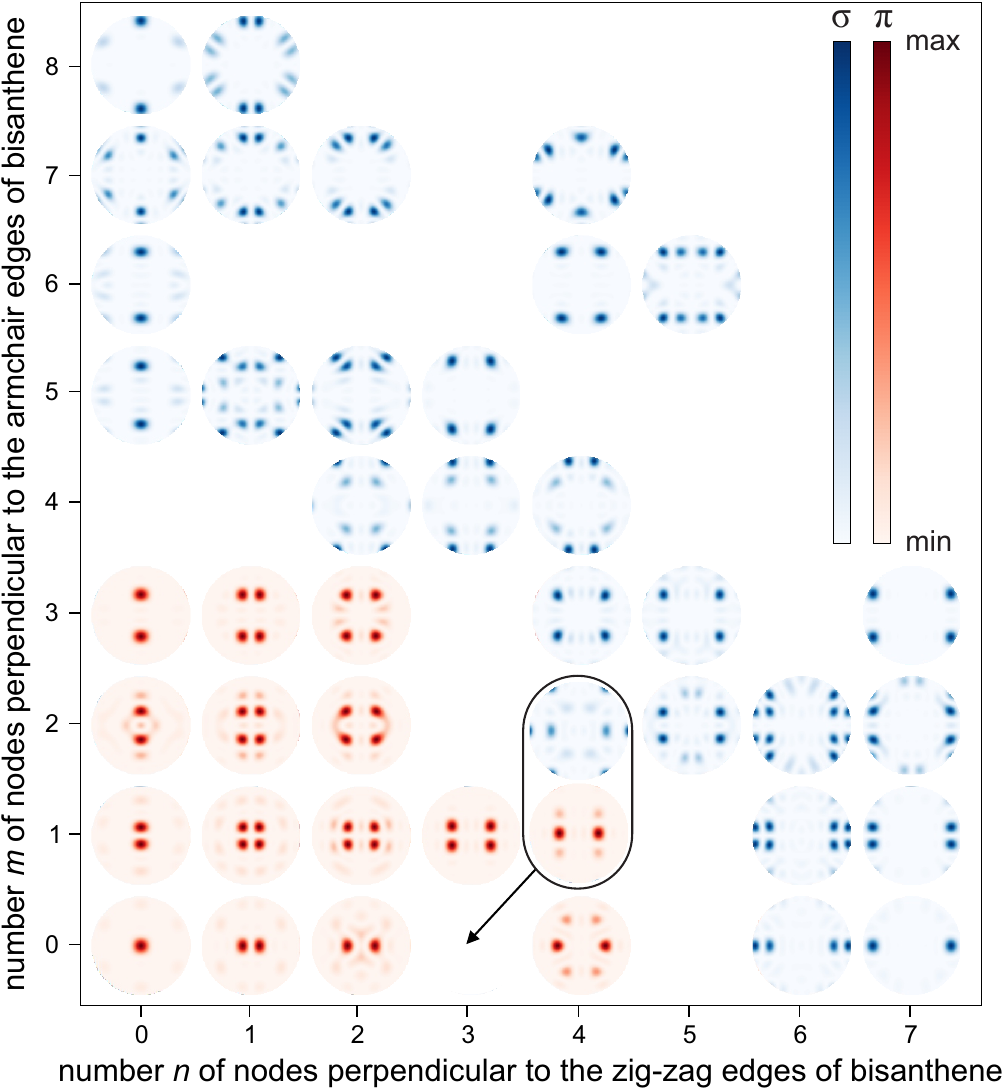}
\end{center}	\caption{\label{fig:momentummaps}  Theoretical momentum maps $|\tilde{\Psi}_{i} (\ve{k})|^2\propto I_i(k_x, k_y)/\left|\ve{A} \cdot \ve{k}\right|^2$ (see Eq.~\ref{eq:PE1}) of the orbitals in Fig.~\ref{fig:orbitals}, obtained by projecting a tomographic cut at constant photoelectron kinetic energy through the three-dimensional Fourier transform of the orbital wave function into the $k_x$, $k_y$ plane (see Eq.~\ref{eq:PE1}). Intensity scales in red and blue are used for $\pi$ and $\sigma$ orbitals, respectively. The momentum maps are positioned in the same way as in Fig.~\ref{fig:orbitals}.}
\end{figure}
As a result, we present the experimental identification of 15 $\pi$ and 23 $\sigma$ Kohn-Sham orbitals of bisanthene, depicted in real and momentum space in Figs.~\ref{fig:orbitals} and  \ref{fig:momentummaps}, respectively, and determine their binding energies for a monolayer of bisanthene on Cu(110). The comparison between the experimental binding energies and those of DFT, using four widely used exchange-correlation potentials, clearly shows that the Kohn-Sham orbitals are good approximators of the Dyson orbitals measured in photoemission, over a wide range of binding energies and \emph{not} just near the Fermi level. This more general demonstration of the physical significance of Kohn-Sham orbitals opens up the possibility of using the rich information contained in these orbitals not only for benchmarking theory, but also for studying the structural, electronic, and chemical properties of molecule/metal interfaces at an unprecedented level of detail.

The remainder of the paper is organized as follows. Section~\ref{sec:exp} contains computational and experimental details on how the orbital-resolved pDOS was obtained from the experimental ARPES data. This includes information on the DFT calculations for isolated bisanthene (used in the deconvolution of the ARPES data) and the bisanthene/Cu(110) interface (to the calculated MOPDOS of which the experimental pDOS is finally compared), as well as sample preparation and a description of the experimental setup and processing of the ARPES data. In particular, in section~\ref{sec:dataanalysis} we present three different numerical approaches to obtain the orbital-projected pDOS from ARPES data, going beyond the original simple fitting approach,\cite{Puschnig2011}, which turned out to be necessary due to the large number of orbitals involved in the deconvolution. The actual results are then presented in section~\ref{sec:results}. There we critically discuss the numerical stability and accuracy of the deconvolution procedure, before the manuscript culminates in a detailed comparison between the experimental and the calculated Kohn-Sham orbital energies for bisanthene/Cu(110). This allows us to evaluate the performance of different exchange-correlation functionals typically used in organic/metal interface calculations on an orbital-by-orbital basis (section~\ref{sec:quantitative_comparison}), and to analyze the mechanisms leading to level alignment at organic/metal interfaces for an energy window of more than 10 eV below the Fermi energy (section~\ref{sec:level_alignment}). After the concluding section~\ref{sec:conclusion}, the appendix~\ref{sec:appendix} contains the complete data set underlying the orbital deconvolution as well as the analysis of the intramolecular band dispersion of the bisanthene orbitals compared to the corresponding band dispersion in the parent graphene layer.

\section{\label{sec:exp}Theoretical and experimental  methods}

\subsection{\label{sec:computational} DFT calculations for the isolated molecule}

The geometry and electronic structure of isolated (gas-phase) bisanthene was calculated in the framework of DFT. We employed the quantum chemistry package NWChem.\cite{Valiev2010}  Four different exchange-correlation functionals were used: (i) the generalized gradient approximation by Perdew, Burke and Ernzerhof (PBE),\cite{Perdew1996} (ii) the range-separated hybrid functional by Heyd, Scuseria and Ernzerhof (HSE),\cite{Heyd2004,Heyd2006} (iii) the global hybrid PBE0 by Perdew, Ernzerhof, and Burke,\cite{Perdew1996a} as well as (iv) the B3LYP hybrid functional by Becke et al. \cite{Becke1993} The resulting orbital energies of isolated bisanthene, which has $D_{2h}$ point group symmetry, are listed in Tables~\ref{tab:piorbitals} and \ref{tab:sigmaorbitals} for $\pi$ and $\sigma$ orbitals, respectively. In addition to the irreducible representations, we also introduced the two integer numbers $n$ and $m$ which count the number of (approximate) nodal surfaces in the respective orbital perpendicular to the $x$ and $y$ directions, \emph{i.e.}, the zig-zag and armchair edges of bisanthene, respectively. In the remainder of the paper, we refer to a specific orbital by its type, $\pi$ or $\sigma$, and its number of nodes $(n,m)$. For instance,  $\pi_{(4,0)}$ refers to the $\pi$ orbital with $n=4$ nodes in $x$ and $m=0$ nodes in $y$ direction, while $\sigma_{(7,3)}$, for instance, denotes the $\sigma$ orbital with $n=7$ and $m=3$. Note that $\pi_{(2,3)}$ and $\pi_{(4,0)}$ are HOMO and LUMO of isolated bisanthene.

\begin{table}
\caption{\label{tab:piorbitals} Energies of the lowest-lying 15 $\pi$ orbitals (in eV) of the isolated (gas-phase) bisanthene molecule (C$_{28}$H$_{14}$) relative to the vacuum level, calculated using four different exchange-correlation functionals (see main text). The corresponding irreducible representations as well as the number of (approximate) nodal surfaces perpendicular to the zig-zag ($n$) and armchair ($m$) edges of bisanthene are also given. The wave functions and momentum maps of the orbitals are displayed in Figs.~\ref{fig:orbitals}  and ~\ref{fig:momentummaps}, respectively.}
\begin{ruledtabular}
\begin{tabular}{ccccccc}
symm.  & $n$ & $m$ & PBE  & HSE & PBE0 & B3LYP \\
 \hline
$b_{1u}$ & 4 & 0 & $-$3.01 & $-$2.84 & $-$2.54 & $-$2.54 \\
$b_{2g}$ & 2 & 3 & $-$4.00 & $-$4.28 & $-$4.62 & $-$4.39 \\
 $a_{u}$ & 1 & 3 & $-$5.41 & $-$5.96 & $-$6.35 & $-$6.05 \\
$b_{1u}$ & 2 & 2 & $-$5.72 & $-$6.28 & $-$6.67 & $-$6.37 \\
 $a_{u}$ & 3 & 1 & $-$5.87 & $-$6.43 & $-$6.82 & $-$6.52 \\
$b_{3g}$ & 3 & 0 & $-$6.03 & $-$6.58 & $-$6.96 & $-$6.66 \\
$b_{2g}$ & 0 & 3 & $-$6.38 & $-$7.08 & $-$7.50 & $-$7.16 \\
$b_{3g}$ & 1 & 2 & $-$6.99 & $-$7.75 & $-$8.17 & $-$7.82 \\
$b_{2g}$ & 2 & 1 & $-$7.15 & $-$7.96 & $-$8.38 & $-$8.02 \\
$b_{1u}$ & 2 & 0 & $-$7.87 & $-$8.74 & $-$9.17 & $-$8.79 \\
$b_{1u}$ & 0 & 2 & $-$7.97 & $-$8.88 & $-$9.31 & $-$8.92 \\
 $a_{u}$ & 1 & 1 & $-$8.53 & $-$9.53 & $-$9.96 & $-$9.55 \\
$b_{3g}$ & 1 & 0 & $-$9.34 & $-$10.42 & $-$10.86 & $-$10.43 \\
$b_{2g}$ & 0 & 1 & $-$9.44 & $-$10.55 & $-$10.99 & $-$10.55 \\
$b_{1u}$ & 0 & 0 & $-$10.31 & $-$11.51 & $-$11.95 & $-$11.49 
\end{tabular}
\end{ruledtabular}
\end{table}

\begin{table}
\caption{\label{tab:sigmaorbitals} Energies of the 27 topmost occupied $\sigma$ orbitals (in eV) of the isolated (gas-phase) bisanthene molecule (C$_{28}$H$_{14}$) relative to the vacuum level, calculated using four different exchange-correlation functionals (see main text). The corresponding irreducible representations as well as the number of (approximate) nodal surfaces perpendicular to the zig-zag ($n$) and armchair ($m$) edges of bisanthene are also given. The wave functions and momentum maps of the orbitals are displayed in Figs.~\ref{fig:orbitals}  and ~\ref{fig:momentummaps}, respectively.}
\begin{ruledtabular}
\begin{tabular}{ccccccc}
symm.  & $n$ & $m$ & PBE  & HSE & PBE0 & B3LYP \\
 \hline
 $a_{g}$ & 0 & 8 & $-$7.24 & $-$8.32 & $-$8.72 & $-$8.43 \\
$b_{1g}$ & 7 & 3 & $-$7.37 & $-$8.48 & $-$8.87 & $-$8.59 \\
$b_{2u}$ & 1 & 8 & $-$8.10 & $-$9.25 & $-$9.66 & $-$9.34 \\
$b_{2u}$ & 7 & 2 & $-$8.22 & $-$9.39 & $-$9.79 & $-$9.48 \\
$b_{3u}$ & 4 & 7 & $-$8.41 & $-$9.58 & $-$9.98 & $-$9.67 \\
$b_{1g}$ & 1 & 7 & $-$8.66 & $-$9.86 & $-$10.26 & $-$9.94 \\
$b_{3u}$ & 0 & 7 & $-$8.74 & $-$9.96 & $-$10.37 & $-$10.04 \\
$b_{3u}$ & 2 & 7 & $-$9.20 & $-$10.45 & $-$10.85 & $-$10.51 \\
 $a_{g}$ & 6 & 2 & $-$9.28 & $-$10.53 & $-$10.94 & $-$10.60 \\
$b_{1g}$ & 7 & 1 & $-$9.40 & $-$10.68 & $-$11.09 & $-$10.73 \\
 $a_{g}$ & 4 & 4 & $-$9.56 & $-$10.82 & $-$11.23 & $-$10.86 \\
 $a_{g}$ & 4 & 6 & $-$9.95 & $-$11.25 & $-$11.65 & $-$11.28 \\
$b_{2u}$ & 7 & 0 & $-$10.03 & $-$11.35 & $-$11.76 & $-$11.38 \\
$b_{1g}$ & 3 & 5 & $-$10.18 & $-$11.51 & $-$11.92 & $-$11.52 \\
$b_{2u}$ & 5 & 6 & $-$10.22 & $-$11.53 & $-$11.95 & $-$11.56 \\
$b_{1g}$ & 5 & 3 & $-$10.54 & $-$11.93 & $-$12.34 & $-$11.95 \\
 $a_{g}$ & 0 & 6 & $-$10.87 & $-$12.26 & $-$12.68 & $-$12.27 \\
$b_{3u}$ & 6 & 1 & $-$10.93 & $-$12.32 & $-$12.74 & $-$12.33 \\
$b_{2u}$ & 3 & 4 & $-$11.24 & $-$12.63 & $-$13.06 & $-$12.62 \\
$b_{3u}$ & 2 & 5 & $-$11.31 & $-$12.71 & $-$13.13 & $-$12.71 \\
 $a_{g}$ & 6 & 0 & $-$11.48 & $-$12.91 & $-$13.33 & $-$12.90 \\
$b_{2u}$ & 5 & 2 & $-$11.65 & $-$13.14 & $-$13.55 & $-$13.12 \\
$b_{1g}$ & 1 & 5 & $-$12.08 & $-$13.53 & $-$13.95 & $-$13.50 \\
$b_{3u}$ & 4 & 3 & $-$12.59 & $-$14.18 & $-$14.60 & $-$14.12 \\
 $a_{g}$ & 2 & 4 & $-$12.79 & $-$14.29 & $-$14.73 & $-$14.23 \\
 $b_{2u}$& 3 & 0 & $-$13.09 & $-$14.62 & $-$15.06 & $-$14.56\\
 $b_{3u}$& 0 & 5 &$-$13.34 & $-$14.96 & $-$15.38 & $-$14.89
\end{tabular}
\end{ruledtabular}
\end{table}

All 15 $\pi$ and 27 $\sigma$ orbitals listed in Tables~\ref{tab:piorbitals} and \ref{tab:sigmaorbitals} are also depicted in Fig.~\ref{fig:orbitals}, arranged according to their $n$ and $m$. For clarity, we use in Fig.~\ref{fig:orbitals} magenta/green and yellow/blue colors to indicate amplitude and phase of the $\pi$ and $\sigma$ orbitals, respectively. The data presented in Tables ~\ref{tab:piorbitals} and \ref{tab:sigmaorbitals} and in Fig.~\ref{fig:orbitals} are also available from the online database using the IDs 415 (PBE), 482 (HSE), 484 (PBE0) and 486 (B3LYP).\cite{Puschnig2020}

\subsection{\label{sec:interface}Interface calculations}

For the DFT calculation of the bisanthene/Cu(110) interface, we employed the repeated-slab approach and the VASP program.\cite{Kresse1996,Kresse1996b,Kresse1999} As described previously,\cite{Yang2019} the Cu(110) substrate was modeled with five atomic layers, a lattice parameter of $a~=~3.61$~{\AA} and a vacuum layer of at least 17~{\AA} between the slabs to avoid spurious electric fields.\cite{Neugebauer1992} The most favorable adsorption site for bisanthene was determined by testing several high-symmetry adsorption sites (hollow, top, short bridge, long bridge) in a local geometry optimization approach, allowing all molecular degrees of freedom and the topmost two Cu-layers to relax until forces were below 0.01\,eV/{\AA}. For these geometry optimizations, we have used the PBE exchange-correlation functional \cite{Perdew1996} with the D3 correction for van der Waals interactions.\cite{Grimme2010} The projector augmented wave (PAW) method \cite{Bloechl1994,Kresse1999} was employed with a plane wave cutoff of 500 eV and a $3 \times 3 \times 1$ Monkhorst-Pack $k$-point grid with a first-order Methfessel-Paxton smearing of 0.2\,eV. 

Based on the relaxed adsorption geometry, which turned out to be the short-bridge site, the electronic structure was analyzed in terms of the calculated MOPDOS. This MOPDOS was calculated by projecting the Kohn-Sham orbitals of the interacting bisanthene/Cu(110) system onto the orbitals of the free-standing bisanthene layer, as described in more detail in a previous publication.\cite{Lueftner2017} Note that for the MOPDOS analysis, we employed the same set of exchange-correlation functionals already used for the gas-phase calculations, that is,  PBE,\cite{Perdew1996} HSE,\cite{Heyd2004,Heyd2006}  PBE0,\cite{Perdew1996a}, and  B3LYP.\cite{Becke1993} The MOPDOS curves shown in this work contain an artificial broadening of 0.1 eV. 

\subsection{\label{sec:maps}Simulation of momentum maps}

In POT, the momentum maps of molecular orbitals, as for example displayed in Fig.~\ref{fig:momentummaps}, serve as fingerprints which can be used for orbital identification in ARPES experiments. In this section, we review the underlying theory and assumptions. In the one-step model of photoemission, the photoelectron intensity $I(k_x,k_y;E_\mathrm{kin})$ is given by Fermi's golden rule \cite{Feibelman1974}
\begin{eqnarray}
\label{eq:Feibelman}
I(k_x,k_y;E_\mathrm{kin}) & \propto & \sum_{i}
                 \left| \langle \Psi_f(k_x,k_y;E_\mathrm{kin}) |
                 \ve{A} \cdot \ve{p} | \Psi_i \rangle \right|^2 \nonumber \\
      & \times & \delta \left(E_i + \Phi + E_\mathrm{kin} - h \nu \right).
\end{eqnarray}
Here, $k_x$ and $k_y$ are the components of the photoemitted electron's wave vector parallel to the surface, which are related to the polar and azimuthal emission angles $\theta$ and~$\phi$ via
\begin{eqnarray}
k_x & = & k \sin \theta \cos \phi  \label{eq:kx} \\
k_y & = & k \sin \theta \sin \phi \label{eq:ky},
\end{eqnarray}
where $k=|\ve{k}|$ is the wave number of the emitted electron, with its kinetic energy being given by $E_\mathrm{kin} = \frac{\hbar^2 k^2}{2m}$, where $\hbar$ is the reduced Planck constant and $m$ is the electron mass. The photoemission intensity of Eq.~\ref{eq:Feibelman} is given by a sum over all transitions from occupied initial states $i$, described by wave functions $\Psi_i$, to a final state $\Psi_f$, characterized by the direction $(\theta,\phi)$ and the kinetic energy of the emitted electron. The $\delta$ function ensures energy conservation, where $\Phi$ denotes the work function, $E_i$ the binding energy of the initial state orbital $i$, and $h \nu$ the photon energy. The transition matrix element in Eq.~\ref{eq:Feibelman} is given in the dipole approximation, where $\ve{p}$ and $\ve{A}$ denote the momentum operator of the electron and the vector potential of the exciting electromagnetic wave, respectively.\cite{Brandstetter2021}

In POT,\cite{Puschnig2017,Brandstetter2021} the final state $\Psi_f$ is commonly approximated by a plane wave (PW). Thereby, the photoemission intensity $I_i$ arising from one particular initial state $i$ turns out to be proportional to the square modulus of the Fourier transform  $\tilde{\Psi}_{i} (\ve{k})$ of the initial state wave function, corrected by the polarization factor $\ve{A} \cdot \ve{k}$,
\begin{equation}
\label{eq:PE1}
I_i(k_x,k_y; E_\mathrm{kin})  \propto \left|\ve{A} \cdot \ve{k}\right|^2  \cdot \left| \tilde{\Psi}_{i} (\ve{k}) \right|^2. 
\end{equation}
Note that the $k_z$ component of $\ve{k}$ in this equation is defined by $(k_x,k_y,E_\mathrm{kin}$). A more detailed discussion regarding the applicability of the PW approximation and its limitations can be found in previous publications.\cite{Puschnig2009a,Haags2022}

We have computed the momentum maps $I_i(k_x,k_y)$ according to Eq.~\ref{eq:PE1} for all bisanthene orbitals depicted in Fig.~\ref{fig:orbitals} and plotted $|\tilde{\Psi}_{i} (\ve{k})|^2\propto I_i(k_x,k_y)/\left|\ve{A} \cdot \ve{k}\right|^2$ in Fig.~\ref{fig:momentummaps} (in the experimental geometry used in this paper, see section \ref{sec:photoemissionexperiments}, the difference between $I_i(k_x,k_y)$ and $|\tilde{\Psi}_{i} (\ve{k})|^2$ is small). It is evident that the nodal patterns of the orbitals, that is the number of nodes $n$ and $m$ along the two principal directions, are also reflected in the momentum maps (more nodes lead to a shorter "wavelength" of the pattern, \textit{i.e.}, with increasing $n,m$ the principal lobes appear at larger $|\ve{k}|$), which therefore can serve as fingerprints for specific orbitals. This correspondence will in fact be used in section~\ref{sec:dataanalysis} to deconvolve ARPES data into an experimental pDOS.

\subsection{\label{sec:samplepreparation}Sample preparation}

Our experiments were performed in ultra-high vacuum ($\sim10^{-10}$\,mbar). The Cu(110) single crystal was cleaned by several cycles of sputtering by Ar$^+$ ions at 1\,keV and subsequent annealing at 800\,K. A film of the 10,10'-dibromo-9,9'-bianthracene  precursor (Sigma-Aldrich, CAS number 121848-75-7) was deposited by evaporation from a molecular evaporator (Kentax GmbH) onto the crystal surface held at room temperature. Subsequently, the sample was annealed at 525\,K to trigger the chemical reaction to bisanthene as described elsewhere.\cite{Yang2019}

\subsection{\label{sec:photoemissionexperiments}Photoemission experiments: 2D band maps and 3D data cubes}

Photoemission experiments were conducted at the insertion device beamline of the Metrology Light Source at the Physikalisch-Technische Bundesanstalt (Berlin, Germany).\cite{Gottwald2019} $p$ polarized ultraviolet light with an incidence angle of 40$^\circ$ to the surface normal was used. In this geometry, the $\ve{A} \parallel \ve{k}$ condition, where $\ve{A}$ is the vector potential of the incident light and $\ve{k}$ the wave vector of the photoelectrons, is approximately fulfilled for most molecular emissions in forward direction, which is favorable for applying the plane-wave approximation for the final state.\cite{Puschnig2017} A toroidal electron energy analyzer was used to detect the photoelectrons. \cite{Broekman2005} 

Two different types of photoemission experiments were conducted.\cite{Broekman2005} First, we measured the photoemission intensity $I$ over a large binding energy window of $\sim13$\,eV for the two emission planes along the principal azimuths of the Cu(110) substrate. In the toroidal electron energy analyzer, this is accomplished by recording emission angles ranging from $-85^\circ$ to $+85^\circ$. After conversion to parallel momenta $k_x$ and $k_y$ and binding energy $E_\mathrm{b}=h\nu - E_\mathrm{kin} -\Phi$, band maps $I_\mathrm{exp}(k_x;E_{\rm b})$ and $I_\mathrm{exp}(k_y;E_{\rm b})$ along the $[001]$ ($x$) and $[1\overline{1}0]$ ($y$) directions of Cu(110) were generated. These band maps were measured with photon energies of $h\nu=35$, $45$ and $57$\,eV. From these band maps, the binding energies of 5 $\pi$ orbitals were determined (Table \ref{tab:experiment_pi_orbitals} and Fig.~\ref{fig:expbandmaps_pi}). Note that in tables and figures throughout this paper we reference the energy axis to the Fermi level. An energy $E<0$, \textit{i.e.} below the Fermi level, thus corresponds to a binding energy $E_\mathrm{b}=-E>0$).

Because of their nodal structures, many molecular orbitals will not be visible along the principal azimuths of the substrate. Thus, in the second type of ARPES experiment, we measured momentum maps: at fixed binding energies $E_{\rm b}$, intensity maps covering the entire half-space above the sample, \emph{i.e.}, with full $\ve{k}_\parallel$, were obtained. In the toroidal electron energy analyzer, the momentum maps are measured by collecting photoelectrons in a given emission plane (polar angle from $-85^\circ$ to $+85^\circ$) and rotating the sample around its normal in 1$^\circ$ steps. In this way, the full photoemission intensity distribution in the $\ve{k}_\parallel$-plane perpendicular to the sample normal can be recorded. This leads to a three-dimensional data cube $I_\mathrm{exp}(k_x, k_y;E_{\rm b})$, \emph{i.e.}, the photoemission intensity  $I_\mathrm{exp}$ as a function of binding energy $E_{\rm b}$ and the momenta $k_x$ and $k_y$ parallel to the surface. 

For the momentum maps we mostly used a photon energy of $57$\,eV, the reason being that the emission signatures of the $\sigma$ orbitals appear at large $k_\parallel$ values and therefore high enough photon energies are needed to allow for a sufficiently large photoemission horizon. For instance, the topmost occupied $\sigma$ orbital ($\sigma_{(0,8)}$, cf.~Fig.~\ref{fig:orbitals}), has its maximum emission intensity at a $k_\parallel$ value of about 2.9~{\AA}$^{-1}$, which corresponds to a final-state kinetic energy of approximately 32\,eV. Taking into account the binding energy of this state and the work function of the sample, the minimal photon energy required to reach this emission is 42\,eV. In fact, to capture the entire maximum of the lobe and to avoid measurement artifacts at grazing emission (\emph{i.e.}, close to 90$^\circ$ polar emission angles) even somewhat ($\sim 10$\,eV) larger photon energies are required.

In this context we note that the so-called universal curve of the electron mean free path has its minimum at $\sim40$\,eV.\cite{Seah1979} Increasing the photon energy far beyond this value will thus result in enhanced emission from the substrate bulk, including the inelastically scattered photoelectrons of Cu. At the same time, the photoemission cross section from the molecular states derived from the atomic orbitals of carbon quickly drops with increasing photon energy. For instance, a factor of 2 decrease in intensity was reported for PTCDA frontier orbitals when increasing the final-state kinetic energy from 25 to 50\,eV.\cite{Weiss2015} Both effects will significantly affect the experimental data, in particular reducing the relative intensity of molecular emissions and thus making it more difficult to identify and analyze the orbitals. Hence, a good compromise to resolve $\sigma$ orbitals is found by choosing a photon energy between $50$ and $60$\,eV.

For technical reasons, the data cube $I_\mathrm{exp}(k_x, k_y;E_{\rm b})$ at $h\nu=57$\,eV was recorded in six overlapping data sets, with energies $E\in$
$[-4.26, -6.01]$\,eV (set~1),   
$[-5.26, -7.01]$\,eV (set 2), 
$[-6.26, -8.01]$\,eV (set 3), 
$[-7.26, -9.01]$\,eV (set 4), 
$[-8.26, -10.01]$\,eV (set 5), 
and $[-9.26, -11.01]$\,eV (set 6). 
Prior to analysis, these separate data sets were merged into a single data cube that spans a binding energy range of 6.75\,eV. Due to experimental conditions, the total intensity within each data set decreases with increasing absolute energy. We averaged all $k$ maps at identical $E_{\rm b}$ in the overlapping sections of two neighboring data sets with relative weights $\in [0,1]$ that depend linearly on $E_{\rm b}$ and become zero at the boundaries of each data set. Thus, at each boundary of an overlapping energy section the averaged map seamlessly merges into the adjacent non-overlapping section. In this way, a single data cube of 226 $k$ maps with energy steps of 30\,meV was created. To remove the intensity modulations mentioned above, we normalized the integrated intensity of each of the 226 maps to a fixed value. Finally, we removed the inhomogeneous  background intensity that is present in the data, maximal in the center of each map and fading out at the edges. We therefore conditioned the data by subtracting a radial intensity profile of the form $I(r) \propto \sqrt{c^2-r^2}$, where $r$ is the distance from the map center and $c\simeq R$, $R$ being the radius of the map. Near the center of the map and in regions where $r > c$, divergent or imaginary values for $I$ were replaced by a constant. Overall, this conditioning resulted in a significantly reduced inhomogeneous background (in the rare cases in which this background subtraction led to negative values, we set them to zero). The resulting data cube, with $E_{\rm b}\in [4.26, 11.01]$\,eV, was employed to identify 11 $\pi$ and 23 $\sigma$ orbitals and determine their binding energies (Tables~\ref{tab:experiment_pi_orbitals} and \ref{tab:experiment_sigma_orbitals} and Figs.~\ref{fig:expdeconvolution_pi_part1}--\ref{fig:expdeconvolution_pi_part3} and Figs. \ref{fig:expdeconvolution_sigma_part1}--\ref{fig:expdeconvolution_sigma_part5}). 
 
In addition, eleven separate data cubes $I_\mathrm{exp}(k_x, k_y;E_{\rm b})$ recorded at $h\nu=35$\,eV in the energy ranges $E\in$
$[+0.60, -0.40]$\,eV (set 7),  
$[+0.05, -0.95]$\,eV (set 8), 
$[-0.60, -1.60]$\,eV (set 9), 
$[-1.70, -2.70]$\,eV (set 10), 
$[-2.20, -3.20]$\,eV (set 11), 
$[-3.60, -4.60]$\,eV (set 12), 
$[-4.30, -5.30]$\,eV (set 13),  
$[-6.10, -7.10]$\,eV (set 14), 
$[-7.60, -8.60]$\,eV (set 15), 
$[-8.70, -9.70]$\,eV (set 16), 
and $[-10.60, -11.60]$\,eV (set 17) 
were used to deconvolve 15 $\pi$ orbitals as will be discussed in Section~\ref{sec:results_band_maps}.

\subsection{\label{sec:dataanalysis}Deconvolution of experimental data cubes}

The principal aim of the data analysis is the deconvolution of the measured data cubes  $I_\mathrm{exp}(k_x, k_y;E_{\rm b})$ into contributions of individual molecular orbitals. 
Mathematically, this requires the minimization of the expression
\begin{equation}
\label{eq:chi2a}
\chi^2 = \sum_{k_x,k_y} \left[ I_\mathrm{exp}(k_x,k_y;E_{\rm b}) - \sum_{i=1}^N w_i(E_{\rm b}) I_i(k_x,k_y) \right]^2 ,
\end{equation}
where the $I_i(k_x,k_y)$ are given by Eq.~\ref{eq:PE1} and the $w_i(E_{\rm b})$ are the corresponding weight functions that are adjusted in the fitting process. The latter provide an orbital-by-orbital decomposition of the experimental data cube into an orbital-resolved pDOS that can be readily compared to the calculated MOPDOS. We employed three different implementations of the fitting process, which will be described in detail below.

\begin{figure}[t]
\begin{center}
	\includegraphics[width=0.9\columnwidth]{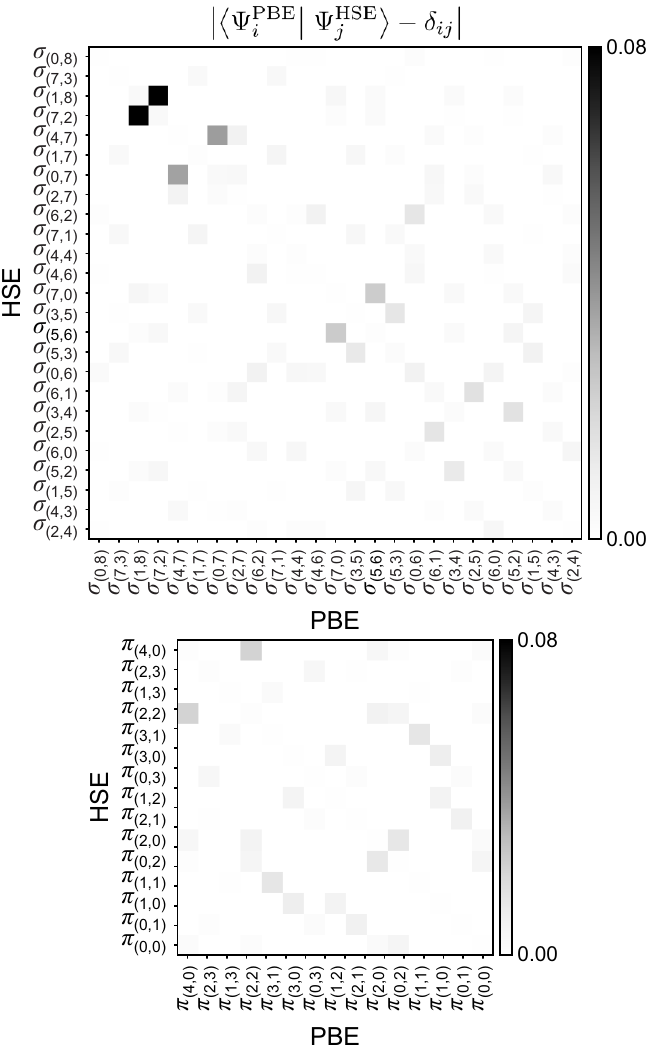}
\end{center}
	\caption{\label{fig:overlaps}  Deviations from the identity matrix of the overlap matrices among $\sigma$ (top) and $\pi$ (bottom) orbitals of bisanthene computed with the PBE and HSE functionals, according to Eq.~\ref{eq:overlap}. } 
\end{figure}

From Eq.~\ref{eq:chi2a} it is clear that the so-obtained experimental pDOS $w_i(E_{\rm b})$ will also depend on the set of simulated momentum maps $I_i(k_x,k_y)$ that are used in the deconvolution procedure. It is therefore important to check how sensitive these computed momentum maps are with respect to the choice of the exchange-correlation  functional. We found that the momentum maps are robust and remain almost unaffected by the choice of the exchange-correlation functional. As an example, this is illustrated for the orbitals obtained from PBE and HSE functionals in Fig.~\ref{fig:overlaps}, in which \begin{equation}
\label{eq:overlap}
\left| \left\langle \Psi_i^\mathrm{PBE} \right| \left. \Psi_j^\mathrm{HSE} \right\rangle  - \delta_{ij} \right|
\end{equation}
is plotted using a gray-scale density map, thus illustrating the degree by which the overlap matrix between PBE and HSE orbitals deviates from the identity matrix. We note that the maximum deviation is about 9\%  in two cases, but much less for the great majority of orbitals. Thus,  also the resulting momentum maps from these two functionals will be essentially indistinguishable. Also note that a similar agreement is found when comparing the PBE orbitals with those of the other two exchange-correlation functionals used in this work.

\subsubsection{\label{sec:linearfitting}Linear fitting}
In the first approach, Eq.~\ref{eq:chi2a} is minimized separately and independently for each $E_{\rm b}$. This results in a linear fitting problem. The most natural ansatz here is the inclusion of all possible $\pi$ and $\sigma$ orbitals in a single fitting process, plus a background that at a given $E_{\rm b}$ is constant for all $k_\parallel$, but may vary with $E_{\rm b}$. However, with a large number of fitting parameters (in the present case 43 for the merged data sets 1$-$6, if all orbitals in the relevant energy range are included), this can lead to the unphysical emulation of patterns that are not due to the molecules, but arise from, \emph{e.g.}, substrate-related photoemission intensities. For this reason, we have also carried out a minimization of Eq.~\ref{eq:chi2a} using only $\sigma$ orbitals, which allowed us to reduce momentum maps from circular disks to rings with $k_\parallel\geq 1.6$~\AA$^{-1}$, thereby cutting out the inner part where many of the substrate-related features (and the $\pi$ orbitals) reside. Moreover, a linear fitting approach for a large number of orbitals is prone to overfitting, where orbitals not present at a certain binding energy are assigned significant negative weights to compensate the aforementioned substrate-related features or even each other. These negative weights, although unphysical in nature, cannot be avoided in what is in essence a matrix inversion. Therefore, we have applied a third version of the linear fitting in which we minimized Eq.~(\ref{eq:chi2a}) for each orbital separately, replacing the sum over $i$ by a single term for the orbital in question. While this approach is not expected to lead to a good overall fit of the experimental data cube, it does allow to identify the binding energy region(s) in which a particular orbital pattern is found in the data. Evidently, this single-orbital fit will not lead to negative weights.

\subsubsection{\label{sec:MonteCarlo}Monte Carlo fitting}
In the linear fits discussed above, the weight functions $w_i(E_{\rm b})$ are unconstrained. While this is computationally efficient and advantageous because it allows for maximum flexibility in the resulting pDOS and thus does not introduce any bias, it can also lead to overly broad and unfocused peaks that are not even positive definite and therefore unphysical. For this reason, we also used an alternative fitting strategy involving regularization. As a regularization constraint, we required that the functions $w_i(E_{\rm b})$ have a single maximum and be positive definite over the entire energy range. In particular, we used a Gaussian   
\begin{equation}
		w_i(E_{\rm b}) = a_i\, e^{-(E_{\rm b}-E_i)^2 / \sigma_i^2}
\label{eq:gaussian}
\end{equation}
for each of the 27 $\sigma$ orbitals $i$ ($\pi$ orbitals were not included in the fit), with peak position $E_i$, width $\sigma_i$, and amplitude $a_i$. We additionally allowed for an asymmetry that is quantified by the skew parameter $s_i$. Since the Gaussians were pre-computed on a linear grid $n = 0,\ldots 800$ with a spacing of $30$\,meV, 
we applied a non-linear scaling to the energy axis by replacing $n$ 
with 
\begin{equation}
		n_{s_i}(n)=400+\frac{(1+s_i)^n-(1+s_i)^{400}}{(1+s_i)^{401}-(1+s_i)^{400}} + \frac{n}{10}-40,
\label{eq:skew}
\end{equation}
in order to introduce the asymmetry. $s_i$ can take values between $\pm 1/1500$ and $\pm1/15$. For positive $s_i$ the grid spacing $n_{s_i}(n)-n_{s_i}(n-1)$ increases with increasing $n$, while it decreases with $n$ if $s_i$ is negative. Around $n=400$ the spacing is always close to 1 and $n_{s_i}(400)=400$. For small absolute $s_i$ the Gaussian in Eq.~\ref{eq:gaussian} is practically symmetric, while it is skewed towards small $n$ for (large) positive $s_i$ and towards large $n$ for (large) negative $s_i$. The last two terms in Eq.~\ref{eq:skew} ensure that also for $s_i\gg 0$ ($s_i\ll 0$) the Gaussian approaches zero for $E_{\mathrm{b}}\ll E_i$ ($E_{\mathrm{b}}\gg E_i$)  .

The introduction of physically motivated boundary conditions regularizes the results of the fit, but this can only be attained at the cost of solving a non-linear optimization problem in a high-dimensional parameter space. We therefore needed to reduce the dimensionality of the data to facilitate fitting with a reasonable computational effort. To this end, we subsampled the experimental maps in both $k_\parallel$ and $E_{\rm b}$, thereby reducing the energy and momentum resolution. This was achieved by reducing the $k_\parallel$ resolution from $234\times 234$ to $47 \times 47$ pixels in the full $I_\mathrm{exp}(k_x,k_y)$ disk, converting the map into a vector, and decreasing the energy resolution from $30$\,meV to $300$\,meV. In addition, all pixels in the vector for which none of the simulated maps had an intensity greater than $5 \%$ of the maximum intensity were excluded from the fit. Finally, every tenth of the remaining pixels in the subsampled vector was used to compute the fit quality $\chi^2$ (Eq.~\ref{eq:chi2a}). In the end, 147 pixels per experimental or simulated momentum map were considered to fit 23 experimental momentum maps with a superposition of 28 component maps (27\,$\sigma$ orbitals, cf.~Fig.~\ref{fig:momentummaps}, plus one additional map for any background that survived the initial background subtraction mentioned in section \ref{sec:photoemissionexperiments} above). Since the weight function of each orbital is characterized by four parameters (peak position, width, amplitude, and skew), the  search had a total of 112 free parameters.

Since the shape of $\chi^2$ on this high-dimensional space is unknown, any optimization designed to locate the nearest minimum in $\chi^2$, such as gradient-based approaches, cannot be applied here. Instead, we used an adaptive Monte Carlo (MC) search in which the search interval for each parameter was adapted to the result of the search.\cite{Ruiz2023} This allows for an unbiased search that still has good convergence properties. 

The MC search was performed in two steps. The first step consisted of 50 rounds (of $N=2\times 10^7$ samples each) and a subsequent gradient descent. In the first round, the search interval for all orbital binding energies $E_i$ was set to $[0.54, 11.04]$\,eV. This broad interval guaranteed an unbiased search. After each round, the search intervals for all 112 parameters were adapted. If a parameter had changed little from its previous best value (found at the end of the previous round, \textit{i.e.} $N$ samples earlier), the search interval was narrowed. If the parameter had changed significantly, the search interval was widened. In both cases, the new search interval was centered on the best parameter value from the previous round. In this way, the effective dimensionality of the searched parameter space decreased over time, as the search intervals of more and more parameters became smaller and smaller, and the respective parameters thus dropped out of the optimization. It should be noted, however, that this procedure principally allows the search to find parameters outside the initially defined search intervals. After the 50 MC search rounds had been completed, the best set of parameters was further optimized in a simple gradient-based procedure to converge to the nearest local minimum. This mainly affected those parameters that had not fully converged even after the 50 search rounds. The complete search procedure up to this point (\textit{i.e.} 50 MC rounds plus subsequent gradient descent) was repeated several times from scratch, in order to check for reproducibility. We found that at this point the energy ordering of the orbitals was indeed recovered reliably for about 50 \% of the orbitals, while the rest still showed a larger scatter in $E_i$. 

In the second step of the MC search we therefore extrapolated the overall energy ordering determined in step 1 across the complete set of orbitals and narrowed the initial search intervals around this prior to $400$\,meV for the $E_i$ of all orbitals. Then, we performed another 50 rounds of  $N=2\times 10^7$ MC samples each, using the same procedure to adapt the search intervals after each round as in step 1. As before, this was followed by a final gradient descent. 

\subsubsection{\label{sec:FMLA}Multilayer averaging}

In a third approach, we recontextualized orbital deconvolution as an example of a more general pattern recognition problem. This allowed us to leverage established methods from the field of computer vision. Specifically, we employed a multilayer averaging (MLA) algorithm \cite{Alemzadeh2006} to supplement the standard orbital deconvolution method \cite{Puschnig2011} and the MC approach discussed above. The MLA method was chosen for its noise resistance. Moreover, its implementation was eased by the fact that in the present case invariance regarding translation, scaling or rotation of the patterns compared to the input data is neither necessary nor desirable.

In its original formulation, the MLA was demonstrated for scalar-valued time-series signals, specifically  in the field of speech recognition. We extended it to 2D momentum maps (full $I_\mathrm{exp}(k_x,k_y)$ disks) with a dependence on the binding energy $E_{\rm b}$ instead of time. At its core, MLA defines a cost function $P_i(E_{\rm b})$ that is evaluated separately for each combination $(i, E_{\rm b})$ of simulated orbital patterns $I_i(k_x,k_y)$ and experimental momentum maps $I_{\mathrm{exp}}(k_x,k_y;E_{\rm b})$. The key idea behind MLA is to divide both the signal $I_{\mathrm{exp}}$ and the orbital patterns $I_i$ into smaller and smaller sub-images and compare them at different levels (layers) of resolution, from rough shapes to fine details. 

In the first step, we defined $N_{\rm L} = \log_2(N_{\rm K})+1$ layers, where $N_{\rm K}$ (an integer to the power of two) is the number of bins on each of the two momentum axes. The layers are labeled by the integer $j\in[0,N_{\rm L}-1]$. Next, we interpolated the simulated orbital patterns $I_i$ and the experimental momentum maps $I_{\rm exp}$ to a common $N_{\rm K} \times N_{\rm K}$ grid of $k$ points. For our case, $N_{\rm K}=64$ was sufficient to yield converged results. Then, at each layer $j$ we divided the orbital patterns $I_i(k_x,k_y)$ and the momentum maps $I_{\mathrm{exp}}(k_x,k_y;E_{\rm b})$ on a regular $2^j \times 2^j$ grid into $4^j$ sub-images, denoted $I_i^{j,\kappa}$ and $I^{j,\kappa}_{\rm exp}(E_{\rm b})$ respectively. Here, the integer $\kappa\in[1,4^j]$ labels the sub-images. Next, we calculated the cost function $P_i^{j,\kappa}(E_{\rm b})$, at each layer $j$, for each such sub-image $\kappa$ and each binding energy $E_{\rm b}$ by taking the absolute value of the difference between the orbital pattern and the experimental momentum map,
\begin{equation}
\label{eq:metric}
    P_i^{j,\kappa}(E_{\rm b}) =  |\langle I_i^{j,\kappa}\rangle_{k_x,k_y} - \langle I^{j,\kappa}(E_{\rm b})\rangle_{k_x,k_y}|.
\end{equation}
Here, $\langle \dots \rangle_{k_x,k_y}$ indicates the average over all pixels of the sub-image $\kappa$. As outlined in the original publication by Alemzadeh et al.,\cite{Alemzadeh2006} one can use different metrics for the calculation of the cost function, for example, the fuzzy multilayer averaging (FMLA) method that uses a trapezoidal-shaped metric. However, in our testing, we found that the MLA method with an absolute difference in Eq.~\ref{eq:metric} performed equally well as FMLA or a mean square difference. 

The total cost function $P_i(E_{\rm })$ for a given orbital pattern $i$ is then given by 
\begin{equation}
    P_i(E_{\rm b}) =  \frac{1}{N_{\rm L}}\sum_{j,\kappa} \frac{1}{4^j}P_i^{j,\kappa}(E_{\rm b}),
\end{equation}
\emph{i.e.}, the average over all layers and sub-images. If the experimental momentum map at a particular binding energy $ E_{\rm b}$ corresponds particularly well to an orbital pattern $i$, then the cost function will have a low value. For the sake of consistency with the linear and MC fits, we consider $-P_i(E_{\rm b})$, such that a maximum corresponds to a good match between the orbital pattern $i$ and the momentum map at that particular $E_{\rm b}$. We note that MLA, like the linear orbital deconvolution method, does not consider cross-correlations between the orbital patterns nor any correlations between subsequent momentum maps on the binding energy axis. The observation of a clear local maximum itself is therefore a robust indicator of the orbital's detection in the corresponding binding energy range.

\begin{figure}[bt]
\begin{center}
	\includegraphics[width=\columnwidth]{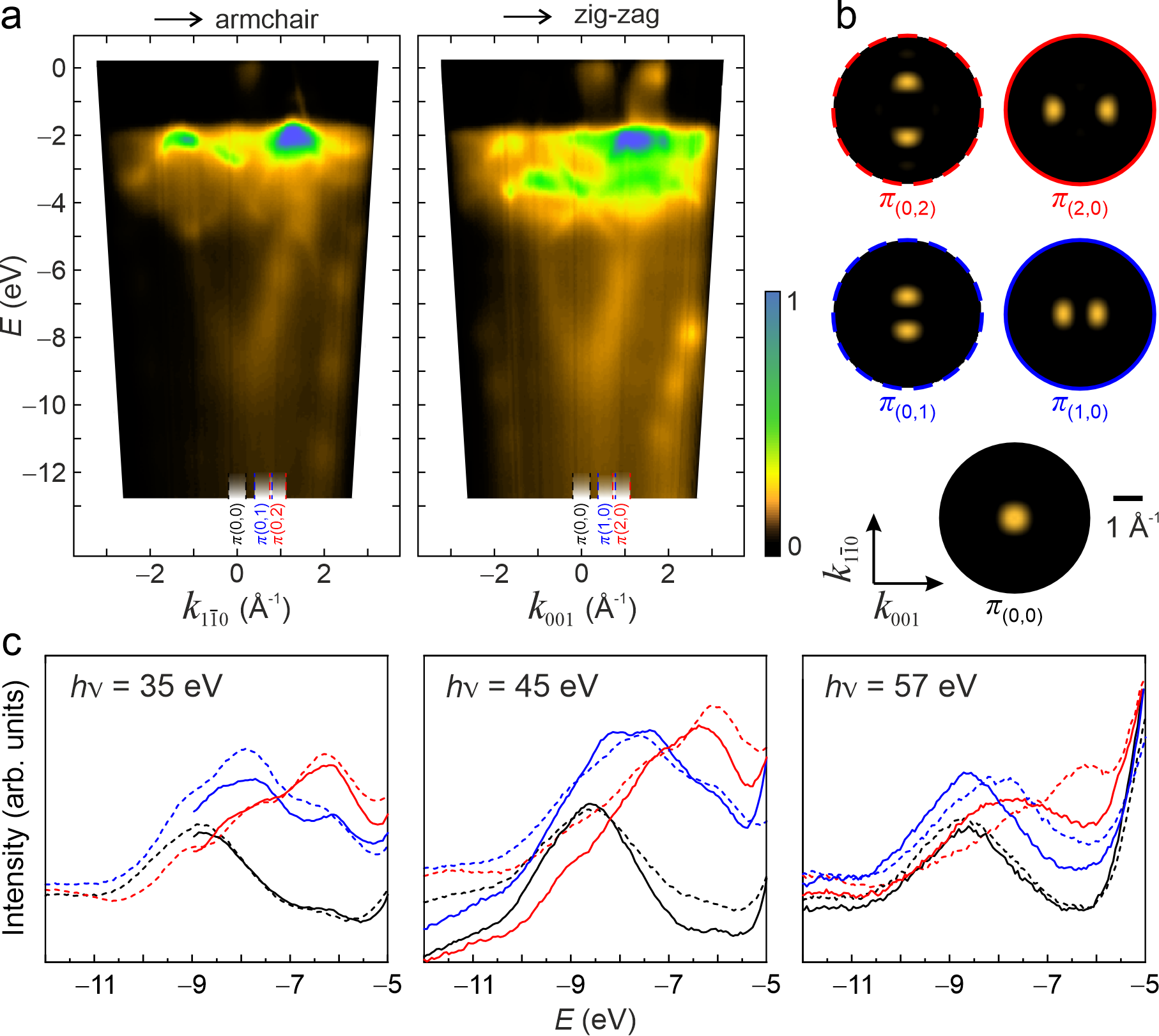}
\end{center}
	\caption{\label{fig:expbandmaps_pi} (a) Experimental band maps along the $[1\overline{1}0]$ and $[001]$ directions of the Cu(110) substrate, measured  with $p$ polarized light with photon energy $h\nu=45$\,eV  and 40$^\circ$ incidence angle relative to the sample normal. (b) Theoretical momentum maps of $\pi_{(0,2)}$, $\pi_{(2,0)}$, $\pi_{(0,1)}$, $\pi_{(1,0)}$, and $\pi_{(0,0)}$ orbitals. (c) Energy distribution curves (EDC) obtained by integrating the photoemission intensity over the $k_\parallel$-ranges marked and labeled in panel a. Positions and widths of these ranges were chosen to coincide with the lobes in the momentum maps in panel b. The colors and line styles of the curves correspond to the circles around the momentum maps in panel b. In addition to the band maps shown in panel a, band maps recorded with the photon energies $h\nu=35$\,eV and $h\nu=57$\,eV were analyzed in an analogous way.}
\end{figure}

\section{\label{sec:results}Results and Discussion}
\subsection{\label{sec:results_band_maps}Analysis of experimental band maps}

Fig.~\ref{fig:expbandmaps_pi}a displays experimental band maps of bisanthene/Cu(110) along the two substrate azimuths $[001]$ ($x$) and $[1\overline{1}0]$ ($y$). The strong intensity between $E\simeq-2$ and $-5$\,eV originates from $d$ states of the Cu substrate. Above the $d$ band, one can clearly discern molecular emissions that are visible only in the $[001]$ direction. Likewise, two distinct bands of emissions below the $d$ band that faintly decompose into separate beads are clearly observable. They originate from the $\pi$ ($k_\parallel< 1.6$~\AA$^{-1}$) and $\sigma$ ($k_\parallel\geq 1.6$~\AA$^{-1}$) orbitals of bisanthene.\cite{Haags2022} The asymmetry of the band maps in Fig.~\ref{fig:expbandmaps_pi}a arises because of the experimental geometry with $p$ polarized light at an angle of incidence of 40$^\circ$ relative to the sample normal. The negative and positive $k_\parallel$ values correspond to backward and forward photoelectron emission (relative to the incidence direction of light). We only used forward-emission data (positive $k_\parallel$) to construct experimental momentum maps (see Section~\ref{sec:results_momentum_maps}). 

The bottom of the $\pi$ band in Fig.~\ref{fig:expbandmaps_pi}a appears at $k_\parallel=0$ and $E\simeq -9$\,eV. Because of its appearance at the $\bar{\Gamma}$ point, this photoemission intensity should correspond to the $\pi_{(0,0)}$ orbital, and we therefore tried to extract the binding energies of the five lowest-lying orbitals from these two bands. According to Table \ref{tab:piorbitals}, apart from $\pi_{(0,0)}$ these are $\pi_{(1,0)}$ and $\pi_{(2,0)}$ along the $[001]$ direction, and $\pi_{(0,1)}$ and $\pi_{(0,2)}$ along $[1\overline{1}0]$. Their patterns are shown in Fig.~\ref{fig:expbandmaps_pi}b. Note that due to its nodal structure, $\pi_{(1,1)}$ does not appear in either of the two $k_\parallel$-space directions. We obtained the energy distribution curves (EDC) in Fig.~\ref{fig:expbandmaps_pi}c by integrating the photoemission intensity over the $k_\parallel$-ranges marked and labeled in Fig.~\ref{fig:expbandmaps_pi}a. Positions and widths of these ranges have been chosen to coincide with the lobes in the momentum maps in Fig.~\ref{fig:expbandmaps_pi}b. The colors and line styles of the curves correspond to the circles around the momentum maps. In addition to the band maps shown in Fig.~\ref{fig:expbandmaps_pi}a, band maps recorded with the photon energies $h\nu=35$\,eV and $h\nu=57$\,eV (not shown) were analyzed in an analogous way. The resulting EDC are also shown in Fig.~\ref{fig:expbandmaps_pi}c. The peak positions of the five lowest $\pi$ orbitals derived in this way are listed in columns 4 to 6 of Table \ref{tab:experiment_pi_orbitals}. The agreement between the $h\nu=35$\,eV and $h\nu=45$\,eV data sets is excellent, only the $h\nu=57$\,eV data set, for which the relative intensity of molecular emissions is lower (see Section \ref{sec:photoemissionexperiments}), deviates substantially, with the exception of the binding energies of $\pi_{(0,0)}$ and $\pi_{(0,2)}$. At any rate, we can unambiguously assign an experimental binding energy of approximately $8.7$\,eV to the bottom of the $\pi$ band ($\pi_{(0,0)}$). 

\begin{figure}[H]
\begin{center}
	\includegraphics[width=\columnwidth]{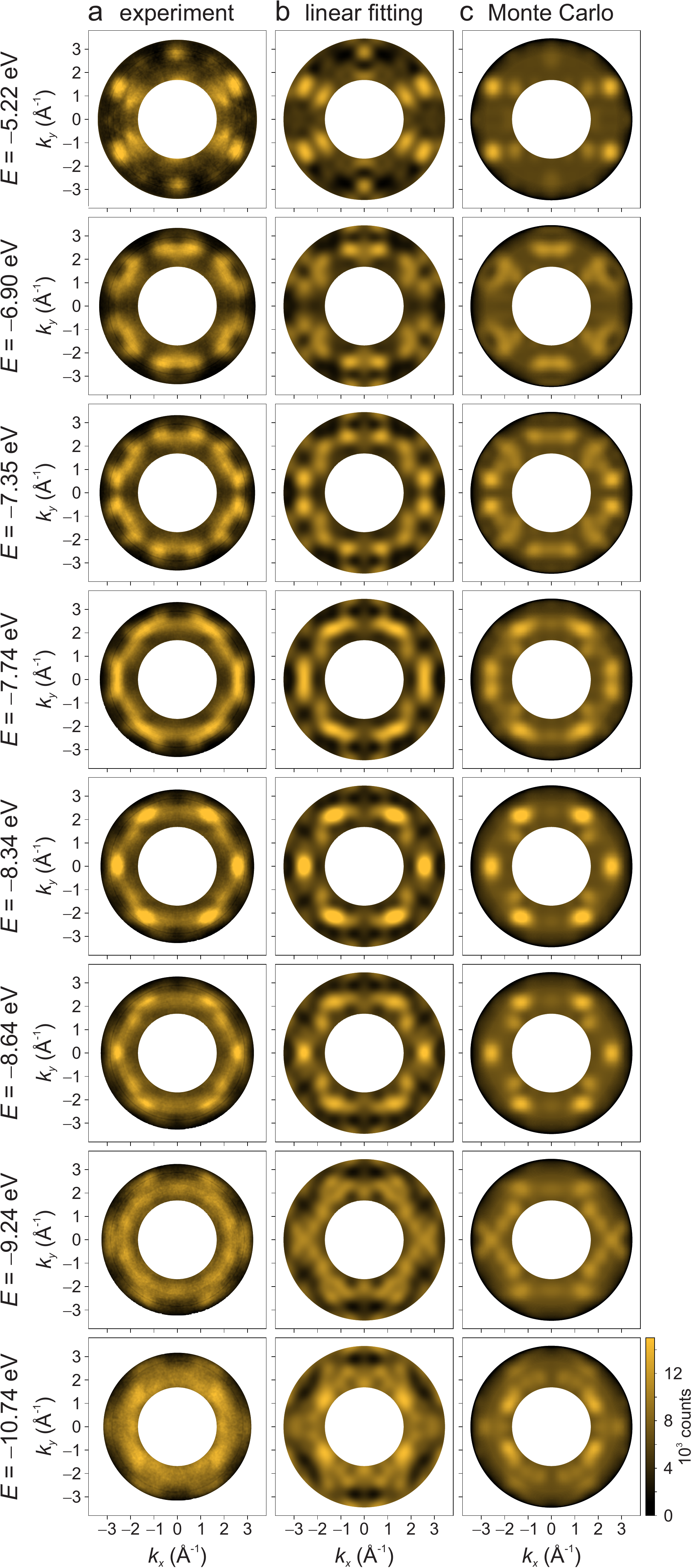} 
\end{center}
	\caption{\label{fig:expkmaps} (a) Momentum maps $I_{\rm exp}(k_x, k_y;E_{\rm b})$ measured at different energies $E=-E_{\rm b}$ (photon energy $h\nu=57$\,eV). To focus on $\sigma$~orbitals, a cut-off of $k_\parallel \geq 1.6$~\AA$^{-1}$ was applied. (b) Simulated maps constructed as a sum over $I_i(k_x,k_y)$, weighted with $w_i(E_{\rm b})$, see Eq.~\ref{eq:chi2a}, as determined from linear fitting with all 27 $\sigma$ orbitals displayed in Fig.~\ref{fig:orbitals}. The peak energies of this fit are essentially the same as the ones for the all-$\pi$-and-$\sigma$ fit listed in Table~\ref{tab:experiment_sigma_orbitals}. The $k_\parallel$ cut-off was applied before the linear fit. (c) As panel b, but with weights from the Monte Carlo fit (Table~\ref{tab:experiment_sigma_orbitals}). The $k_\parallel$ cut-off was applied after the Monte Carlo fit. }
\end{figure}

\begin{table*}[htb]
\caption{\label{tab:experiment_pi_orbitals} Energies $E=-E_\mathrm{b}$ in eV of 15 $\pi$ orbitals of bisanthene/Cu(110), measured relative to $E_{\rm F}$. The first three columns report experimentally determined energies, derived by linear fitting from data cubes $I(k_x, k_y;E_{\rm b})$ measured at $h\nu = 57$\,eV and $h\nu = 35$\,eV, using different fitting strategies as indicated (see Figs.~\ref{fig:expdeconvolution_pi_part1} to \ref{fig:expdeconvolution_pi_part3} for the deconvolved pDOS and main text for details). In some cases, when the deconvolution yielded two maxima of similar strengths, two values are listed in one column. The next three columns report energies derived from band maps recorded with three different photon energies, see Fig.~\ref{fig:expbandmaps_pi}. The seventh column displays the averages of all experimental values (except for those of band maps with $h\nu=45$\,eV and 57 eV, marked with an asterisk, which are excluded because of their relatively low intensities). Columns 8 to 11 give DFT-calculated energies for bisanthene/Cu(110), using four different exchange-correlation functionals (see main text for details). The last column lists the differences between the experimental values and the HSE calculation. For reasons of consistency, we used the values from the fit of the $h\nu= 35$\,eV data cube with all $\pi$ orbitals (column 3) for this purpose, because this is the only data set which offers a value for all 15 $\pi$ orbitals.
} 

\begin{tabular}{ccccccccccc}
\hline
\hline
\multicolumn{1}{c}{}   & \multicolumn{5}{c}{}                                                                                                                                                                    & \multicolumn{4}{c}{} & \multicolumn{1}{c}{}                                                     \\
\multicolumn{1}{c}{}   & \multicolumn{5}{c}{\textbf{Experiment}}                                                                                                                                                                    & \multicolumn{4}{c}{\textbf{Theory}}  & \multicolumn{1}{c}{}                                                    \\

\multicolumn{1}{c}{}   & \multicolumn{1}{c}{}     & \multicolumn{1}{c}{}                         & \multicolumn{1}{c}{}      & \multicolumn{1}{c}{} & \multicolumn{1}{c}{}        & \multicolumn{1}{c}{}    & \multicolumn{1}{c}{}      & \multicolumn{1}{c}{}     & &    \\

\hline

\multicolumn{1}{c}{}   & \multicolumn{3}{c}{momentum maps}                                                                                                   & \multicolumn{1}{c}{band maps}        & \multicolumn{1}{c}{} & \multicolumn{1}{c}{HSE} & \multicolumn{1}{c}{B3LYP} & \multicolumn{1}{c}{PBE0} & PBE & \multicolumn{1}{c}{$\Delta$} \\
\multicolumn{1}{c}{}   & \multicolumn{3}{c}{linear fitting}                                                                                                  & \multicolumn{1}{c}{$h\nu$ = 35 / 45 / 57 eV}                 & \multicolumn{1}{c}{}        & \multicolumn{1}{c}{}    & \multicolumn{1}{c}{}      & \multicolumn{1}{c}{}     &    &  exp. all \\

\cline{2-4}

\multicolumn{1}{c}{}   & \multicolumn{1}{c}{single } & \multicolumn{1}{c}{all $\pi$ and $\sigma$ } & \multicolumn{1}{c}{all $\pi$ } & \multicolumn{1}{c}{}                 & \multicolumn{1}{c}{}        & \multicolumn{1}{c}{}    & \multicolumn{1}{c}{}      & \multicolumn{1}{c}{}     &  &$\pi$ orbitals   \\

\multicolumn{1}{c}{}   & \multicolumn{1}{c}{ orbital} & \multicolumn{1}{c}{orbitals} & \multicolumn{1}{c}{orbitals} & \multicolumn{1}{c}{}                 & \multicolumn{1}{c}{\cellcolor{Gray}average}        & \multicolumn{1}{c}{}    & \multicolumn{1}{c}{}      & \multicolumn{1}{c}{}     &  &$-$   \\

\multicolumn{1}{c}{}   & \multicolumn{1}{c}{$h\nu$ = 57 eV}     & \multicolumn{1}{c}{$h\nu$ = 57 eV}                         & \multicolumn{1}{c}{$h\nu$ = 35 eV}      & \multicolumn{1}{c}{} & \multicolumn{1}{c}{}        & \multicolumn{1}{c}{}    & \multicolumn{1}{c}{}      & \multicolumn{1}{c}{}     &  & HSE   \\

\hline

\multicolumn{1}{c}{}   & \multicolumn{1}{c}{}     & \multicolumn{1}{c}{}                         & \multicolumn{1}{c}{}      & \multicolumn{1}{c}{} & \multicolumn{1}{c}{}        & \multicolumn{1}{c}{}    & \multicolumn{1}{c}{}      & \multicolumn{1}{c}{}     &  &   \\

\multicolumn{1}{c}{$\pi_{(4,0)}$ $b_{1u}$} & 
\multicolumn{1}{c}{} & 
\multicolumn{1}{c}{} & 
\multicolumn{1}{c}{$-$0.50} & 
\multicolumn{1}{c}{} & 
\multicolumn{1}{c}{\cellcolor{Gray}$-$0.50} & 
\multicolumn{1}{c}{$-$0.46} & 
\multicolumn{1}{c}{$-$0.64} & 
\multicolumn{1}{c}{$-$0.72} & 
$-$0.04 & 
$-$0.04 \\

\multicolumn{1}{c}{$\pi_{(2,3)}$ $b_{2g}$} & 
\multicolumn{1}{c}{} & 
\multicolumn{1}{c}{} & 
\multicolumn{1}{c}{$-$1.12} & 
\multicolumn{1}{c}{} & 
\multicolumn{1}{c}{\cellcolor{Gray}$-$1.12} & 
\multicolumn{1}{c}{$-$1.28} & 
\multicolumn{1}{c}{$-$1.48} & 
\multicolumn{1}{c}{$-$1.61} & 
$-$0.79 & 
$+$0.16 \\

\multicolumn{1}{c}{$\pi_{(1,3)}$ $a_{u}$} & 
\multicolumn{1}{c}{} & 
\multicolumn{1}{c}{} & 
\multicolumn{1}{c}{$-$2.13} & 
\multicolumn{1}{c}{} & 
\multicolumn{1}{c}{\cellcolor{Gray}$-$2.13} & 
\multicolumn{1}{c}{$-$2.27} & 
\multicolumn{1}{c}{$-$2.44} & 
\multicolumn{1}{c}{$-$2.62} & 
$-$2.90 & 
$+$0.14 \\

\multicolumn{1}{c}{$\pi_{(2,2)}$ $b_{1u}$} & 
\multicolumn{1}{c}{} & 
\multicolumn{1}{c}{} & 
\multicolumn{1}{c}{$-$2.41} & 
\multicolumn{1}{c}{} & 
\multicolumn{1}{c}{\cellcolor{Gray}$-$2.41} & 
\multicolumn{1}{c}{$-$2.58} & 
\multicolumn{1}{c}{$-$2.75} & 
\multicolumn{1}{c}{$-$2.94} & 
$-$3.45 & 
$+$0.17 \\

\multicolumn{1}{c}{$\pi_{(3,1)}$ $a_{u}$} & 
\multicolumn{1}{c}{$-$4.64} & 
\multicolumn{1}{c}{$-$4.64} & 
\multicolumn{1}{c}{$-$4.54} & 
\multicolumn{1}{c}{} & 
\multicolumn{1}{c}{\cellcolor{Gray}$-$4.61} & 
\multicolumn{1}{c}{$-$4.57} & 
\multicolumn{1}{c}{$-$4.69} & 
\multicolumn{1}{c}{$-$4.94} & 
$-$3.60& 
$+$0.03  \\

\multicolumn{1}{c}{$\pi_{(3,0)}$ $b_{3g}$} & 
\multicolumn{1}{c}{$-$4.69} & 
\multicolumn{1}{c}{$-$4.67} & 
\multicolumn{1}{c}{$-$4.69} & 
\multicolumn{1}{c}{} & 
\multicolumn{1}{c}{\cellcolor{Gray}$-$4.68} & 
\multicolumn{1}{c}{$-$4.57} & 
\multicolumn{1}{c}{$-$4.69} & 
\multicolumn{1}{c}{$-$4.95} & 
$-$3.69 & 
$-$0.12 \\

\multicolumn{1}{c}{$\pi_{(0,3)}$ $b_{2g}$} & 
\multicolumn{1}{c}{$-$5.30} & 
\multicolumn{1}{c}{$-$5.19} & 
\multicolumn{1}{c}{$-$4.86} & 
\multicolumn{1}{c}{} & 
\multicolumn{1}{c}{\cellcolor{Gray}$-$5.12} & 
\multicolumn{1}{c}{$-$4.37} & 
\multicolumn{1}{c}{$-$4.51} & 
\multicolumn{1}{c}{$-$4.74} & 
$-$4.11& 
$-$0.49  \\

\multicolumn{1}{c}{$\pi_{(1,2)}$ $b_{3g}$} & 
\multicolumn{1}{c}{$-$6.80 / $-$5.79} &
\multicolumn{1}{c}{$-$6.43} &
\multicolumn{1}{c}{$-$6.26} & 
\multicolumn{1}{c}{} & 
\multicolumn{1}{c}{\cellcolor{Gray}$-$6.32} & 
\multicolumn{1}{c}{$-$5.31} & 
\multicolumn{1}{c}{$-$5.35} & 
\multicolumn{1}{c}{$-$5.69} & 
$-$4.31 & 
$-$0.95 \\

\multicolumn{1}{c}{$\pi_{(2,1)}$ $b_{2g}$} & 
\multicolumn{1}{c}{$-$6.98 / $-$5.73} &
\multicolumn{1}{c}{$-$6.83 / $-$6.03} &
\multicolumn{1}{c}{$-$6.27} & 
\multicolumn{1}{c}{} & 
\multicolumn{1}{c}{\cellcolor{Gray}$-$6.37} & 
\multicolumn{1}{c}{$-$5.82} & 
\multicolumn{1}{c}{$-$5.93} & 
\multicolumn{1}{c}{$-$6.20} & 
$-$4.88 & 
$-$0.45 \\

\multicolumn{1}{c}{$\pi_{(2,0)}$ $b_{1u}$} & 
\multicolumn{1}{c}{$-$7.00 / $-$5.72} & 
\multicolumn{1}{c}{$-$7.71 / $-$5.49} &
\multicolumn{1}{c}{$-$6.73} & 
\multicolumn{1}{c}{$-$6.18 / $-$6.39* / $-$7.64*} & 
\multicolumn{1}{c}{\cellcolor{Gray}$-$6.47} & 
\multicolumn{1}{c}{$-$6.66} & 
\multicolumn{1}{c}{$-$6.76} & 
\multicolumn{1}{c}{$-$7.05} & 
$-$5.73& 
$-$0.07  \\

\multicolumn{1}{c}{$\pi_{(0,2)}$ $b_{1u}$} & 
\multicolumn{1}{c}{$-$6.86} &
\multicolumn{1}{c}{$-$6.98} &
\multicolumn{1}{c}{$-$6.58} & 
\multicolumn{1}{c}{$-$6.28 / $-$6.16* / $-$6.25*} & 
\multicolumn{1}{c}{\cellcolor{Gray}$-$6.68} & 
\multicolumn{1}{c}{$-$6.36} & 
\multicolumn{1}{c}{$-$6.45} & 
\multicolumn{1}{c}{$-$6.75} & 
$-$5.41& 
$-$0.22  \\

\multicolumn{1}{c}{$\pi_{(1,1)}$ $a_{u}$} & 
\multicolumn{1}{c}{$-$8.71 / $-$7.37} &
\multicolumn{1}{c}{$-$8.17} &
\multicolumn{1}{c}{$-$6.93} & 
\multicolumn{1}{c}{} & 
\multicolumn{1}{c}{\cellcolor{Gray}$-$7.80} & 
\multicolumn{1}{c}{$-$6.94} & 
\multicolumn{1}{c}{$-$7.03} & 
\multicolumn{1}{c}{$-$7.33} & 
$-$5.84 & 
$+$0.01 \\

\multicolumn{1}{c}{$\pi_{(1,0)}$ $b_{3g}$} & 
\multicolumn{1}{c}{$-$8.76} &
\multicolumn{1}{c}{$-$8.87} &
\multicolumn{1}{c}{$-$7.90} &
\multicolumn{1}{c}{$-$7.76 / $-$7.40* / $-$8.71*} & 
\multicolumn{1}{c}{\cellcolor{Gray}$-$8.32} & 
\multicolumn{1}{c}{$-$7.70} & 
\multicolumn{1}{c}{$-$7.79} & 
\multicolumn{1}{c}{$-$8.08} & 
$-$6.57 & 
$-$0.20 \\

\multicolumn{1}{c}{$\pi_{(0,1)}$ $b_{2g}$} & 
\multicolumn{1}{c}{$-$8.77} &
\multicolumn{1}{c}{$-$8.84} &
\multicolumn{1}{c}{$-$7.96} & 
\multicolumn{1}{c}{$-$7.88 / $-$7.63* / $-$8.20*} & 
\multicolumn{1}{c}{\cellcolor{Gray}$-$8.36} & 
\multicolumn{1}{c}{$-$7.88} & 
\multicolumn{1}{c}{$-$7.91} & 
\multicolumn{1}{c}{$-$8.27} & 
$-$6.78 & 
$-$0.08 \\

\multicolumn{1}{c}{$\pi_{(0,0)}$ $b_{1u}$} & 
\multicolumn{1}{c}{$-$8.80} &
\multicolumn{1}{c}{$-$8.84} &
\multicolumn{1}{c}{$-$9.04} &
\multicolumn{1}{c}{$-$8.78 / $-$8.63* / $-$8.66*} & 
\multicolumn{1}{c}{\cellcolor{Gray}$-$8.87} & 
\multicolumn{1}{c}{$-$8.95} & 
\multicolumn{1}{c}{$-$8.80} & 
\multicolumn{1}{c}{$-$9.35} & 
$-$7.63 & 
$-$0.09 \\

\multicolumn{1}{c}{}   & \multicolumn{1}{c}{}     & \multicolumn{1}{c}{}                         & \multicolumn{1}{c}{}      & \multicolumn{1}{c}{} & \multicolumn{1}{c}{}        & \multicolumn{1}{c}{}    & \multicolumn{1}{c}{}      & \multicolumn{1}{c}{}     &     \\

\hline
\hline

\end{tabular}
\end{table*}

\subsection{\label{sec:results_momentum_maps}Deconvolution of experimental momentum maps}
 
Above an energy of approximately $-6$\,eV, we enter the $d$ band of Cu (Fig.~\ref{fig:expbandmaps_pi}a), and therefore the fitting of band maps in order to determine binding energies of higher orbitals becomes impossible. In this range, it is necessary to use momentum maps instead, as they display distinctive fingerprints of individual orbitals that contrast against the background of the $d$ band emissions.

Fig.~\ref{fig:expkmaps}a displays slices through the $h\nu=57$\,eV data cube at selected  energies. The full data cube is available as a movie in the supplemental material. For clarity, we restricted the maps to the range of the $\sigma$ orbitals, $k_\parallel\geq 1.6$~\AA$^{-1}$. One observes a strong energy dependence of the momentum pattern. We attribute this to the contribution of different orbitals at each of the corresponding  binding energies. The fitting results to be discussed below corroborate this inference.

Momentum maps for bisanthene/Cu(110) have been analyzed previously in the low binding energy range, where they have revealed fingerprints of $\pi_{(4,0)}$ (filled LUMO), $\pi_{(2,3)}$ (HOMO) and $\pi_{(1,3)}$ (HOMO$-$1),\cite{Yang2019} and for selected energies below the Cu $d$ band, where they have shown the emissions around $-5.2$~eV to originate from the $\sigma_{(7,3)}$, $\sigma_{(0,8)}$ and $\pi_{(0,3)}$ orbitals.\cite{Haags2022} Here, we attempt a more complete analysis of $\sigma$ and $\pi$~orbitals below the $d$ band. Specifically, we aim for identifying the binding energies of as many $\sigma$ and $\pi$ orbitals as possible in the broad energy range $\sim -4$ to $-11$\,eV. To this end, we applied the orbital deconvolution procedures \cite{Puschnig2011,Puschnig2017,Brandstetter2021} described in section \ref{sec:dataanalysis}. These provided an orbital-by-orbital decomposition of the experimental data cube into orbital-resolved pDOS that can be readily compared to the computed MOPDOS. The complete deconvolution results for 15 $\pi$ and 27 $\sigma$ orbitals are shown in the appendix in  Figs.~\ref{fig:expdeconvolution_pi_part1} to \ref{fig:expdeconvolution_pi_part3} and \ref{fig:expdeconvolution_sigma_part1} to \ref{fig:expdeconvolution_sigma_part5}, respectively.  For most orbitals, the pDOS from several distinct deconvolution procedures are plotted. In general, a good agreement between the different deconvolution methods was found. 

We now turn to a detailed discussion of the individual fits. Generally speaking, the obtained experimental pDOS exhibit rather broad structures, which nevertheless in most cases display a clear maximum that often coincides for the different methodologies, such that a clear binding energy can be assigned ($>90$ \% of the 42 considered orbitals). Only for $<10$\% of the orbitals a binding energy cannot be assigned. A general problem appeared between the energies of approximately $-5.5$ to $-3$\,eV. Here many of the fits exhibit spurious peaks which occur because they try to emulate substrate-related structures deriving from the $d$ band of Cu. It is also not clear to which extent the broadness of the fitted pDOS reflects the true broadening of the orbitals, \emph{e.g.}, due to hybridization. Here it must be noted that the energy resolution of the analyzer is only $250$\,meV.\cite{Emtsev2006}

\begin{table*}[ht]
\centering
\caption{\label{tab:experiment_sigma_orbitals} Energies $E=-E_\mathrm{b}$ in eV of 27 $\sigma$ orbitals of bisanthene/Cu(110), measured relative to $E_{\rm F}$. The first four columns report experimentally determined energies, derived from the data cube $I( k_x, k_y;E_{\rm b})$ measured at $h\nu = 57$\,eV using four different fitting strategies (see Fig.~\ref{fig:expdeconvolution_sigma_part1} to \ref{fig:expdeconvolution_sigma_part5} for the deconvolved pDOS and main text for details). Note that the fit results for the all-$\sigma$ and all-$\pi$-and-$\sigma$ fits are essentially identical and therefore only the latter have been included in the table. The fifth column displays the average of all experimental values (except for those marked with an asterisk, which are excluded from the average because of the poor quality of their fitted pDOS). The sixth column classifies the overall fit quality in four categories from 1 (best) to 4 (fit essentially fails) (see text). The next four columns give DFT-calculated energies for bisanthene/Cu(110), using four different exchange-correlation functionals (see main text for details). The last column lists the differences between the experimental average and the HSE calculation.}
\begin{tabular}{cccccccccccc}
\hline
\hline

\multicolumn{1}{c}{} & \multicolumn{6}{c}{}                                                                                                                              & \multicolumn{4}{c}{}        &
\multicolumn{1}{c}{}\\

\multicolumn{1}{c}{} & \multicolumn{6}{c}{\textbf{Experiment}}                                                                                                                              & \multicolumn{4}{c}{\textbf{Theory}}  
& 
\multicolumn{1}{c}{}\\

\multicolumn{1}{c}{}      & \multicolumn{1}{c}{} & \multicolumn{1}{c}{} & \multicolumn{1}{c}{}     & \multicolumn{1}{c}{}            & \multicolumn{1}{c}{}        & \multicolumn{1}{c}{}    & \multicolumn{1}{c}{}      & \multicolumn{1}{c}{}     &  
\multicolumn{1}{c}{} &\multicolumn{1}{c}{} &\\ 

\hline

\multicolumn{1}{c}{}      & \multicolumn{6}{c}{momentum maps, $h\nu$ = 57 eV}                                                                                                                       & \multicolumn{1}{c}{HSE} & \multicolumn{1}{c}{B3LYP} & \multicolumn{1}{c}{PBE0} & PBE    & \multicolumn{1}{c}{$\Delta$}                                                               \\

\multicolumn{1}{c}{}      & \multicolumn{2}{c}{linear fitting}                                     & \multicolumn{1}{c}{} & \multicolumn{1}{c}{} & \multicolumn{1}{c}{\cellcolor{Gray}exp.}&\multicolumn{1}{c}{} & \multicolumn{4}{c}{} & \multicolumn{1}{c}{exp. averaged }\\

\cline{2-3}
\multicolumn{1}{c}{}      & \multicolumn{1}{c}{single } & \multicolumn{1}{c}{all $\pi$ and $\sigma$ } & \multicolumn{1}{c}{MLA}     & \multicolumn{1}{c}{Monte}            & \multicolumn{1}{c}{\cellcolor{Gray}aver-}  & \multicolumn{1}{c}{category}      & \multicolumn{1}{c}{}    & \multicolumn{1}{c}{}      & \multicolumn{1}{c}{}     &  &
\multicolumn{1}{c}{$-$ } \\ 
\multicolumn{1}{c}{}      & \multicolumn{1}{c}{ $\sigma$ orbital} & \multicolumn{1}{c}{orbitals} & \multicolumn{1}{c}{}     & \multicolumn{1}{c}{Carlo}            & \multicolumn{1}{c}{\cellcolor{Gray}age}& 
\multicolumn{1}{c}{(fit quality)}      &  \multicolumn{1}{c}{}&\multicolumn{1}{c}{}    & \multicolumn{1}{c}{}           &  &
\multicolumn{1}{c}{ HSE} \\

\hline

\multicolumn{1}{c}{}& \multicolumn{1}{c}{} & 
                \multicolumn{1}{c}{} & 
                \multicolumn{1}{c}{}  &  
                \multicolumn{1}{c}{} & 
                \multicolumn{1}{c}{}&
                \multicolumn{1}{c}{} & 
                \multicolumn{1}{c}{} & 
                \multicolumn{1}{c}{}  &  
                \multicolumn{1}{c}{} &  
                \multicolumn{1}{c}{} & 
                \multicolumn{1}{c}{}\\ 

\multicolumn{1}{c}{$\sigma_{(0,8)}$ $a_g$ }& \multicolumn{1}{c}{$-$5.28} & 
                \multicolumn{1}{c}{$-$5.25} & 
                \multicolumn{1}{c}{$-$5.25}  &  
                \multicolumn{1}{c}{$-$5.35} & 
                \multicolumn{1}{c}{\cellcolor{Gray}$-$5.28}&
                \multicolumn{1}{c}{1} & 
                \multicolumn{1}{c}{$-$5.17} & 
                \multicolumn{1}{c}{$-$5.44} & 
                \multicolumn{1}{c}{$-$5.53}  &  
                \multicolumn{1}{c}{$-$4.12} & $-$0.11\\ 
\multicolumn{1}{c}{$\sigma_{(7,3)}$ $b_{1g}$}& 
                \multicolumn{1}{c}{$-$5.26} & 
                \multicolumn{1}{c}{$-$5.26} & 
                \multicolumn{1}{c}{$-$5.24}  &  
                \multicolumn{1}{c}{$-$5.38} & 
                \multicolumn{1}{c}{\cellcolor{Gray}$-$5.29} & 
                \multicolumn{1}{c}{1} &
                \multicolumn{1}{c}{$-$5.22} & 
                \multicolumn{1}{c}{$-$5.47} & 
                \multicolumn{1}{c}{$-$5.58}  &  
                \multicolumn{1}{c}{$-$4.14} &
                $-$0.07\\                
\multicolumn{1}{c}{$\sigma_{(1,8)}$ $b_{2u}$}& 
                \multicolumn{1}{c}{$-$6.22} & 
                \multicolumn{1}{c}{$-$6.01} & 
                \multicolumn{1}{c}{$-$6.24}  &  
                \multicolumn{1}{c}{} & 
                \multicolumn{1}{c}{\cellcolor{Gray}$-$6.16}&
                \multicolumn{1}{c}{2} &
                \multicolumn{1}{c}{$-$6.05} & 
                \multicolumn{1}{c}{$-$6.28} &
                \multicolumn{1}{c}{$-$6.42}  & 
                \multicolumn{1}{c}{$-$4.91} & 
                $-$0.11\\ 
\multicolumn{1}{c}{$\sigma_{(7,2)}$ $b_{2u}$}&  \multicolumn{1}{c}{$-$6.20} & 
                \multicolumn{1}{c}{$-$6.17} & 
                \multicolumn{1}{c}{$-$6.22}  &  
                \multicolumn{1}{c}{$-$6.19} & 
                                       \multicolumn{1}{c}{ \cellcolor{Gray}$-$6.20}   &   
                                       \multicolumn{1}{c}{1} &
                \multicolumn{1}{c}{$-$6.10} & 
                \multicolumn{1}{c}{$-$6.32} & 
                \multicolumn{1}{c}{$-$6.46}  &  
                \multicolumn{1}{c}{$-$4.96}& 
                $-$0.10 \\ 
\multicolumn{1}{c}{$\sigma_{(4,7)}$ $b_{3u}$}&  \multicolumn{1}{c}{$-$6.31} &              
                \multicolumn{1}{c}{$-$6.39} &
                \multicolumn{1}{c}{$-$6.27}  &  
                \multicolumn{1}{c}{$-$6.43} &                                         
                                        \multicolumn{1}{c}{\cellcolor{Gray}$-$6.35} &
                                        \multicolumn{1}{c}{1} &
                \multicolumn{1}{c}{$-$6.26} & 
                \multicolumn{1}{c}{$-$6.48} & 
                \multicolumn{1}{c}{$-$6.63}  &  
                \multicolumn{1}{c}{$-$5.12} & 
                $-$0.09\\ 
\multicolumn{1}{c}{$\sigma_{(1,7)}$ $b_{1g}$}&  
\multicolumn{1}{c}{$-$6.50} &
                \multicolumn{1}{c}{$-$6.60} & 
                \multicolumn{1}{c}{$-$6.30}  &  
                \multicolumn{1}{c}{$-$6.61} & 
 
 \multicolumn{1}{c}{\cellcolor{Gray}$-$6.50} &
 \multicolumn{1}{c}{1} &
                \multicolumn{1}{c}{$-$6.43} & 
                \multicolumn{1}{c}{$-$6.65} & 
                \multicolumn{1}{c}{$-$6.80}  &  
                \multicolumn{1}{c}{$-$5.27}& 
                $-$0.07 \\ 
\multicolumn{1}{c}{$\sigma_{(0,7)}$ $b_{3u}$}&  \multicolumn{1}{c}{$-$6.38} & 
                \multicolumn{1}{c}{$-$6.82} & 
                \multicolumn{1}{c}{$-$6.26}  &  
                \multicolumn{1}{c}{$-$6.76} & 
                                      
                                    \multicolumn{1}{c}{\cellcolor{Gray}$-$6.56} &
                                    \multicolumn{1}{c}{2} &
                \multicolumn{1}{c}{$-$6.56} & 
                \multicolumn{1}{c}{$-$6.77} & 
                \multicolumn{1}{c}{$-$6.93}  &  
                \multicolumn{1}{c}{$-$5.36}& 
                $+$0.00 \\ 
\multicolumn{1}{c}{$\sigma_{(2,7)}$ $b_{3u}$}& %
\multicolumn{1}{c}{$-$8.09} &
                \multicolumn{1}{c}{$-$7.25} & 
                
               \multicolumn{1}{c}{$-$7.73} &
                \multicolumn{1}{c}{$-$7.39} & 
                                 
                                  \multicolumn{1}{c}{\cellcolor{Gray}$-$7.62} &
                                  \multicolumn{1}{c}{2} &
                \multicolumn{1}{c}{$-$7.00} & 
                \multicolumn{1}{c}{$-$7.19} & 
                \multicolumn{1}{c}{$-$7.36}  &  
                \multicolumn{1}{c}{$-$5.79}& 
                $-$0.62 \\ 
\multicolumn{1}{c}{$\sigma_{(6,2)}$ $a_{g}$}& 
\multicolumn{1}{c}{$-$7.25 / $-$6.46* } &
                \multicolumn{1}{c}{$-$7.28} & 
                 
                \multicolumn{1}{c}{$-$7.22 / $-$6.24*} &
                \multicolumn{1}{c}{$-$7.24} & 
                                      \multicolumn{1}{c}{\cellcolor{Gray}$-$7.25    }  &  
                                      \multicolumn{1}{c}{2} &
                \multicolumn{1}{c}{$-$7.23} & 
                \multicolumn{1}{c}{$-$7.42} & 
                \multicolumn{1}{c}{$-$7.60}  &  
                \multicolumn{1}{c}{$-$6.00}& 
                $-$0.02 \\ 
\multicolumn{1}{c}{$\sigma_{(7,1)}$ $b_{1g}$}&  \multicolumn{1}{c}{$-$7.49} & 
                \multicolumn{1}{c}{$-$7.47} & 
                \multicolumn{1}{c}{$-$7.42}  &  
                \multicolumn{1}{c}{$-$7.42} & 
                                     \multicolumn{1}{c}{ \cellcolor{Gray}$-$7.45}      &    
                                     \multicolumn{1}{c}{1} &
                \multicolumn{1}{c}{$-$7.48} & 
                \multicolumn{1}{c}{$-$7.65} & 
                \multicolumn{1}{c}{$-$7.85}  &  
                \multicolumn{1}{c}{$-$6.27}& 
                $+$0.03 \\ 
\multicolumn{1}{c}{$\sigma_{(4,4)}$ $a_{g}$}&  \multicolumn{1}{c}{$-$7.47 / $-$6.26* } & 
                \multicolumn{1}{c}{$-$7.51} & 
                \multicolumn{1}{c}{$-$7.32 / $-$6.26*}  &  
                \multicolumn{1}{c}{} & 
                                       
                                      \multicolumn{1}{c}{\cellcolor{Gray}$-$7.43} &
                                       \multicolumn{1}{c}{2} &
                \multicolumn{1}{c}{$-$7.46} & 
                \multicolumn{1}{c}{$-$7.64} & 
                \multicolumn{1}{c}{$-$7.83}  &  
                \multicolumn{1}{c}{$-$6.23}& 
                $+$0.03 \\ 
\multicolumn{1}{c}{$\sigma_{(4,6)}$ $a_{g}$}&  \multicolumn{1}{c}{$-$8.17} & 
               
               \multicolumn{1}{c}{$-$8.08} &
                \multicolumn{1}{c}{$-$8.19}  &  
                \multicolumn{1}{c}{$-$8.08} & 
                                       
                                    \multicolumn{1}{c}{\cellcolor{Gray}$-$8.13} &
                                     \multicolumn{1}{c}{1} &
                \multicolumn{1}{c}{$-$7.88} & 
                \multicolumn{1}{c}{$-$8.05} & 
                \multicolumn{1}{c}{$-$8.25}  &  
                \multicolumn{1}{c}{$-$6.61}& 
                $-$0.25 \\ 
\multicolumn{1}{c}{$\sigma_{(7,0)}$ $b_{2u}$}&  \multicolumn{1}{c}{$-$8.21} & 
                \multicolumn{1}{c}{$-$8.22} & 
                \multicolumn{1}{c}{$-$8.22}  &  
                \multicolumn{1}{c}{$-$8.20} & 
                                     
                                     \multicolumn{1}{c}{\cellcolor{Gray}$-$7.71} &
                                     \multicolumn{1}{c}{1} &
                \multicolumn{1}{c}{$-$8.24} & 
                \multicolumn{1}{c}{$-$8.34} & 
                \multicolumn{1}{c}{$-$8.62}  &  
                \multicolumn{1}{c}{$-$7.00}& 
                $+$0.53 \\ 
\multicolumn{1}{c}{$\sigma_{(3,5)}$ $b_{1g}$}&  \multicolumn{1}{c}{$-$8.17} & 
                \multicolumn{1}{c}{$-$8.20} & 
                \multicolumn{1}{c}{$-$8.19}  &  
                \multicolumn{1}{c}{$-$8.32*} & 
                                     \multicolumn{1}{c}{ \cellcolor{Gray}$-$8.19 }     &     
                                     \multicolumn{1}{c}{2} &
                \multicolumn{1}{c}{$-$8.20} & 
                \multicolumn{1}{c}{$-$8.35} & 
                \multicolumn{1}{c}{$-$8.57}  &  
                \multicolumn{1}{c}{$-$6.88}& 
                $+$0.01 \\ 
\multicolumn{1}{c}{$\sigma_{(5,6)}$ $b_{2u}$}&  \multicolumn{1}{c}{$-$8.18} & 
                 
                \multicolumn{1}{c}{$-$8.37} &
               
               \multicolumn{1}{c}{} &
                \multicolumn{1}{c}{} & 
                                    \multicolumn{1}{c}{\cellcolor{Gray}$-$8.28 }      &      
                                    \multicolumn{1}{c}{2} &
                \multicolumn{1}{c}{$-$8.17} & 
                \multicolumn{1}{c}{$-$8.34} & 
                \multicolumn{1}{c}{$-$8.54}  &  
                \multicolumn{1}{c}{$-$6.86}& 
                $-$0.11 \\ 
\multicolumn{1}{c}{$\sigma_{(5,3)}$ $b_{1g}$}&  \multicolumn{1}{c}{$-$8.73} & 
                \multicolumn{1}{c}{$-$8.69} & 
                \multicolumn{1}{c}{}  &  
                \multicolumn{1}{c}{} & 
                                     \multicolumn{1}{c}{  \cellcolor{Gray}$-$8.71 }    &      
                                     \multicolumn{1}{c}{3} &
                \multicolumn{1}{c}{$-$8.53} & 
                \multicolumn{1}{c}{$-$8.67} & 
                \multicolumn{1}{c}{$-$8.91}  &  
                \multicolumn{1}{c}{$-$7.17} & 
                $-$0.18\\ 
\multicolumn{1}{c}{$\sigma_{(0,6)}$ $a_{g}$}&  \multicolumn{1}{c}{$-$9.22} & 
                \multicolumn{1}{c}{$-$9.22} & 
                \multicolumn{1}{c}{$-$9.27}  &  
                \multicolumn{1}{c}{$-$9.07} & 
                                   \multicolumn{1}{c}{  \cellcolor{Gray}$-$9.20  }    &     
                                   \multicolumn{1}{c}{2} &
                \multicolumn{1}{c}{$-$8.88} & 
                \multicolumn{1}{c}{$-$9.01} & 
                \multicolumn{1}{c}{$-$9.26}  &  
                \multicolumn{1}{c}{$-$7.50}& 
                $-$0.32 \\ 
\multicolumn{1}{c}{$\sigma_{(6,1)}$ $b_{3u}$}&  \multicolumn{1}{c}{$-$9.28} & 
                \multicolumn{1}{c}{$-$9.19} & 
                \multicolumn{1}{c}{$-$9.28}  &  
                \multicolumn{1}{c}{$-$9.28} & 
                                  \multicolumn{1}{c}{ \cellcolor{Gray}$-$9.26   }     &  
                                  \multicolumn{1}{c}{1} &
                \multicolumn{1}{c}{$-$8.95} & 
                \multicolumn{1}{c}{$-$9.08} & 
                \multicolumn{1}{c}{$-$9.32}  &  
                \multicolumn{1}{c}{$-$7.59}& 
                $-$0.31 \\ 
\multicolumn{1}{c}{$\sigma_{(3,4)}$ $b_{2u}$}&  \multicolumn{1}{c}{$-$9.33} & 
                \multicolumn{1}{c}{$-$9.27} & 
                \multicolumn{1}{c}{$-$9.30}  &  
                \multicolumn{1}{c}{$-$10.87*} & 
                                  \multicolumn{1}{c}{ \cellcolor{Gray} $-$9.30  }     &      
                                  \multicolumn{1}{c}{2} &
                \multicolumn{1}{c}{$-$9.38} & 
                \multicolumn{1}{c}{$-$9.49} & 
                \multicolumn{1}{c}{$-$9.76}  &  
                \multicolumn{1}{c}{$-$7.84}& 
                $+$0.08 \\ 
\multicolumn{1}{c}{$\sigma_{(2,5)}$ $b_{3u}$}&  \multicolumn{1}{c}{$-$9.37} & 
                \multicolumn{1}{c}{$-$9.45} & 
                \multicolumn{1}{c}{$-$9.29}  &  
                \multicolumn{1}{c}{} & 
                                 \multicolumn{1}{c}{ \cellcolor{Gray}$-$9.37   }      &      
                                 \multicolumn{1}{c}{2} &
                \multicolumn{1}{c}{$-$9.22} & 
                \multicolumn{1}{c}{$-$9.33} & 
                \multicolumn{1}{c}{$-$9.59}  &  
                \multicolumn{1}{c}{$-$7.99} & 
                $-$0.15\\ 
\multicolumn{1}{c}{$\sigma_{(6,0)}$ $a_{g}$}&  \multicolumn{1}{c}{$-$9.65} & 
                \multicolumn{1}{c}{$-$9.79} & 
                \multicolumn{1}{c}{$-$9.30}  &  
                \multicolumn{1}{c}{} & 
                                   \multicolumn{1}{c}{ \cellcolor{Gray}$-$9.58  }     &        
                                   \multicolumn{1}{c}{3} &
                \multicolumn{1}{c}{$-$9.60} & 
                \multicolumn{1}{c}{$-$9.70} & 
                \multicolumn{1}{c}{$-$9.98}  &  
                \multicolumn{1}{c}{$-$8.19}& 
                $+$0.02 \\ 
\multicolumn{1}{c}{$\sigma_{(5,2)}$ $b_{2u}$}&  \multicolumn{1}{c}{$-$10.29} & 
                \multicolumn{1}{c}{$-$10.29} & 
                
                \multicolumn{1}{c}{} &
              
                \multicolumn{1}{c}{$-$10.66} &

                                   \multicolumn{1}{c}{\cellcolor{Gray}$-$10.41} &
                                    \multicolumn{1}{c}{2} &
                \multicolumn{1}{c}{$-$9.81} & 
                \multicolumn{1}{c}{$-$9.90} & 
                \multicolumn{1}{c}{$-$10.18}  &  
                \multicolumn{1}{c}{$-$8.31}& 
                $-$0.60 \\ 
\multicolumn{1}{c}{$\sigma_{(1,5)}$ $b_{1g}$}& 
\multicolumn{1}{c}{} &
                \multicolumn{1}{c}{$-$10.27*} & 
              
               \multicolumn{1}{c}{} &
                \multicolumn{1}{c}{} & 
                                    
                                   \multicolumn{1}{c}{} &
                                    \multicolumn{1}{c}{4} &
                \multicolumn{1}{c}{$-$10.08} & 
                \multicolumn{1}{c}{$-$10.17} & 
                \multicolumn{1}{c}{$-$10.46}  &  
                \multicolumn{1}{c}{$-$8.64}& 
               \\ 
\multicolumn{1}{c}{$\sigma_{(4,3)}$ $b_{3u}$}&  \multicolumn{1}{c}{} & 
                \multicolumn{1}{c}{} & 
                
                \multicolumn{1}{c}{} &
                \multicolumn{1}{c}{} & 
                                            &   
                                            \multicolumn{1}{c}{4} &
                \multicolumn{1}{c}{$-$10.87} & 
                \multicolumn{1}{c}{$-$10.93} & 
                \multicolumn{1}{c}{$-$11.25}  &  
                \multicolumn{1}{c}{$-$9.28} & \\ 
\multicolumn{1}{c}{$\sigma_{(2,4)}$ $a_{g}$}&  \multicolumn{1}{c}{$-$9.43 } & 
                \multicolumn{1}{c}{$-$9.94*} & 
                \multicolumn{1}{c}{$-$9.33}  &  
                \multicolumn{1}{c}{} & 
                                  
                                 \multicolumn{1}{c}{\cellcolor{Gray}$-$9.38} &
                                    \multicolumn{1}{c}{3} &
                \multicolumn{1}{c}{$-$11.00} & 
                \multicolumn{1}{c}{$-$11.04} & 
                \multicolumn{1}{c}{$-$11.38}  &  
                \multicolumn{1}{c}{$-$9.49}& 
                $+$1.62 \\ 
\multicolumn{1}{c}{$\sigma_{(3,0)}$ $b_{2u}$}&  \multicolumn{1}{c}{$-$9.52*} & 
                \multicolumn{1}{c}{} & 
                
                \multicolumn{1}{c}{$-$9.33*} &
                \multicolumn{1}{c}{} & 
                                    \multicolumn{1}{c}{  }      &     
                                    \multicolumn{1}{c}{4} &
                \multicolumn{1}{c}{$-$11.25} & 
                \multicolumn{1}{c}{$-$11.29} & 
                \multicolumn{1}{c}{$-$11.63}  &  
                \multicolumn{1}{c}{$-$9.73}& \\ 
\multicolumn{1}{c}{$\sigma_{(0,5)}$ $b_{3u}$}&  \multicolumn{1}{c}{} & 
                \multicolumn{1}{c}{} & 
                \multicolumn{1}{c}{}  &  
               
                \multicolumn{1}{c}{} &
                                            &         
                                            \multicolumn{1}{c}{4} &
                \multicolumn{1}{c}{$-$11.57} & 
                \multicolumn{1}{c}{$-$11.61} & 
                \multicolumn{1}{c}{$-$11.95}  &  
                \multicolumn{1}{c}{$-$9.95}& \\ 

\multicolumn{1}{c}{}      & \multicolumn{1}{c}{} & \multicolumn{1}{c}{} & \multicolumn{1}{c}{}     & \multicolumn{1}{c}{}            & \multicolumn{1}{c}{}        & \multicolumn{1}{c}{}    & \multicolumn{1}{c}{}      & \multicolumn{1}{c}{}     &   \multicolumn{1}{c}{} &  \\ 

\hline
\hline
\end{tabular}
\end{table*}

Focusing on the $\pi$ orbitals (Figs.~\ref{fig:expdeconvolution_sigma_part1} to \ref{fig:expdeconvolution_sigma_part3} in the appendix), we find that all four of them above the $d$ band ($\pi_{(4,0)}$, $\pi_{(2,3)}$, $\pi_{(1,3)}$, and $\pi_{(2,2)}$) yield sharply peaked pDOS that allow an unambiguous assignment of the respective binding energies with high accuracy. This is not surprising, since in this range POT and orbital deconvolution have been proven many times \cite{Puschnig2011, Willenbockel2013, Puschnig2017, Willenbockel2015, Stadtmueller2012}. But even below the $d$ band, the $h\nu=35$\,eV data cube still yields sharp peaks, while for the $h\nu=57$\,eV data cube (for which the inelastic mean free path of the photoelectrons is longer and which is therefore more susceptible to substrate features) the situation is more complex: the experimental pDOS either exhibits shoulders on a fast-rising background that stems from the onset of the Cu $d$~band around $-4.5$\,eV ($\pi_{(3,1)}$, $\pi_{(3,0)}$), or broad multiple-peak structures, in some cases close to the sharp peak of the $h\nu=35$\,eV data cube ($\pi_{(0,3)}$, $\pi_{(1,2)}$, $\pi_{(2,1)}$, and $\pi_{(0,0)}$), in other cases ($\pi_{(1,1)}$, $\pi_{(2,0)}$, $\pi_{(0,2)}$, $\pi_{(1,0)}$, and $\pi_{(0,1)}$) in only moderate agreement. Therefore, we conclude that with regard to the binding energies of the $\pi$ orbitals, the $h\nu=35$\,eV data cube and the band map fits (at the same photon energy) are the most reliable ones. This result clearly illustrates the extent to which higher photon energies make the identification and analysis of molecular orbitals more difficult (see section \ref{sec:photoemissionexperiments}). Unfortunately, however, for $\sigma$ orbitals, to which we now turn, this data cube does not offer a sufficient photoelectron horizon. But, as we will see below, because of the larger $k_\parallel$, there is less signal from the substrate, and thus the reduction of the molecular photoemission intensities in the $h\nu=57$\,eV data cube is less of a problem for $\sigma$ than for $\pi$~orbitals.  

Looking at the $\sigma$ orbital fits in detail (Fig.~\ref{fig:expdeconvolution_sigma_part1} to \ref{fig:expdeconvolution_sigma_part5}), there are several clear cases where all five fitting methods discussed in section \ref{sec:dataanalysis} lead to well-defined peaks with coinciding peak energies. Classifying the fit qualities for all $\sigma$ orbitals, we assign these cases to category 1 (best). In order of decreasing energy, these are:  $\sigma_{(0,8)}$, $\sigma_{(7,3)}$,  $\sigma_{(7,2)}$, $\sigma_{(4,7)}$, $\sigma_{(1,7)}$, $\sigma_{(7,1)}$, $\sigma_{(4,6)}$, $\sigma_{(7,0)}$, and $\sigma_{(6,1)}$. These make up for a third of all 27 $\sigma$ orbitals in the considered energy range. Another 40 \% are cases which, with the exception of minor inconsistencies, are also essentially clear (category 2). Examples of these inconsistencies are negative weights (outside the maximum) in one or more of the linear fit methodologies ($\sigma_{(1,8)}$, $\sigma_{(0,7)}$, $\sigma_{(6,2)}$, $\sigma_{(3,5)}$, $\sigma_{(0,6)}$, $\sigma_{(3,4)}$), failure of the regularized MC fit ($\sigma_{(1,8)}$, $\sigma_{(4,4)}$, $\sigma_{(3,5)}$,  $\sigma_{(5,6)}$, $\sigma_{(3,4)}$, $\sigma_{(2,5)}$---notably, nearly all failures of the MC fit occur for binding energies $E_\mathrm{b}$ above $8.2$\,eV) or a disagreement between the MLA or a single orbital fit and the other methodologies ($\sigma_{(0,7)}$, $\sigma_{(2,7)}$, $\sigma_{(5,2)}$). Nevertheless, for these cases still a rather clear picture emerges if all methodologies are considered together. In 11 \% of the cases ($\sigma_{(5,3)}$, $\sigma_{(6,0)}$, $\sigma_{(2,4)}$) the fits are less convincing (category 3): For $\sigma_{(5,3)}$, MC and MLA fail to produce a clear peak and there is a problem with negative weights for the all-$\sigma$ fit, although all linear fits consistently produce a weak maximum; for $\sigma_{(6,0)}$, while the three linear fits agree, the MLA exhibits a peak at lower binding energy and MC fails altogether; for $\sigma_{(2,4)}$, MC fails while two pairs of fits match but disagree with each other. Yet, in these three cases it is still possible to  assign binding energies. Only in 15 \% percent of the cases ($\sigma_{(1,5)}$, $\sigma_{(4,3)}$, $\sigma_{(3,0)}$, and $\sigma_{(0,5)}$) the fit essentially fails (category 4). As we will see below, these orbitals are predicted to have binding energies very close or even above $11$\,eV, essentially out of the experimental range. To summarize, for the majority (20 out of 27) of $\sigma$ orbitals considered here the clear assignment of an energy in the measured range $E=[-4.26, -11.01]$\,eV is possible, with another three allowing an assignment with some qualification. 

The reconstruction of the data cube from the theoretical momentum maps and the fitted weights constitutes an independent way to gauge the overall quality of the fitting process. Fig.~\ref{fig:expkmaps}b-c displays cuts through reconstructed data cubes at selected  energies (for $k_\parallel \geq 1.6$~\AA$^{-1}$) (the reconstructed cube at all  energies is shown in a movie in the supplemental material). Both the data cubes from linear and the Monte Carlo fitting are in excellent agreement with the experimental one, showing that the structure of the latter can be reconstructed in the approximation of a superposition of emission features from single-particle Kohn-Sham orbitals. Fig.~\ref{fig:expkmaps} also reveals that the remarkable quasi-hexagonal appearance of the momentum maps in the data cube is a result of the overlap of many orbitals of the set shown in Fig.~\ref{fig:momentummaps}.

\begin{figure*}[bt]
\begin{center}
	\includegraphics[width=0.7\textwidth]{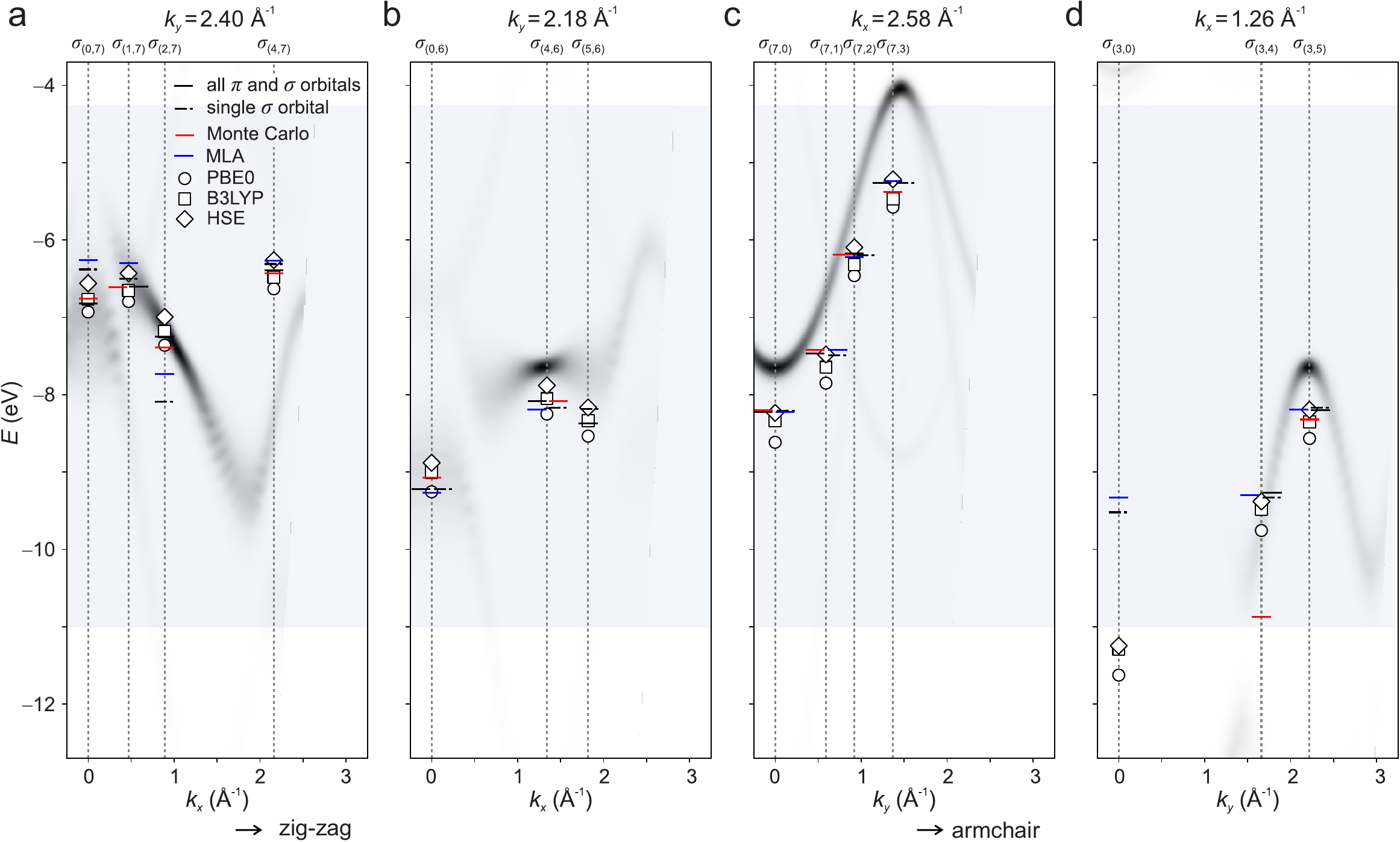}
\end{center}
\caption{\label{fig:intramoleculardispersion}  Intramolecular dispersion of $\sigma$ bands. (a) $\sigma_{(n,7)}$, (b) $\sigma_{(n,6)}$, (c) $\sigma_{(7,m)}$, (d) $\sigma_{(3,m)}$. Horizontal bars denote the orbital energies derived from the experimental data cube $I_{\rm exp}(k_x, k_y;E_{\rm b})$ measured at $h\nu$ = 57 eV, using the following fitting methodologies: linear fitting with the theoretical momentum map of the single corresponding $\sigma$\,orbital (\textit{black dash-dotted}), linear fitting with theoretical momentum maps of all-$\pi$-and-$\sigma$ orbitals (\textit{black solid}), MLA pattern recognition (\textit{blue}), and Monte Carlo fitting with theoretical momentum maps of all $\sigma$ orbitals (\textit{red}). Note that the fit results for the linear all-$\sigma$ and all-$\pi$-and-$\sigma$ fits are essentially identical (see Fig.~\ref{fig:expdeconvolution_sigma_part1} to \ref{fig:expdeconvolution_sigma_part5} in the appendix) and therefore only the latter have been included in the figure. Symbols denote the DFT-calculated orbital energies of bisanthene/Cu(110), using the following exchange-correlation functionals: PBE0 (\textit{open circles}), B3LYP (\textit{open squares}), HSE (\textit{open diamonds}). The bars and symbols appear at the $k_x$, $k_y$ position of the strongest emission lobe of the corresponding orbital in the dispersion direction of the respective band (\textit{dotted vertical lines}). The gray boxes mark the  energy range in which the experimental data were fitted. Corresponding plots for all $\sigma$ and $\pi$ bands are shown in Figs.~\ref{fig:expdispersion_pi_part1} to \ref{fig:expdispersion_sigma_part4} in the appendix. Band maps of free-standing graphene simulated with HSE are underlaid.\cite{Puschnig2015} The band maps are cut at varying $k_x$, $k_y$ to slice through the main lobes of the respective orbitals, namely (a) $k_y = 2.40$~\AA$^{-1}$, (b) $k_y = 2.18$~\AA$^{-1}$, (c) $k_x = 2.58$~\AA$^{-1}$, and (d) $k_x =1.26$~\AA$^{-1}$.}
\end{figure*}

While the fitting methodologies that were applied all use uncorrelated orbital momentum patterns, in reality there must exist manifold correlations between the orbitals (both $\sigma$ and $\pi$) regarding their relative energies. This comes from the fact that following their specific nodal patterns, the individual orbitals of bisanthene form discrete bands of intramolecular dispersion (these discrete intramolecular bands derive by confinement from the band structure of graphene; in the limit of very large nanographenes, their intramolecular band structure approaches that of graphene). If we are to explain the observed momentum patterns as deriving from single-particle orbitals, these correlations must be present in the experimental data cube. Thus, a crucial question is whether these correlations are indeed present, and if yes, whether the unbiased fits with uncorrelated orbital patterns can retrieve them. In Fig.~\ref{fig:intramoleculardispersion}, the calculated intramolecular dispersion from DFT calculations of bisanthene/Cu(110) is shown for the examples of four bands, together with the orbital energies from the uncorrelated fits. The complete set of bands for the orbitals considered here is shown in the appendix in Figs.~\ref{fig:expdispersion_pi_part1} and \ref{fig:expdispersion_pi_part2} for $\pi$ and \ref{fig:expdispersion_sigma_part1} to \ref{fig:expdispersion_sigma_part4} for $\sigma$ bands. The figures show a remarkable agreement between experiment and theory, which proves that the experimental data cube implicitly contains the intrinsic dispersion structure and that the applied unbiased fitting methodologies are able to successfully retrieve it. Not surprisingly, the figure also illustrates that the intramolecular dispersion in bisanthene derives from the band structure of graphene. The fact that the intramolecular bands of bisanthene/Cu(110) tend to have larger binding energies than the corresponding graphene bands is due to the metallic substrate (cf.~section~\ref{sec:level_alignment})---the graphene bands were calculated for a free-standing layer. 

In conclusion, from the successful fits of individual orbitals, the good agreement between the experimental and reconstructed data cubes, and the retrieval of the dispersion structure we can conclude beyond doubt that the structure imprinted in the photoemission data cube $I_\mathrm{exp}(k_x, k_y;E_{\rm b})$ derives from single-particle initial-state orbitals and that the Kohn-Sham orbitals of DFT which were used for the retrieval are a good approximation to these. This allows benchmarking the orbital binding energies predicted by different exchange-correlation functionals of DFT against the experimental data (section~\ref{sec:quantitative_comparison}).

\subsection{\label{subsec:dos}Comparing DFT calculations with different exchange-correlation functionals}
 
\begin{figure*}[bt]
\begin{center}
	\includegraphics[width=0.8\textwidth]{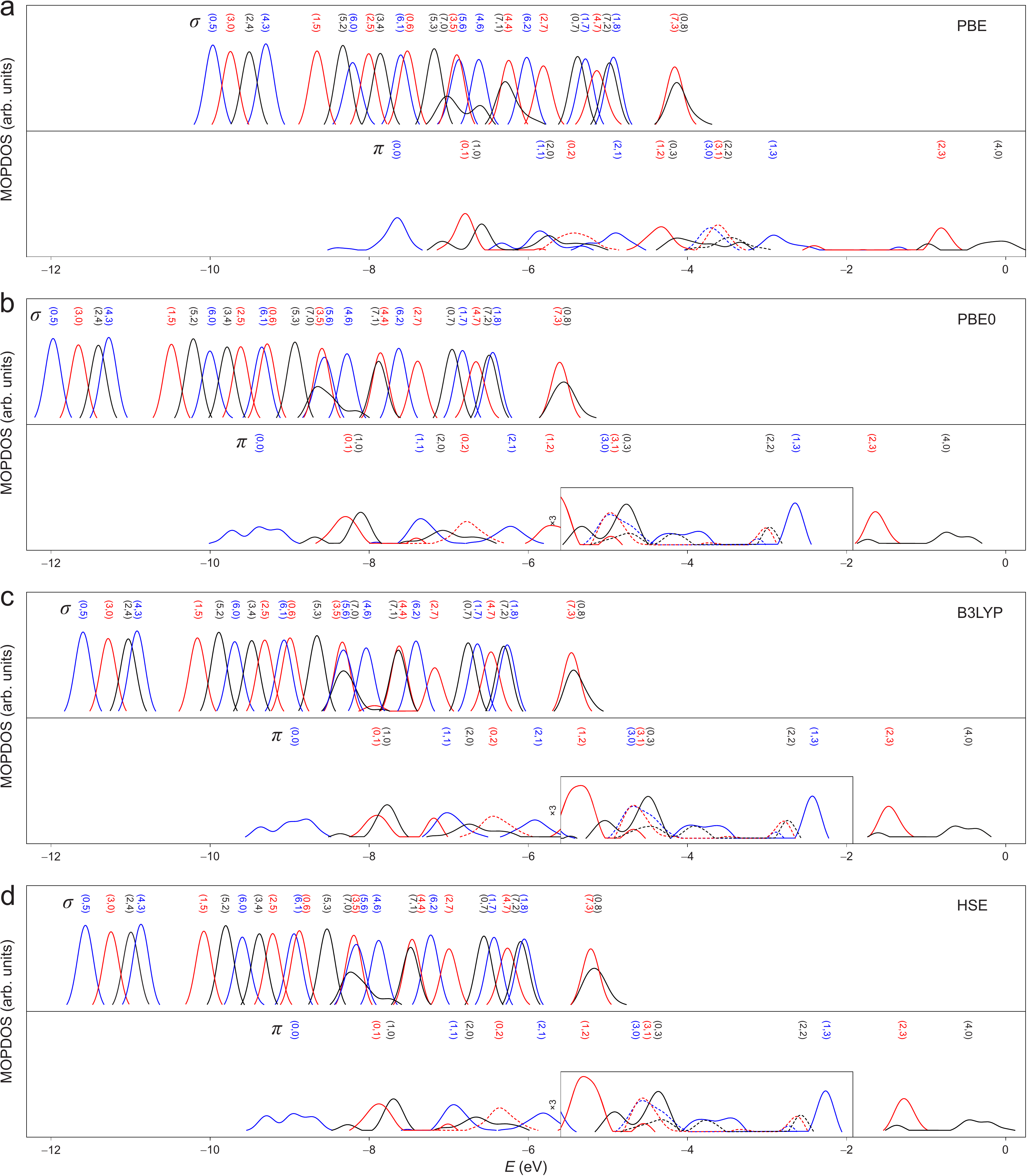}
\end{center}
	\caption{\label{fig:mopdos_pi_sigma}  Molecular-orbital-projected density of states (MOPDOS) for $\sigma$ (top of each panel) and $\pi$ (bottom of each panel) orbitals of bisanthene/Cu(110) calculated with DFT  using the following exchange-correlation functionals: (a) PBE, (b) PBE0, (c) B3LYP, and (d) HSE. Individual MOPDOS curves are only drawn above a threshold value. The artificial broadening of the MOPDOS curves is 0.1 eV and thus much smaller than the smallest observed intrinsic width ($\sim 0.25$\,eV for most of the $\sigma$~orbitals). For clarity, the color of curves alternates in a cyclic pattern from black to red to blue, following the orbital energies of the isolated molecule. Sometimes, dashed lines are used to aid the eye in distinguishing MOPDOS curves of different orbitals. Also, in panels b to d the MOPDOS curves for the $\pi$ orbitals in the energy range of the Cu $d$ band are magnified in vertical direction by a factor of three (black rectangle).  }  
\end{figure*}

Before turning to the benchmark of different exchange-correlation potentials against the experimental data, we compare in this section the results of the DFT calculations among each other. Already in Fig.~\ref{fig:pDOS_theory}, which depicts the computational results for PBE (top panel) and HSE (bottom panel), we could observe interesting trends. First, we notice that the energetic position of the copper $d$ band (green shaded areas) depends sensitively on the exchange-correlation functional. For PBE, its onset is $\sim1.2$~eV below $E_\mathrm{F}$, while the hybrid functional HSE yields a significantly deeper $d$ band onset at about $-2.3$\,eV. Thus, the inclusion of exact exchange leads to an improvement when comparing to the experimentally observed onset at approximately $-2.0$\,eV (Fig.~\ref{fig:expbandmaps_pi}). However, this correction overshoots somewhat for the range-separated HSE and even more so for the global hybrid PBE0 (not shown). It should also be noted that the computed band width of the $d$ band,  $\sim 3.7$\,eV, remains almost unchanged for all exchange-correlation functionals tested in this work and that this value is almost 25\% larger than the experimentally observed band width of $\sim 3$\,eV (Fig.~\ref{fig:expbandmaps_pi}). This overestimation could be a consequence of the fact that we use five atomic layers to model the Cu substrate (see section \ref{sec:interface}).

Similar trends are also observed when comparing the calculated DOS originating from the adsorbed organic molecule. To this end, we projected the DOS of the bisanthene/Cu(110) system onto the carbon atoms of bisanthene (gray shaded areas in Fig.~\ref{fig:pDOS_theory}) and also computed the MOPDOS for all 15 $\pi$ and 27 $\sigma$ orbitals introduced in Tables~\ref{tab:piorbitals}  and \ref{tab:sigmaorbitals}(Fig.~\ref{fig:mopdos_pi_sigma}). We find that the overall band width of occupied $\pi$ bands shows some variation with the exchange-correlation functional. For PBE, the deepest $\pi$~orbital, $\pi_{(0,0)}$, peaks at $-7.63$ eV, while the onset of the $\pi$~bands for the hybrid functionals is at $-9.35$~eV (PBE0), $-8.95$~eV (HSE), and $-8.80$~eV (B3LYP), see Table~\ref{tab:experiment_pi_orbitals}.

The choice of the exchange-correlation functional also affects the calculated amount of charge transfer into the LUMO ($\pi_{(4,0)}$) and influences the degree of hybridization between molecular and metallic states. For PBE, the LUMO is partially occupied, for HSE it is almost entirely below $E_\mathrm{F}$, while for PBE0 it is fully occupied. A pronounced difference of PBE compared to HSE and PBE0 can, for instance, also be observed for the HOMO-1 ($\pi_{(1,3)}$). Due to the high-lying $d$ band in PBE, $\pi_{(1,3)}$ overlaps entirely with the $d$ band and PBE predicts a strong hybridization, as indicated by the broad MOPDOS curve. The hybridization is predicted to be less strong in HSE and PBE0, which both yield a $\pi_{(1,3)}$-related MOPDOS peak right above the Cu $d$ band, in agreement with the experimental observation,\cite{Yang2019} although the calculations also predict a minor maximum within the $d$ band.

Regarding the influence of the exchange-correlation functional on the $\sigma$ orbitals, we recognize the same trend as for the $\pi$ orbitals, as Fig.~\ref{fig:mopdos_pi_sigma} shows. There is an overall shift to larger binding energies when going from PBE via HSE and B3LYP to PBE0, accompanied by an increase in the band width. 

\subsection{\label{sec:quantitative_comparison}Benchmarking DFT functionals against experiment}
In this section, we use the experimental binding energies to quantitatively benchmark the DFT calculations with four commonly used exchange-correlation potentials. First, we focus on the onsets and widths of the $\pi$ and $\sigma$ bands. In experiment, the bottom of the $\pi$ band is found at $-8.87\,(-9.04)$\,eV, cf.~Table \ref{tab:experiment_pi_orbitals} (column \textit{average} and, in brackets, column \textit{all $\pi$ orbitals $h\nu=35$\,eV}). According to the table, the B3LYP and HSE hybrid functionals predict onset energies of $-8.80$ and  $-8.95$\,eV, respectively, and thus agree within $+0.07\,(+0.24) $ or $-0.08\,(+0.09)$\,eV with experimental values. For PBE ($\pi_{(0,0)}$ at $-7.63$\,eV) and PBE0 ($\pi_{(0,0)}$ at $-9.35$\,eV), the agreement is significantly worse, with deviations of $+1.24\,(+1.41)$\,eV ($\pi$ band onset predicted too high) or $-0.48\,(-0.31)$\,eV ($\pi$ band onset predicted too low) from the experimental value. Thus, the predicted width of the occupied $\pi$ band (difference between the binding energies of $\pi_{(4,0)}$ and $\pi_{(0,0)}$) comes out far too small in PBE ($7.59$\,eV), while PBE0 ($8.63$\,eV) slightly overestimates it, in comparison with the experimental value $8.37\,(8.54)$\,eV. In contrast, HSE ($8.49$\,eV)  yields a band width in almost perfect agreement (within $+120\,(-50)$\,meV) with experiment. Similarly, the energy of the highest occupied orbital for bisanthene/Cu(110), the former LUMO $\pi_{(4,0)}$ of the isolated molecule, is best predicted by HSE, to within $40$\,meV of the experimental value. In B3LYP, both the band width ($8.16$\,eV) and the $\pi_{(4,0)}$ orbital come out too low. 

The HSE functional also shows the best agreement with the experimental binding energies of the two topmost $\sigma$ orbitals, $\sigma_{(0,8)}$ and $\sigma_{(7,3)}$, the deviation is only $+0.07$ and $+0.11$\,eV, respectively (Table~\ref{tab:experiment_sigma_orbitals}). The lowest $\sigma$ orbital that can be reliably assigned ($\sigma_{(5,2)}$) appears at $-10.41$\,eV in experiment, \textit{i.e.},~$5.13$\,eV below the highest occupied $\sigma$ orbital. In all four calculations, the binding energy difference between $\sigma_{(0,8)}$ and $\sigma_{(5,2)}$ is considerably smaller, with PBE0 ($4.65$\,eV) and HSE ($4.64$\,eV) having the values closest to the experiment. In contrast, PBE, which underestimates all binding energies by $\sim 1$\,eV, has the smallest $\sigma$ band width ($4.19$\,eV). We note, however, that the experimental and calculated band widths are in better agreement if the difference between the topmost ($\sigma_{(0,8)}$) and the lowest but one orbital ($\sigma_{(6,0)}$) is considered. In this case, the three hybrid functionals yield values similar to experiment, with B3LYP ($-0.04$\,eV) significantly closer than HSE ($+0.13$\,eV) and PBE0 ($+0.15$\,eV). 

 To illustrate once more the excellent agreement between experiment and HSE, we note that six $\pi$ orbitals have $|\Delta|=|E_i^{\rm exp}-E_i^{\rm HSE}| \leq 0.1\,{\rm eV}$, six $0.1\,{\rm eV}<|\Delta| \leq 0.22\,{\rm eV}$, three $0.22\,{\rm eV}<|\Delta| \leq 0.95 \,{\rm eV}$ (last column in Table~\ref{tab:experiment_pi_orbitals}). Although the three orbitals belonging to the last group ($\pi_{(0,3)}$, $\pi_{(1,2)}$, and $\pi_{(2,1)}$) show a relatively large deviation from the HSE predicted binding energy, their pDOS and their binding energies resulting from the deconvolution of the data cube are credible---see the discussion in section~\ref{sec:results_momentum_maps}. It is noteworthy that for these three orbitals the PBE0 prediction is closer to the experiment; for $\pi_{(2,1)}$ with a deviation of $+0.17\,(+0.07)$\,eV it is almost exact. A slight advantage for the PBE0 calculation over HSE is also found for $\pi_{(0,2)}$ and $\pi_{(1,0)}$. 
 
 The situation is similarly good for the $\sigma$ orbitals (last column in Table~\ref{tab:experiment_sigma_orbitals}): For eleven $\sigma$ orbitals the absolute deviation  is $|\Delta|\leq0.1\,{\rm eV}$, for a further five $\sigma$ orbitals $0.1\,{\rm eV}<|\Delta| \leq 0.2\,{\rm eV}$ holds, while three $\sigma$ orbitals have $0.2\,{\rm eV}<|\Delta| \leq 0.32\,{\rm eV}$ and another three have $0.32\,{\rm eV}<|\Delta| \leq 0.62\,{\rm eV}$, and only one outlier with  $\Delta=+1.62$\,eV is found. As we will see below, this outlier originates from a systematic error of the deconvolution procedure. Apart from this outlier, there are---similar to the situation regarding the $\pi$ orbitals---several orbitals for which PBE0 provides better agreement with experiment. These are $\sigma_{(2,7)}$, $\sigma_{(4,6)}$, $\sigma_{(0,6)}$, $\sigma_{(6,1)}$, and $\sigma_{(5,2)}$. All of them belong to category 1 or 2 regarding the quality of the deconvolution. A special case is that of $\sigma_{(7,0)}$, which apart from the most probably erroneous outlier $\sigma_{(2,4)}$ is the only case in which the experimental binding energy is much  smaller than the HSE prediction (and that of the other two hybrid functionals), although it is still considerably larger than the prediction of PBE.
 
In order to benchmark the quality of the binding energy predictions more systematically, we computed the mean errors (ME) and mean absolute errors (MAE), defined as 
\begin{eqnarray}
\mathrm{ME} & = & \frac{1}{N} \sum_i (E_i^\mathrm{exp} - E_i^\mathrm{DFT}) \label{eq:ME}\\
\mathrm{MAE} & = & \frac{1}{N} \sum_i | E_i^\mathrm{exp} - E_i^\mathrm{DFT} | \label{eq:MAE},
\end{eqnarray} 
for the DFT calculations with four exchange-correlation potentials, separately for $\pi$ and $\sigma$ orbitals. Here, $E_i^\mathrm{exp}$ is the energy of orbital $i$, measured with respect to the Fermi level, as determined from the deconvolution of the photoemission data cubes; for $\sigma$ orbitals, the average values in Table~\ref{tab:experiment_sigma_orbitals} were used for $E_i^\mathrm{exp}$, while for $\pi$ orbitals the deconvolution results of the all-$\pi$-orbitalsfit with $h\nu=35$\,eV (column 3 in Table~\ref{tab:experiment_pi_orbitals}) was employed. The calculated orbital energies $E_i^\mathrm{DFT}$ are the peak positions of the MOPDOS curves in Fig.~\ref{fig:mopdos_pi_sigma}.

\begin{figure}[bt]
\begin{center}
	\includegraphics[width=\columnwidth]{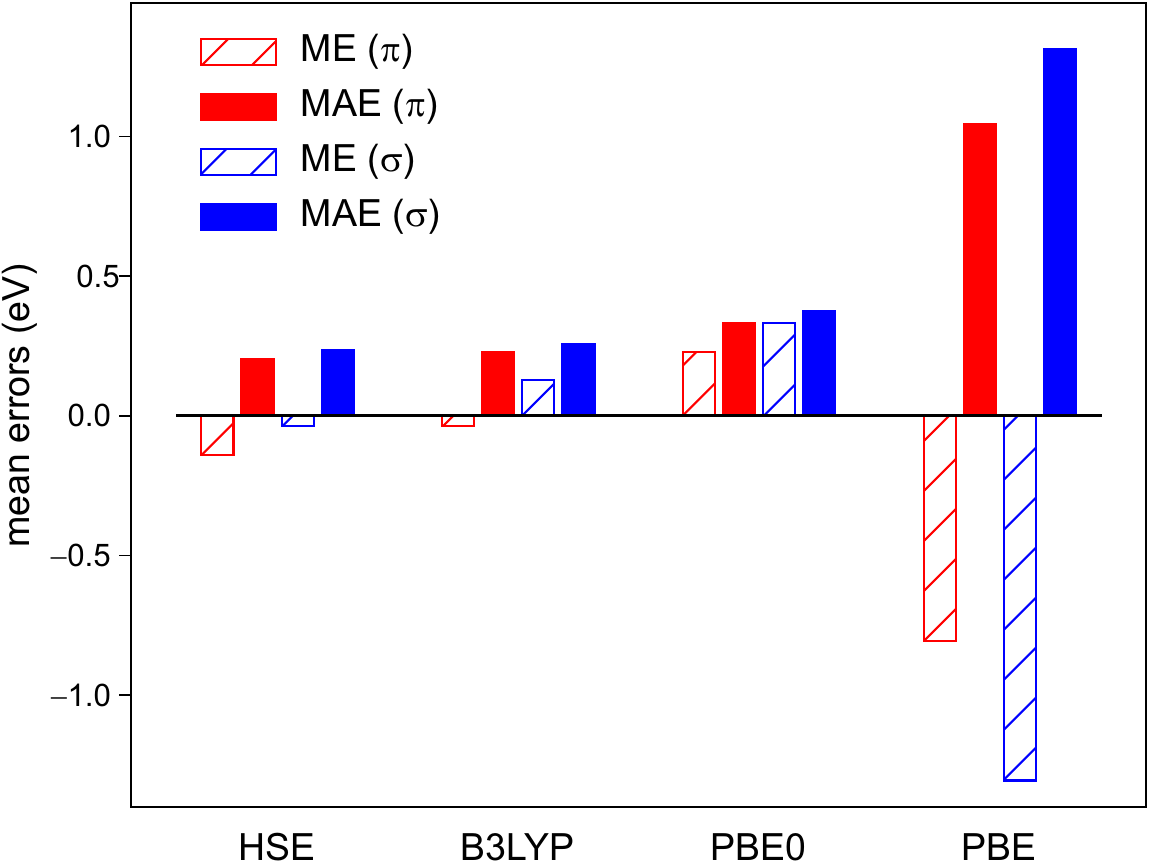}
\end{center}
	\caption{\label{fig:errors} Mean errors (ME) and mean absolute errors (MAE) of $\pi$ (red) and $\sigma$ (blue) orbital energies as obtained by using four different exchange-correlation functionals. For the $\sigma$ orbitals, average energies (Table~\ref{tab:experiment_sigma_orbitals}, column 5) were used to calculate the ME and MAE. For the $\pi$ orbitals, binding energies from the all-$\pi$-orbitals fit of the $h\nu=35$\,eV data cube (Table~\ref{tab:experiment_pi_orbitals}, column 3) were used.
 }
\end{figure}	

The ME (hatched bars) and MAE (filled bars) are displayed in Fig.~\ref{fig:errors}. By far the largest errors are found for PBE. The reason is the self-interaction of the electrons, which destabilizes their single-particle levels and leads to erroneously small binding energies (negative ME). In PBE, the ME for the $\sigma$ orbitals are about 0.4~eV larger than for $\pi$ orbitals. This is expected, because the $\sigma$~orbitals are more localized than $\pi$ orbitals and therefore more vulnerable to self-interaction errors. The self-interaction error is 
mitigated in the hybrid functionals (PBE0, B3LYP, HSE) by incorporating a fraction $\alpha$ of Hartree-Fock (HF) exchange. Specifically, $\alpha$ is 0.25 for PBE0. We find that for both $\pi$ and $\sigma$ orbitals the self-interaction correction in PBE0 in fact overshoots, leading to an overestimation of the binding energies (positive ME). Again, this overshooting is more dramatic for the $\sigma$ orbitals, because exchange is generally more relevant for more strongly localized electrons and therefore the correction by exact exchange works more effectively for the $\sigma$ orbitals.  In B3LYP, $\alpha$ is reduced to 0.2, with the result that the self-interaction correction for the $\pi$ orbitals is now spot on ($\mathrm{ME}(\pi)\approx 0$), while the $\sigma$ orbitals, being more sensitive to exchange, are still overcorrected. In contrast to PBE0 and B3LYP (which are global hybrid functionals), HSE is range-separated: HF exchange is only included in the short range. In the language of optimally-tuned range separated hybrid functionals,\cite{Refaely-Abramson2013,Lueftner2014,Kronik2018} the HSE functional can be viewed as having effectively infinite dielectric screening in the long-range.  Apparently, this is appropriate for the bisanthene/Cu(110) system studied in this work, as the MAE for both $\pi$ an $\sigma$ orbitals is reduced in comparison with B3LYP, making HSE the most accurate functional of the four tested ones for the present purpose. Notably, HSE has the same $\alpha=0.25$ as PBE0. Thus, range separation and the inclusion of HF exchange only in the short range seems to be a more effective way of reducing the overcompensation of the self-interaction error in PBE0 than the global reduction of $\alpha$ from 0.25 to 0.2 (as in B3LYP). Yet, in HSE the binding energies of both the $\pi$ and $\sigma$ orbitals are underestimated, the effect being stronger for the former. 

Finally, we stress that these findings are system-dependent and expected to vary with the type of substrate and the molecule-substrate distance. For other molecule/metal interfaces with larger molecule-metal distances (e.g., on Ag or Au surfaces) the superior performance of HSE over PBE0 may therefore not hold to the same extent, because in those cases the effectively infinite dielectric screening may not be appropriate. Our analysis of 42 orbitals for the bisanthene/Cu(110) system reveals that even for the same material, the various exchange-correlation functionals need to be evaluated separately for each of the orbitals. Several $\pi$ and $\sigma$ orbitals even seem to be better described by PBE0 than by HSE or B3LYP ($\pi_{(0,3)}$, $\pi_{(1,2)}$, $\pi_{(2,1)}$, $\sigma_{(2,7)}$, $\sigma_{(4,6)}$, $\sigma_{(0,6)}$, $\sigma_{(6,1)}$, and $\sigma_{(5,2)}$). One can speculate that the screening for these orbitals is not so efficient and that there is therefore stronger self-interaction, such that the self-interaction correction in PBE0 does not overshoot to the same degree as for other orbitals.

In addition to the peak positions, it is also important to consider the widths of the individual peaks in the experimental pDOS and the calculated MOPDOS. In the calculations, we notice little variation over all $\sigma$ orbitals and almost no influence of the functional, with an overall full width at half maximum (FWHM) of approximately 0.25 eV (Fig.~\ref{fig:mopdos_pi_sigma}), which is larger than the artificial Gaussian broadening of 0.1\,eV. This is to be contrasted with the considerably larger FWHM in the theoretical MOPDOS of the $\pi$ orbitals. The different widths of $\sigma$ and $\pi$ orbitals can be explained by different degrees of hybridization with the substrate. Because of their perpendicular orientation to the substrate surface, $\pi$ orbitals overlap more strongly with the substrate than the parallel oriented $\sigma$ orbitals and hence exhibit an enhanced tendency to hybridize with the latter.

In experiment the situation is less clear. If we consider the most reliable pDOS for the $\pi$ orbitals, derived from the linear fit of momentum maps to the $h\nu=35$\,eV data cube (column 3 in Table ~\ref{tab:experiment_pi_orbitals}), we observe a FWHM of approximately $0.5$\,eV (Fig.~\ref{fig:expdeconvolution_pi_part1} to \ref{fig:expdeconvolution_pi_part3}). This is roughly comparable to the theoretical MOPDOS of the $\pi$ orbitals (Fig.~\ref{fig:mopdos_pi_sigma}). For the $\sigma$ orbitals, however, the widths of the experimental pDOS peaks is substantially larger, in the region of $1.0$ to $1.5$\,eV. Thus, in experiment the $\sigma$ peaks are not only broader than the corresponding theoretical MOPDOS, but also broader then the experimental $\pi$ peaks and even significantly larger than the experimental resolution of the toroidal electron energy analyzer would imply. To explain this discrepancy, we remark that the experimental pDOS is inextricably interwoven with the underlying photoemission process, while the theoretical MOPDOS reflects a pure density of states. Specifically, we suggest that the reason for the much larger FWHM in the experimental data for the $\sigma$ orbitals arises from the short lifetime of the photohole, which leads to a broadening of the spectral signatures. Theoretically, such an effect would be contained in Eq.~\ref{eq:Feibelman} if one were to replace the $\delta$ function by the spectral function, for instance, from a $GW$ calculation.\cite{Gerlach2001,Marini2002,Yi2010} While a calculation of the spectral function is beyond the scope of the present work, we note that the experimentally observed peak width would suggest extremely short lifetimes of only 0.6\,fs. This estimate may appear surprising, but we note that the  binding energy range of the $\sigma$ orbitals, $5$ to $10$~eV, is considerably below the copper $d$ band and that quasiparticle calculations for Cu have predicted hole lifetimes of only $\sim 1$~fs for states 5\,eV below $E_\mathrm{F}$.\cite{Yi2010} 

\subsection{\label{sec:level_alignment}Level alignment at the interface}

\begin{figure*}[bt]
\begin{center}
	\includegraphics[width=0.95\textwidth]{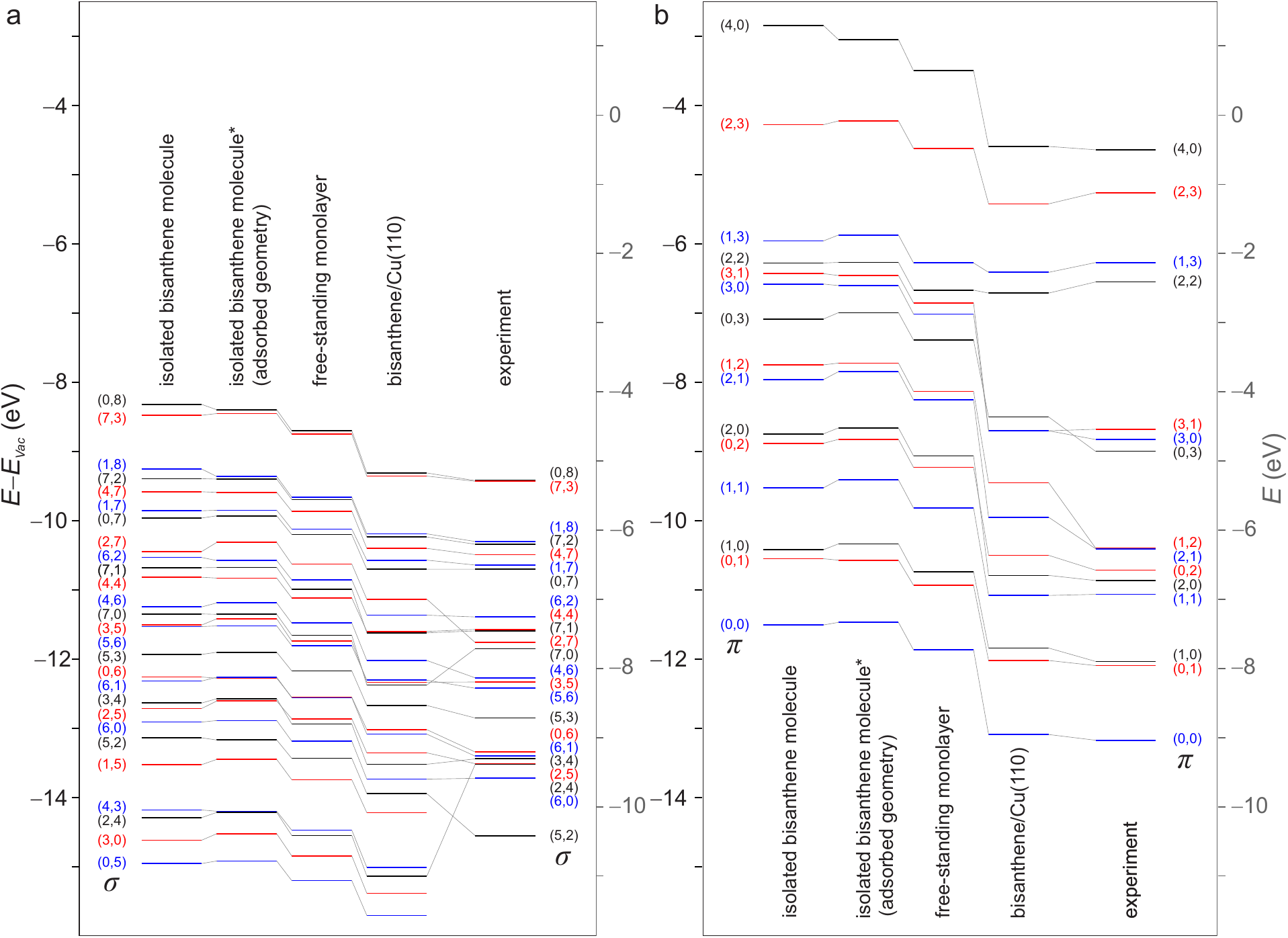}
\end{center}
	\caption{Theoretical and experimental  energies $E=-E_{\rm b}$ of (a) $\sigma$ and (b) $\pi$ orbitals of bisanthene on Cu(110) (two rightmost columns in each panel, energy axis on the right, referenced to the Fermi level). For $\sigma$ orbitals, averaged experimental energies are displayed
  (Table~\ref{tab:experiment_sigma_orbitals}, column 5). For $\pi$ orbitals, the values from the fit of the $h\nu= 35$\,eV data cube with all $\pi$ orbitals (column 3 of Table~\ref{tab:experiment_sigma_orbitals}) are shown for reasons of consistency, because this is the only data set which offers a value for every single one of the 15 $\pi$ orbitals. For the four columns on the left of each panel, the adsorption process was split into successive steps (see main text for details), and orbital energies were calculated for each of the steps (energy axis on the left, referenced to the vacuum level). All theoretical energies were calculated with DFT, employing the HSE exchange-correlation functional. The different sets of calculated orbital energies were aligned at their vacuum levels, using the calculated work function of bisanthene/Cu(110) ($\Phi=4.33$\,eV). For clarity, the color of the orbitals alternates in a cyclic pattern from black to red to blue, following the orbital energies of the isolated molecule. 
  }
 \label{fig:theoretical_orbital_energies} 
\end{figure*}

In this section we analyze the effects that determine the energy alignment of the molecular states with the states of the metal upon adsorption on the surface. To this end, we split the adsorption process into several  steps. In the first stage of this gedankenexperiment, we considered the isolated bisanthene molecule in its native, gas-phase $D_{2h}$ point group symmetry. Its calculated $\pi$ and $\sigma$ orbital energies are listed in Tables~\ref{tab:piorbitals} and \ref{tab:sigmaorbitals}, respectively, and are plotted (for the HSE functional) in the columns labeled \textit{isolated bisanthene molecule} in Fig.~\ref{fig:theoretical_orbital_energies}. The orbital energies of the isolated molecule are naturally referenced to the vacuum level (left axes). 

In the second step, we took the structural distortion of the molecule upon adsorption, as predicted by our van-der-Waals-corrected GGA optimizations, into account,\cite{Yang2019} but still computed orbital energies for an isolated molecule (no charge transfer yet). The resulting HSE orbital energies are shown in Fig.~\ref{fig:theoretical_orbital_energies} in the columns \textit{isolated bisanthene molecule* (adsorbed geometry)}. Overall, the distortion-induced level shifts are small (only in the order of $0.1$\,eV).

In the third step, we calculated the electronic structure of a free-standing layer of bisanthene molecules, \emph{i.e.}, we took the fully relaxed geometry of the bisanthene/Cu(110) interface (with molecules that are distorted relative to their gas-phase structure) and cut away the Cu(110) substrate, but kept the distorted structure of the molecules frozen. The resulting orbital energies are labeled \textit{free-standing monolayer} in Fig.~\ref{fig:theoretical_orbital_energies}. It is important to note that we applied periodic boundary conditions in this calculation and therefore each molecular state developed into a band. However, the intermolecular interactions are small, yielding band widths in the order of 0.1\,eV only. The horizontal lines plotted in Fig.~\ref{fig:theoretical_orbital_energies} represent the centers of these bands. The main result of forming a free-standing monolayer is a downshift of orbitals energies by about $0.25$ to $0.35$\,eV, essentially independent of the specific orbital. It originates from a step in the vacuum potential of the free-standing monolayer that arises from a geometry-induced dipole perpendicular to the plane of the molecule, sometimes also referred to as the bending dipole.\cite{Willenbockel2013}  In fact, the infinitely extended free-standing monolayer of bisanthene has two vacuum potentials, depending on whether one removes an electron in positive or negative $z$ direction. This step in the vacuum potential amounts to about 0.25~eV and is caused by the concave shape of the bisanthene molecule in its adsorbed state, with the terminating hydrogen atoms at the zig-zag edges of the molecule displaced by about 0.15\,{\AA} upwards compared to the central carbon atoms.
		
In the fourth and final step, we investigated how the orbital energies are further affected when the free-standing monolayer is brought into contact with the Cu(110) surface. The results of this last step of our gedankenexperiment are displayed in the columns \textit{bisanthene/Cu(110)} in Fig.~\ref{fig:theoretical_orbital_energies} and should be directly compared with the orbital energies determined (column \textit{experiment}). Inspecting the changes from the free-standing monolayer to bisanthene/Cu(110), first for the $\sigma$ orbitals, we notice an overall shift of roughly 0.5~eV to lower energies. This shift is a consequence of the so-called bonding dipole, which is due to adsorption-induced charge density rearrangements, and its concomitant step in the potential.\cite{Willenbockel2013} As mentioned before, the orbital energies predicted by HSE for the bisanthene/Cu(110) interface are in good agreement with the experimental values obtained from the orbital deconvolution procedure. However, Fig.~\ref{fig:theoretical_orbital_energies}a clearly shows that the four lowest $\sigma$ orbitals ($\sigma_{(4,3)}$, $\sigma_{(2,4)}$, $\sigma_{(3,0)}$, and $\sigma_{(0,5)}$) are pushed out of the binding energy window of our experimental data cube (which ends at $E=-11$\,eV) by the bending and bonding dipoles. With hindsight, it is therefore no surprise that $\sigma_{(4,3)}$, $\sigma_{(3,0)}$, and $\sigma_{(0,5)}$ cannot be retrieved in the fitting process (category 4 in Table~\ref{tab:experiment_sigma_orbitals}). Similarly, the (most likely) erroneous assignment of $\sigma_{(2,4)}$ to a binding energy of $9.38$\,eV is due to the fact that this orbital is expected at a binding energy just at the edge of the experimental data cube.         

In analogy to the $\sigma$ orbitals, an overall energy shift of 0.5\,eV to lower energies also affects the $\pi$ orbitals (Fig.~\ref{fig:theoretical_orbital_energies}b). However, they are additionally stabilized by the hybridization with the underlying Cu surface. This latter effect can, for instance, be observed for $\pi_{(4,0)}$ and $\pi_{(2,3)}$, \emph{i.e.}, the (former) LUMO and HOMO, which exhibit downshifts of 1.1 and 0.8\,eV, respectively. Subtracting the shift of 0.5\,eV due to the overall bonding dipole, we are left with orbital-specific bond stabilizations of approximately 0.6\,eV for the LUMO and 0.3~eV for the HOMO. Similarly, for the lowest-lying $\pi_{(0,0)}$, we can understand the energy shift of 1.25\,eV from the free-standing monolayer to bisanthene/Cu(110) as an overall shift of 0.5\,eV due to the bonding dipole and an additional bond stabilization of 0.75\,eV owing to the molecule-substrate bond.

Finally, a note on the peak energies of the theoretical MOPDOS is in order. The horizontal lines in Fig.~\ref{fig:theoretical_orbital_energies} are drawn at the global maxima of the respective MOPDOS curves in Fig.~\ref{fig:mopdos_pi_sigma}d. However, some of the orbitals exhibit a spread-out MOPDOS with several local maxima. In particular, this is true for those $\pi$ states which overlap in energy with the Cu $d$ band (approximately $-2$ to $-5$\,eV below the Fermi level). For instance, the \mbox{HOMO$-$1} and HOMO$-$2 ($\pi_{(1,3)}$ and $\pi_{(2,2)}$), although having their global MOPDOS maxima above the $d$ band, are spread out over the whole $d$ band region. For simplicity, we nevertheless assigned a single binding energy at the global maximum of each orbital's MOPDOS. In the majority of cases, \textit{i.e.}, above and below the $d$ band, this is a good approximation, as Fig.~\ref{fig:mopdos_pi_sigma} reveals.

\section{\label{sec:conclusion}Conclusion}

In the present study, we have addressed the long-standing challenge of experimentally identifying complete sets of orbitals in surface-adsorbed molecules in wide binding energy ranges by photoemission spectroscopy. Specifically, we used photoemission orbital tomography to pinpoint 15 $\pi$ and 23 $\sigma$ orbitals of bisanthene on Cu(110), covering a binding energy range of more than 10\,eV below the Fermi level. We then proceeded to use the results of our experimental study as benchmarks for density functional calculations of the molecule-metal interface with four different commonly  employed exchange-correlation potentials, showing that for the material system under study the hybrid functional HSE performs marginally better than B3LYP and much better than PBE and PBE0. More importantly, our results provide a far-reaching systematic confirmation of the often-assumed (but not stringently proven) relevance of Kohn-Sham orbitals of density functional theory as good overall approximations for experimentally measured Dyson orbitals over a larger binding energy range, both with regard to their energy and their momentum distribution. A sufficiently large set of such orbitals thus can be used for benchmarking DFT calculations (as we have done here), also opening up the possibility to reveal systematic differences in the quality of predictions between different exchange-correlation functionals on an orbital-by-orbital basis. It has to be kept in mind, though, that for individual orbitals, there may be systematic differences not only between the energies but also between the momentum distributions of Dyson and Kohn-Sham orbitals, as reference \cite{Puschnig2017} shows.   

The main challenge to be overcome in this application of photoemission orbital tomography turned out to be the reliable identification of momentum patterns in the experimental data cube, partly against a highly structured background from the substrate, especially in the energy range of the Cu $d$ band. To overcome this challenge, we used a combination of several methodologies, ranging from linear fitting to a pattern recognition algorithm. Although successful (for example, we were able to track the complete set of the 15 occupied $\pi$ orbitals through the $d$ band), it is clear that the identification of orbitals could benefit from an improved momentum and energy resolution of the photoelectron analyzer. In this respect, modern instruments offer better performance than the toroidal electron energy analyzer used in this study, which on the other hand offers the advantage of a constant light incidence geometry for each $|k|$. We speculate that with an improved analyzer performance one may be able to sharpen the experimental pDOS curves, thus yielding even better benchmarks of theory, and possibly even reveal systematic differences between Kohn-Sham and Dyson orbitals, not only in energy but also $k$ space. Nevertheless, the only tangible failure of orbital assignments even with the present analyzer appears in cases where the energy range of the experimental data cube turns out to be insufficient. Another possible application of our methodology is the retrieval of electron densities in finite energy intervals from experimental data. 

While the present work has focused on the occupied molecular orbitals, first steps to extend photoemission orbital tomography  to unoccupied states have already been undertaken. By transiently exciting electrons into unoccupied orbitals, the measurement of momentum signatures in excited states has recently been demonstrated by pump-probe angle-resolved photoemission experiments.\cite{Wallauer2021,Baumgartner2022,Neef2023} We envision that in future this will provide equally stringent experimental information for benchmarking the performance of electronic structure methods describing optically excited states.

\acknowledgements
This work was supported by the Austrian Science Fund (FWF) through project I3731, the Deutsche Forschungsgemeinschaft (DFG) through projects Po 2226/2-1, 223848855-SFB 1083 and Ri 804/8-1, and by the European Union through the Synergy Grant Orbital Cinema (101071259) by the European Research Council (ERC). C.W.~ acknowledges funding through the European Research
Council (ERC-StG 757634 “CM3”). The experiments have been performed at the Metrological Light Source (MLS) of the Physikalisch-Technische Bundesanstalt (PTB) in Berlin. We thank J. Riley (La Trobe University, Australia) for experimental support. The computations have been performed on the Vienna Scientific Computer (VSC) and the HPC facilities of the University of Graz.

\clearpage
\onecolumngrid 
\newpage
\appendix

\section{Orbital-resolved pDOS for $\pi$ and $\sigma$ orbitals}
\label{sec:appendix}

\begin{figure}[h]
\begin{center}
	\includegraphics[width=0.7\textwidth]{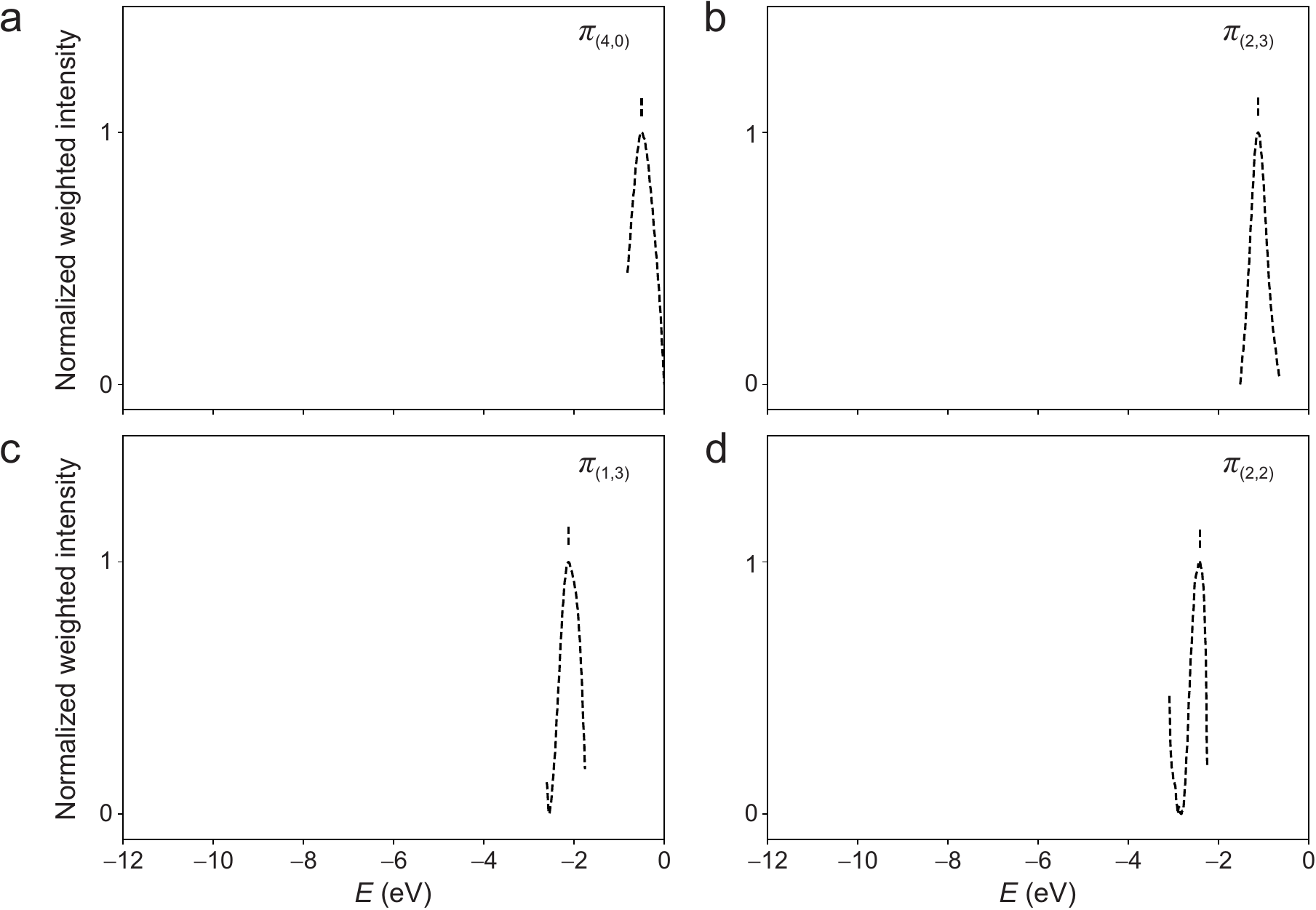}
\end{center}
	\caption{Experimental pDOS of the $\pi$ orbitals of bisanthene, part 1: panels (a) to (d). Curves are derived from the experimental data cube $I_{\rm exp}(k_x, k_y;E_{\rm b})$ measured with $h\nu$ = 35 eV, employing linear fitting with the theoretical momentum maps of all $\pi$ orbitals. Vertical bars denote the peak energies of the respective orbitals as listed in Table~\ref{tab:experiment_pi_orbitals}.}
 \label{fig:expdeconvolution_pi_part1}
\end{figure}	

\begin{figure*}[h]
\begin{center}
	\includegraphics[width=0.7\textwidth]{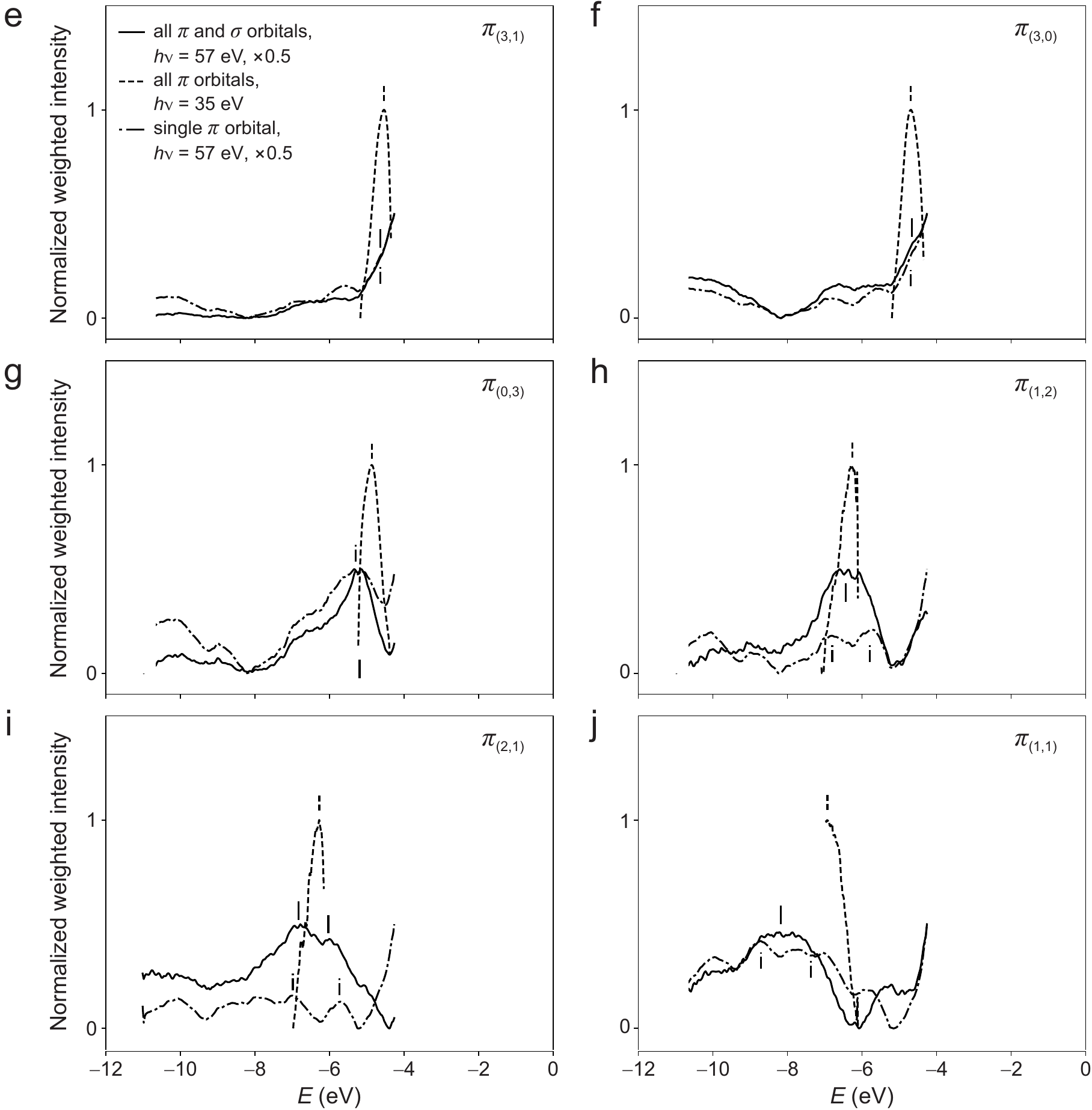}
\end{center}
	\caption{ Experimental pDOS of the $\pi$ orbitals of bisanthene, part 2: panels (e) to (j). \textit{Black solid curves} are derived from the experimental data cube $I_{\rm exp}(k_x, k_y;E_{\rm b})$ measured with $h\nu$ = 57 eV, employing linear fitting with the theoretical momentum maps of all $\pi$ and $\sigma$ orbitals. \textit{Black dash-dotted curves}: $h\nu$ = 57 eV, linear fitting with the one $\pi$ orbital in question only. \textit{Black dashed curve}: $h\nu$ = 35\,eV, linear fitting with the theoretical momentum maps of all $\pi$ orbitals. Vertical bars denote the peak energies of the respective orbitals as listed in Table~\ref{tab:experiment_pi_orbitals}.}
 \label{fig:expdeconvolution_pi_part2}
\end{figure*}

\begin{figure*}[h]
\begin{center}
	\includegraphics[width=0.7\textwidth]{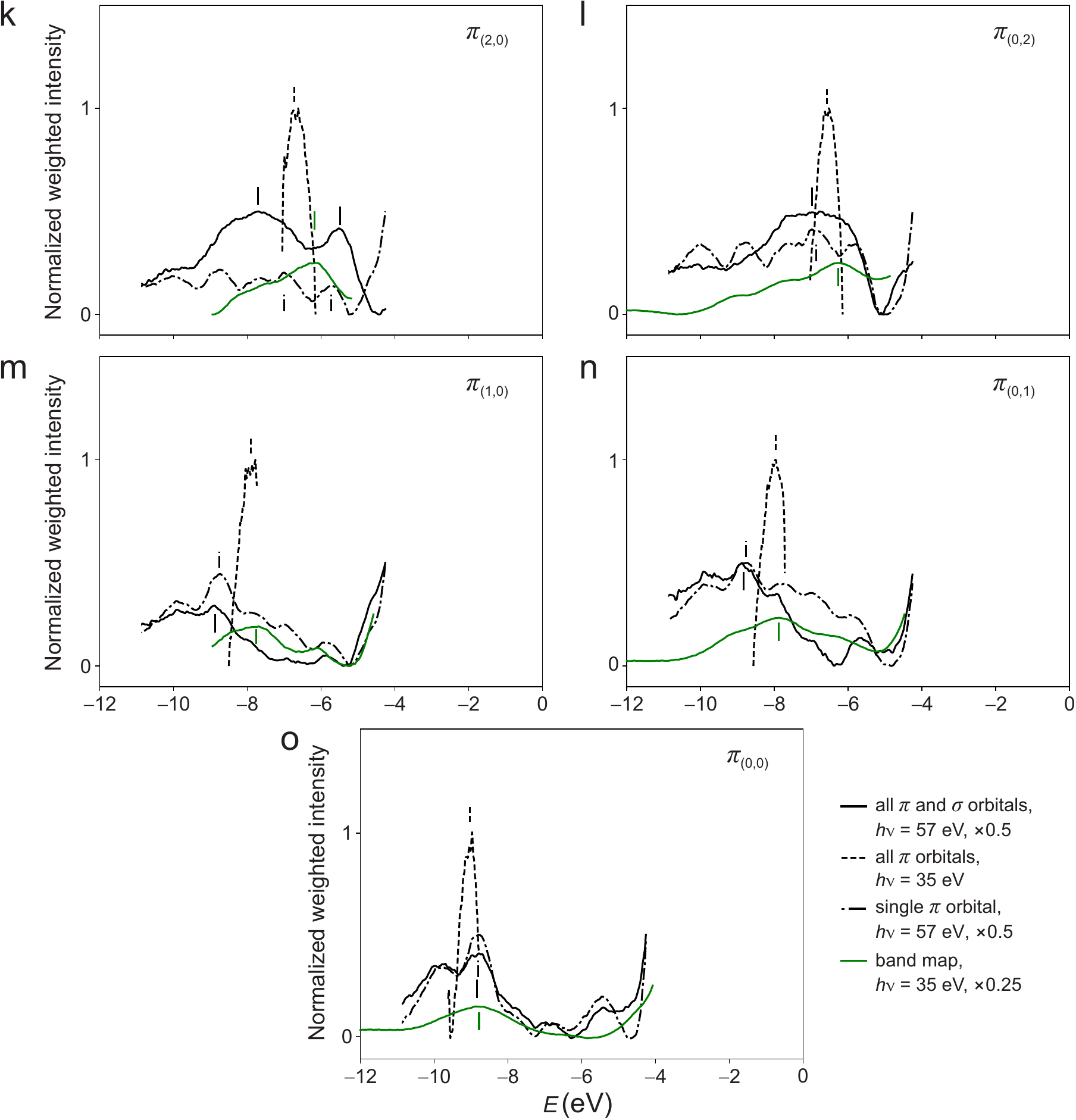}
\end{center}
	\caption{Experimental pDOS of the $\pi$ orbitals of bisanthene, part 3: panels (k) to (o). \textit{Black solid curves} are derived from the experimental data cube $I_{\rm exp}(k_x, k_y;E_{\rm b})$ measured with $h\nu$ = 57 eV, employing linear fitting with the theoretical momentum maps of all $\pi$ and $\sigma$ orbitals. \textit{Black dash-dotted curves}: $h\nu$ = 57 eV, linear fitting with the one $\pi$ orbital in question only. \textit{Black dashed curve}: $h\nu$ = 35 eV, linear fitting with the theoretical momentum maps of all $\pi$  orbitals. \textit{Green solid curves} are derived from band map fits (Fig.~\ref{fig:expbandmaps_pi}c) obtained for $h\nu$ = 35 eV. Vertical bars denote the peak energies of the respective orbitals as listed in Table~\ref{tab:experiment_pi_orbitals}.}
 \label{fig:expdeconvolution_pi_part3} 
\end{figure*}

\begin{figure*}[h]
\begin{center}
	\includegraphics[width=0.7\textwidth]{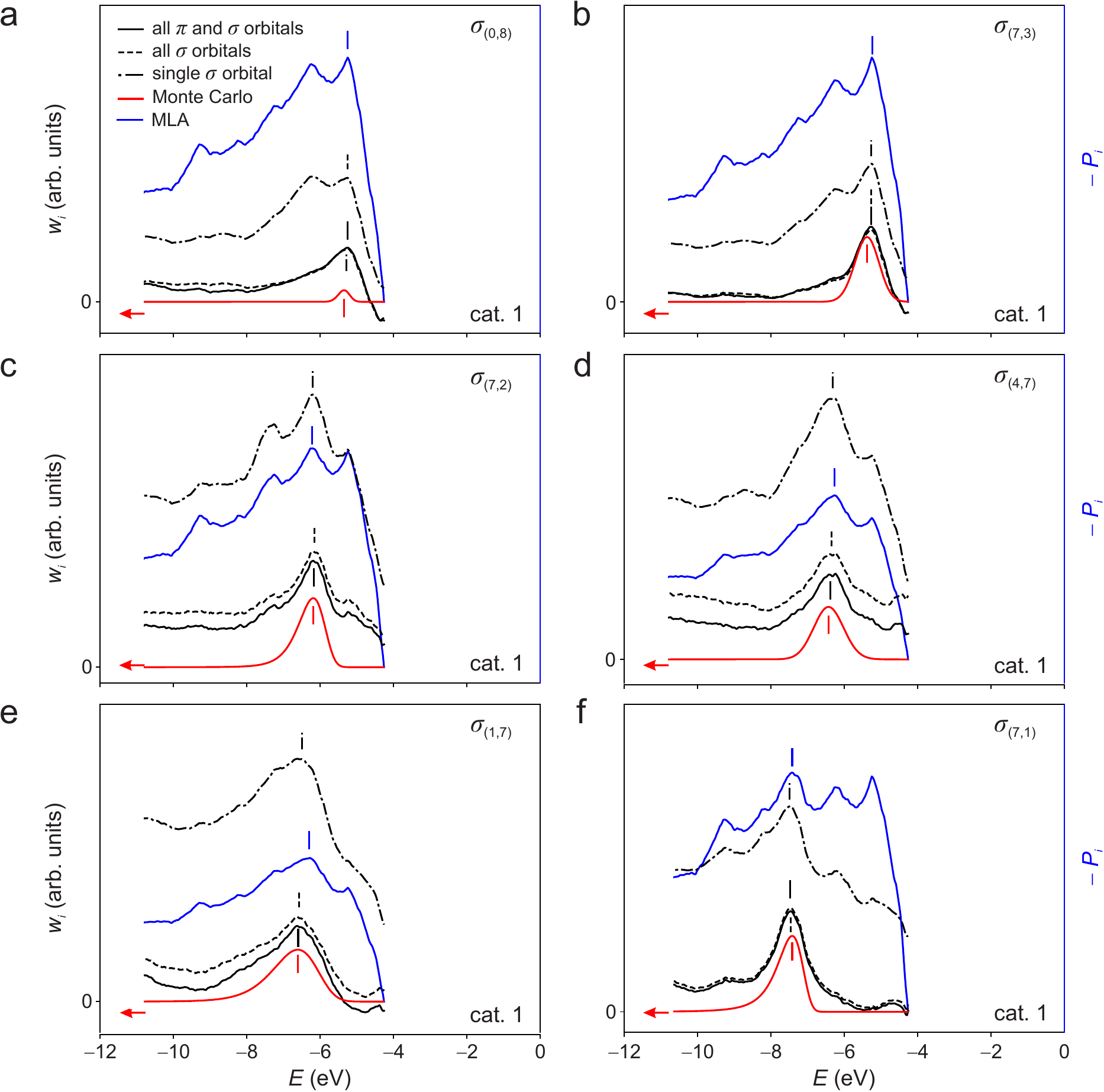}
\end{center}
	\caption{Experimental pDOS of the $\sigma$ orbitals of bisanthene, part 1: panels (a) to (f). All curves are derived from the experimental data cube $I_{\rm exp}(k_x, k_y;E_{\rm b})$ measured with $h\nu$ = 57 eV. \textit{Black solid curves} are the result of linear fitting  with  theoretical momentum maps of all $\pi$ and $\sigma$ orbitals. \textit{Black dashed curves}: linear fitting with theoretical momentum maps of all $\sigma$ orbitals. \textit{Black dash-dotted curves}: linear fitting with the theoretical momentum map of the one $\sigma$ orbital in question only. \textit{Red solid curves}: Monte Carlo (MC) fitting with theoretical momentum maps of all $\sigma$ orbitals. \textit{Blue solid curves}: Negative cost function of the multilayer averaging (MLA) evaluated between the experimental data and the single orbital $\sigma$ orbital in question. Note that linear and Monte Carlo fittings provide results in units of photoemission counts (left axis), whilst MLA delivers a unitless cost function (right axis) that has been scaled to match the other results. Vertical bars denote the peak energies of the respective orbitals as listed in Table~\ref{tab:experiment_sigma_orbitals}.}
 \label{fig:expdeconvolution_sigma_part1} 
\end{figure*}

\begin{figure*}[h]
\begin{center}
	\includegraphics[width=0.7\textwidth]{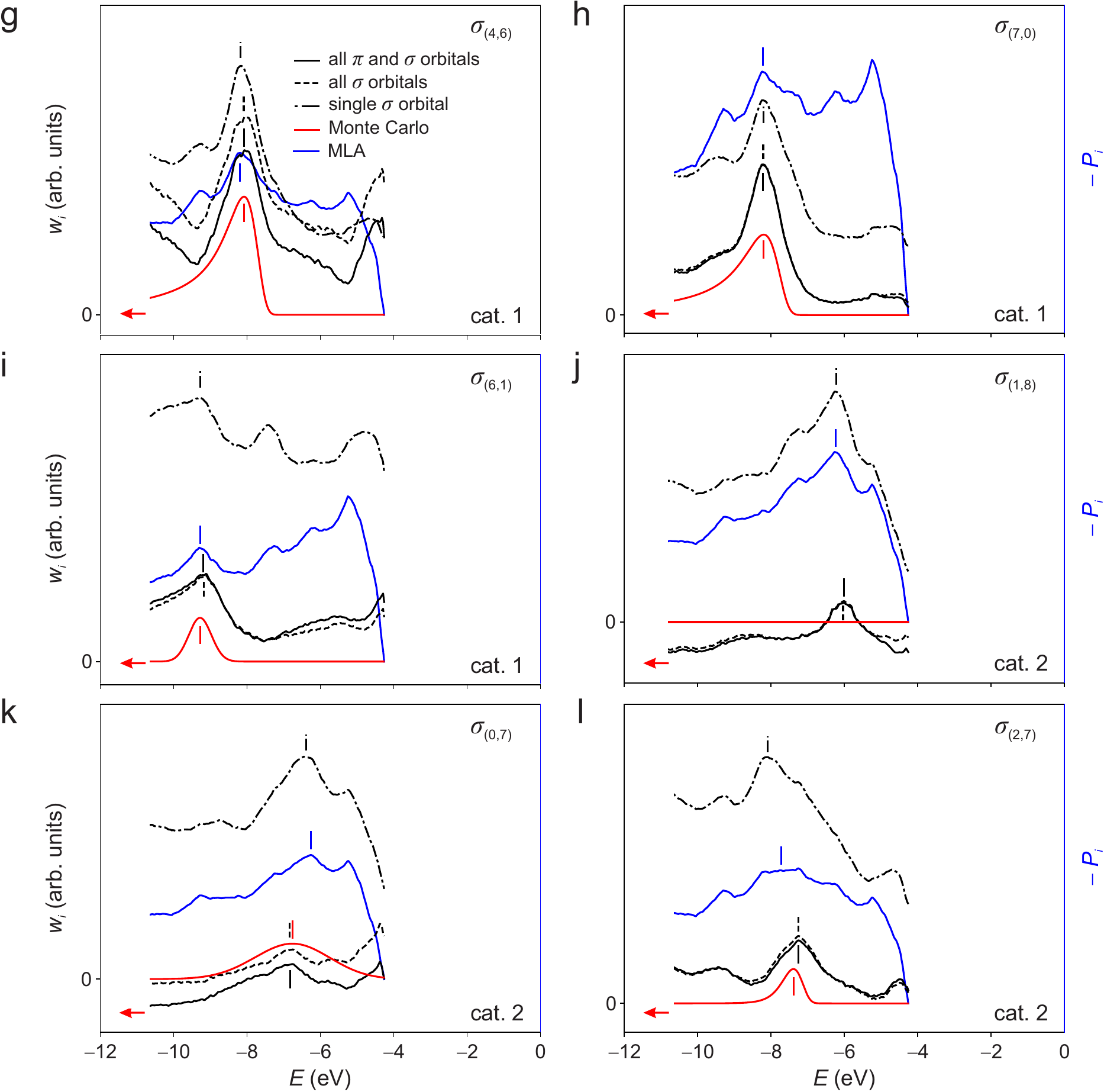}
\end{center}
	\caption{ Experimental pDOS of the $\sigma$ orbitals of bisanthene, part 2: panels (g) to (l). For details see caption of Fig.~\ref{fig:expdeconvolution_sigma_part1}. }
 \label{fig:expdeconvolution_sigma_part2}
\end{figure*}

\begin{figure*}[h]
\begin{center}
	\includegraphics[width=0.7\textwidth]{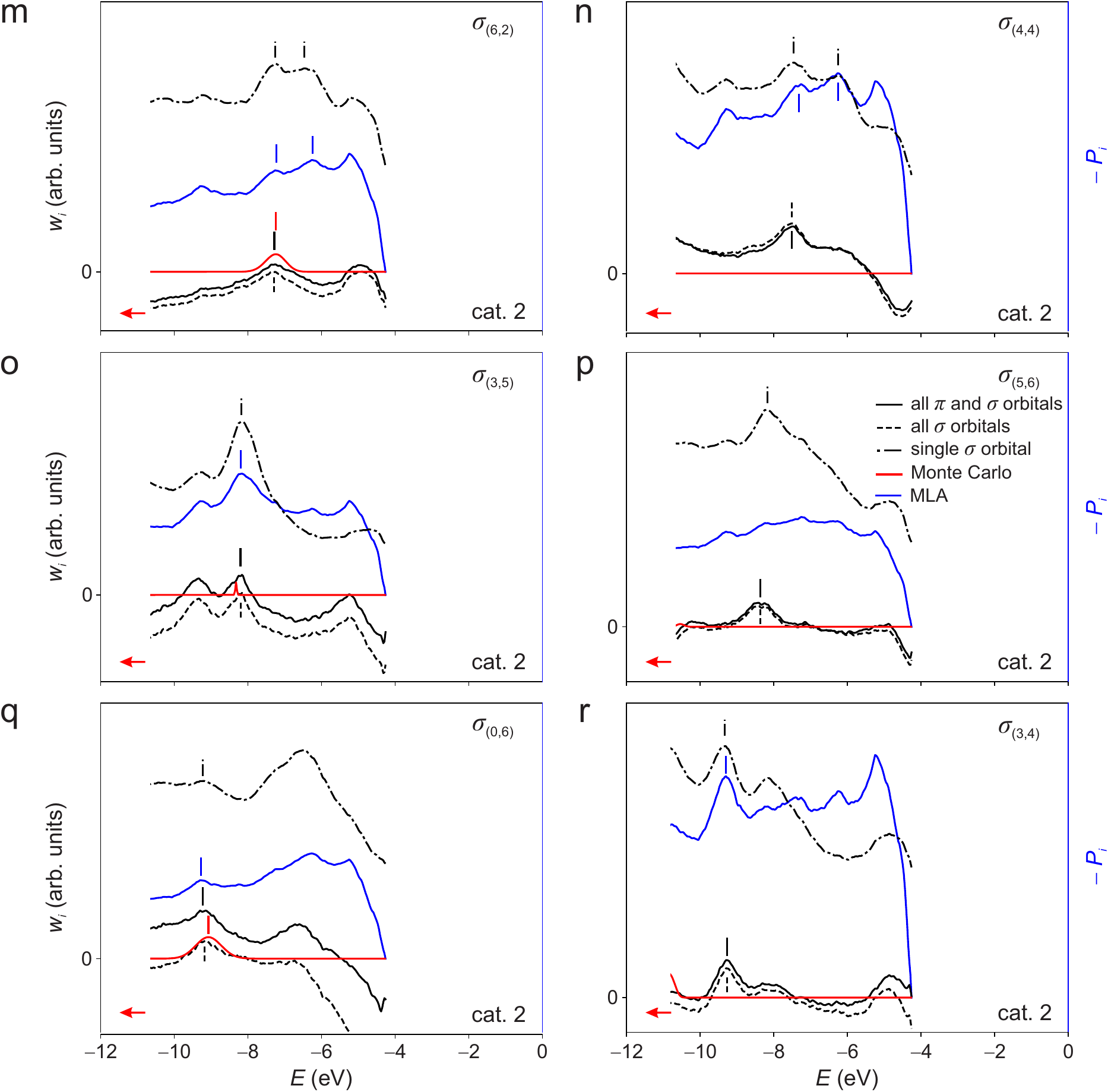}
\end{center}
	\caption{ Experimental pDOS of the $\sigma$ orbitals of bisanthene, part 3: panels (m) to (r).  For details see caption of Fig.~\ref{fig:expdeconvolution_sigma_part1}.}
\label{fig:expdeconvolution_sigma_part3}
\end{figure*}

\begin{figure*}[h]
\begin{center}
	\includegraphics[width=0.7\textwidth]{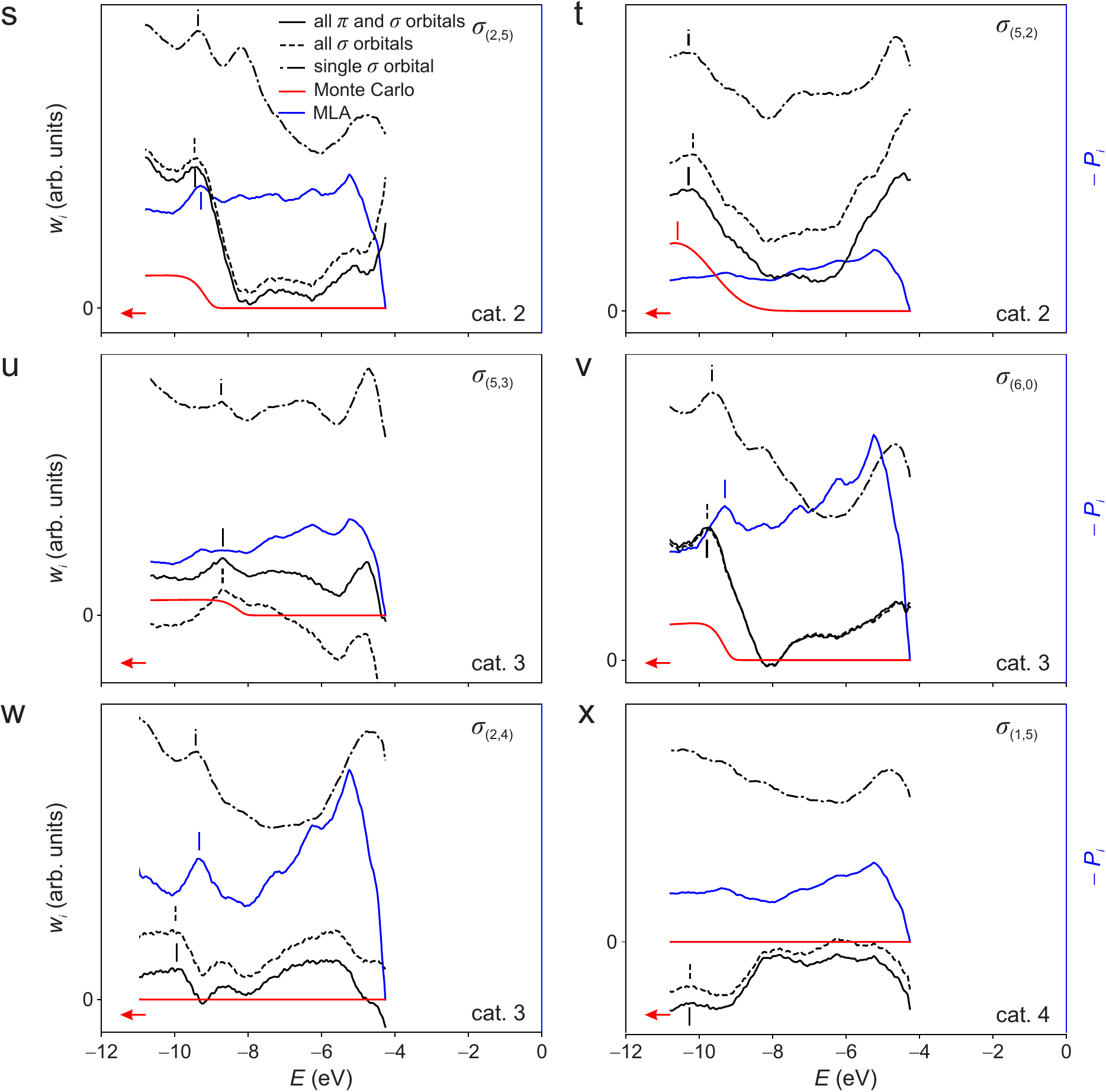}
\end{center}
	\caption{Experimental pDOS of the $\sigma$ orbitals of bisanthene, part 4: panels (s) to (x).  For details see caption of Fig.~\ref{fig:expdeconvolution_sigma_part1}. }
 \label{fig:expdeconvolution_sigma_part4} 
\end{figure*}

\begin{figure*}[h]
\begin{center}
	\includegraphics[width=0.7\textwidth]{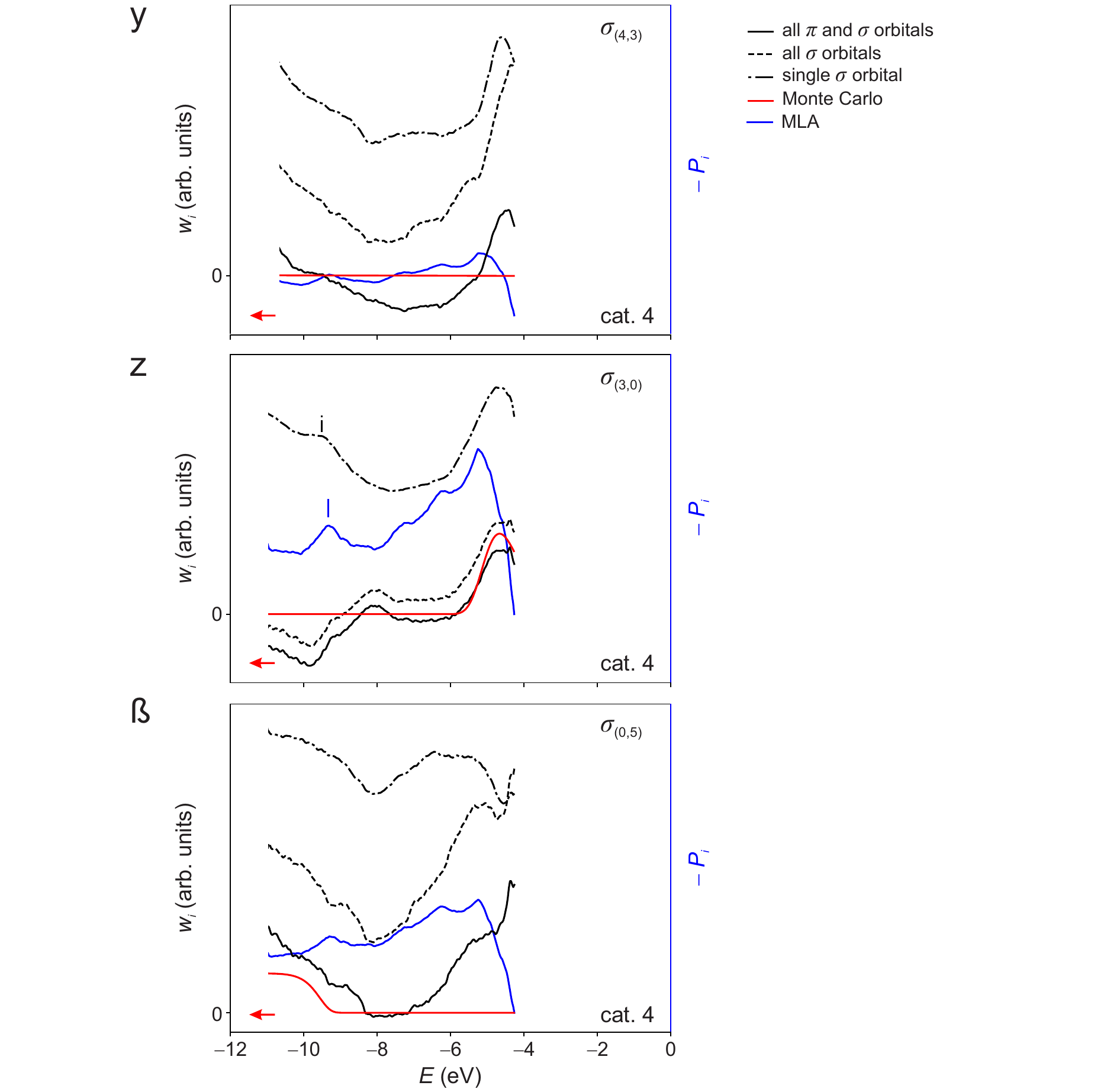}
\end{center}
	\caption{Experimental pDOS of the $\sigma$ orbitals of bisanthene, part 5: panels (y) to (ß).  For details see caption of Fig.~\ref{fig:expdeconvolution_sigma_part1}.}
\label{fig:expdeconvolution_sigma_part5} 
\end{figure*}	

\clearpage
\section{Intramolecular band dispersion}

\begin{figure*}[b]
\begin{center}
	\includegraphics[width=\textwidth]{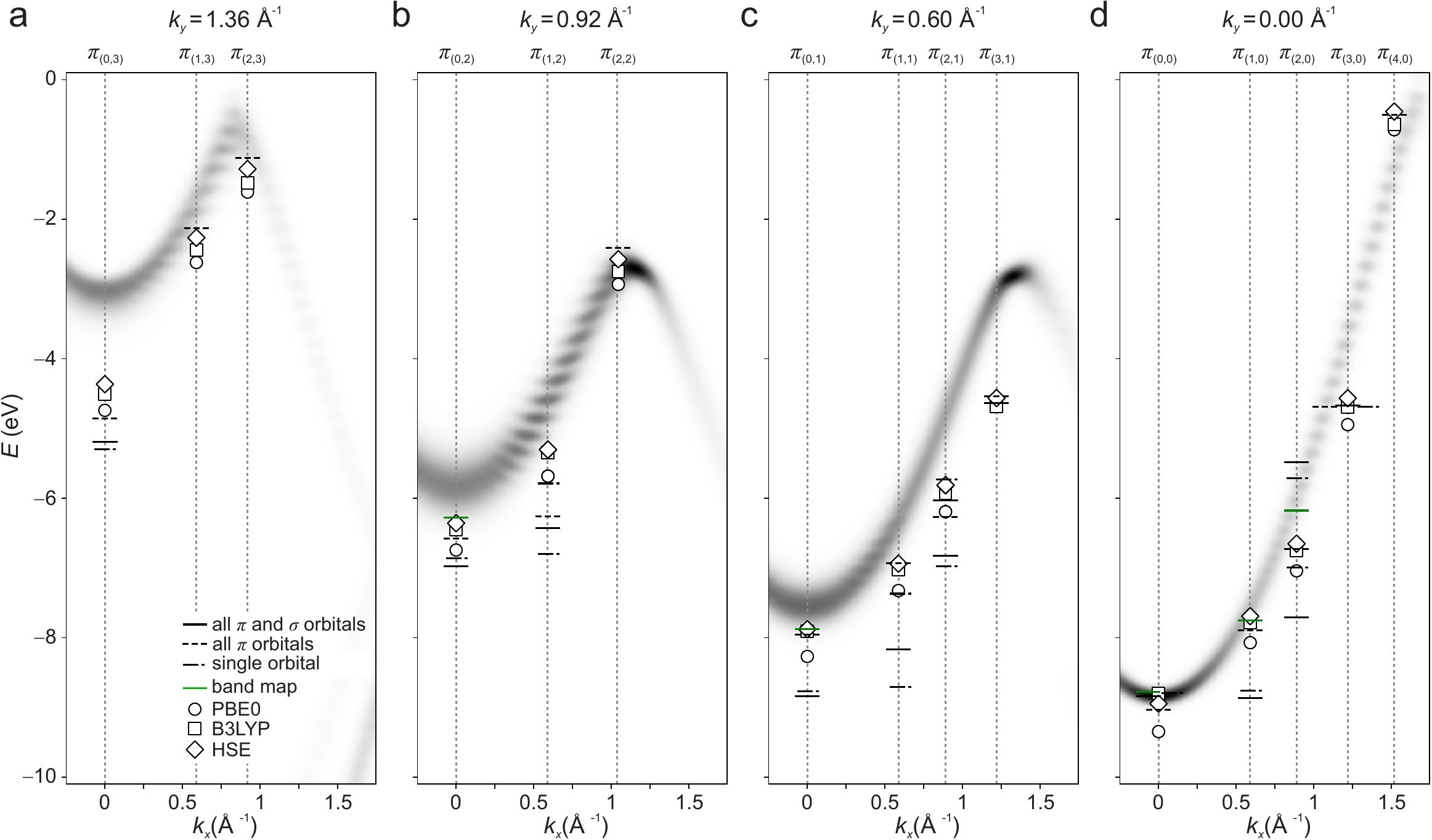}
 \end{center}
	\caption{Intramolecular dispersion of  $\pi$ bands, part 1: panels (a) $\pi_{(n,3)}$, (b) $\pi_{(n,2)}$, (c) $\pi_{(n,1)}$, (d) $\pi_{(n,0)}$.  Horizontal bars denote the orbital energies derived from the corresponding experimental data cubes $I_{\rm exp}(k_x, k_y;E_{\rm b})$, using the following fitting methodologies: linear fitting of the 57-eV data cube with the theoretical momentum map of the single corresponding $\pi$ orbital (\textit{black dash-dotted}), linear fitting of the 57-eV data cube with theoretical momentum maps of all $\pi$ and $\sigma$ orbitals (\textit{black solid}), linear fitting of the 35-eV data cube with theoretical momentum maps of all $\pi$ orbitals (\textit{black dashed}). \textit{Green bars} denote the orbital energies derived from band maps. Symbols denote the DFT-calculated orbital energies of bisanthene/Cu(110), using the following exchange-correlation functionals: PBE0 (\textit{open circles}), B3LYP (\textit{open squares}), HSE (\textit{open diamonds}). The bars and symbols appear at the $k_x$, $k_y$ position of the strongest emission lobe of the corresponding orbital in the dispersion direction of the respective band (\textit{dotted vertical lines}). Note that the $k_x$ or $k_y$ at which the data cube of graphene is sliced to generate these band maps are chosen such that the main lobes of the respective orbitals are located within the band map. These values are: (a) $k_y = 1.36$~\AA$^{-1}$, (b) $k_y = 0.92$~\AA$^{-1}$, (c) $k_y = 0.60$~\AA$^{-1}$, and (d) $k_y = 0.00$~\AA$^{-1}$.}
  \label{fig:expdispersion_pi_part1} 
\end{figure*}	

\begin{figure*}[h]
\begin{center}
	\includegraphics[width=\textwidth]{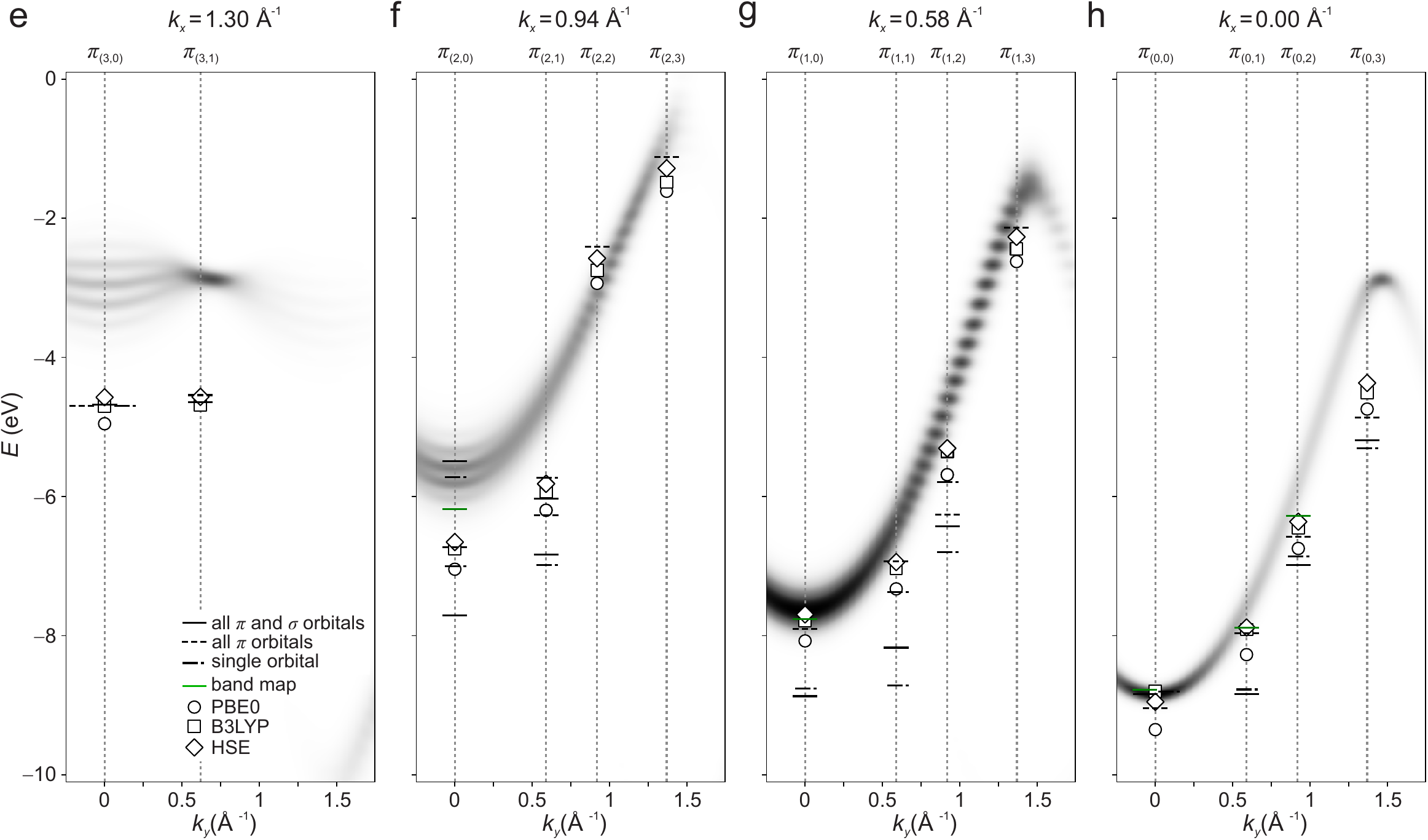}
\end{center}
	\caption{Intramolecular dispersion of  $\pi$ bands, part 2: panels (e) $\pi_{(3,m)}$, (f) $\pi_{(2,m)}$, (g) $\pi_{(1,m)}$, (h) $\pi_{(0,m)}$. Note that the $k_x$ or $k_y$ at which the data cube of graphene is sliced to generate these band maps are chosen such that the main lobes of the respective orbitals are located within the band map. These values are: (e) $k_x = 1.30$~\AA$^{-1}$, (f) $k_x = 0.94$~\AA$^{-1}$, (g) $k_x = 0.58$~\AA$^{-1}$, and (h) $k_x = 0.00$~\AA$^{-1}$. For details see caption of Fig.~\ref{fig:expdispersion_pi_part1}. } 
 \label{fig:expdispersion_pi_part2} 
\end{figure*}	

\begin{figure*}[h]
\begin{center}
	\includegraphics[width=\textwidth]{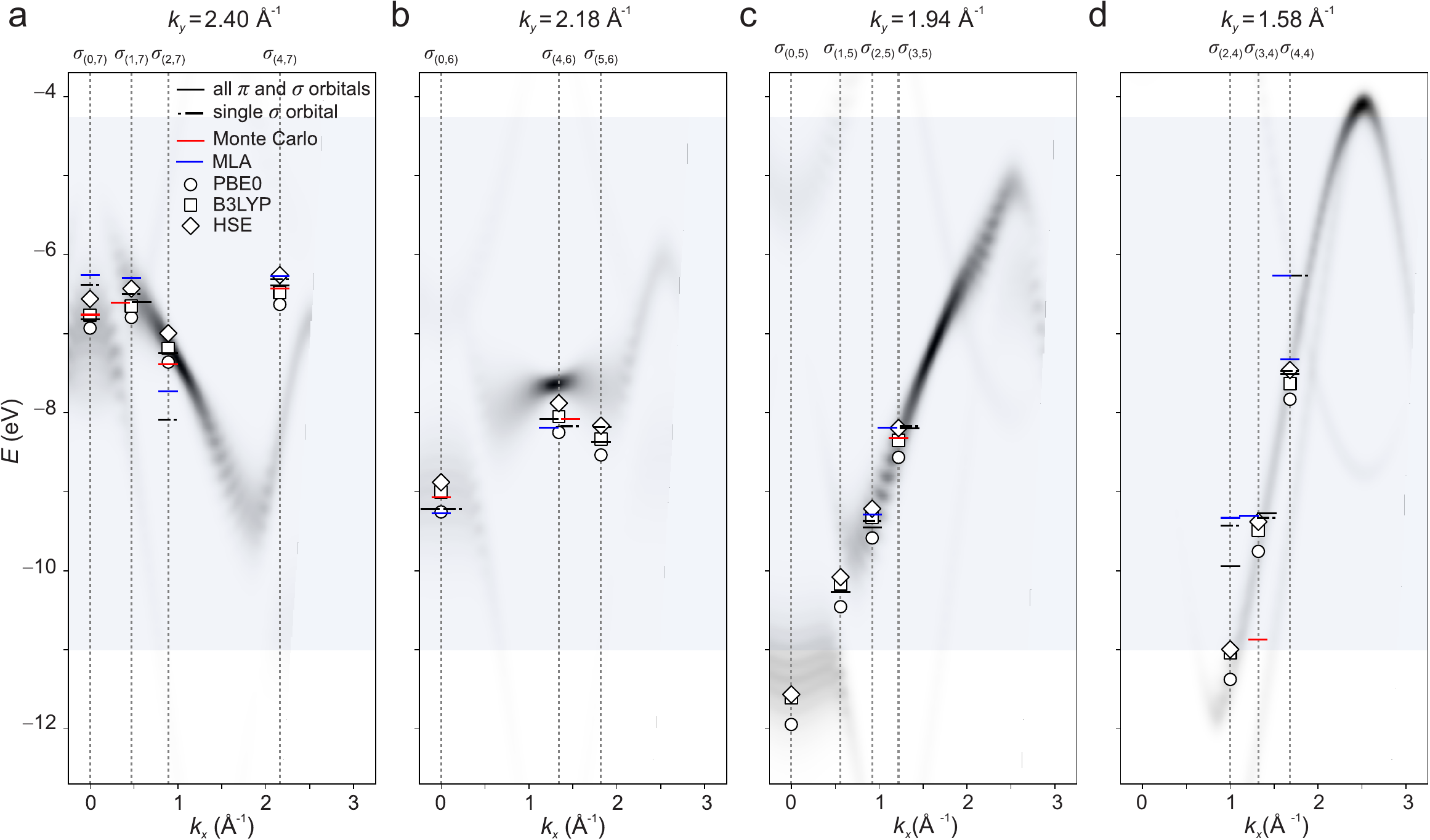}
\end{center}
	\caption{Intramolecular dispersion of $\sigma$ bands, part 1: panels (a) $\sigma_{(n,7)}$, (b) $\sigma_{(n,6)}$, (c) $\sigma_{(n,5)}$, (d) $\sigma_{(n,4)}$. Horizontal bars denote the orbital energies derived from the experimental data cube $I_{\rm exp}(k_x, k_y;E_{\rm b})$ measured at $h\nu$ = 57 eV, using the following fitting methodologies: linear fitting with the theoretical momentum map of the single corresponding $\sigma$ orbital (\textit{black dash-dotted}), linear fitting with theoretical momentum maps of all $\pi$ and $\sigma$ orbitals (\textit{black solid}), MLA pattern recognition (\textit{blue}), and Monte Carlo fitting with theoretical momentum maps of all $\sigma$ orbitals (\textit{red}). The gray boxes mark the binding energy range in which the experimental data were fitted. Note that the fit results for the linear all-$\sigma$ and all-$\pi$-and-$\sigma$ fits are essentially identical (see Fig.~\ref{fig:expdeconvolution_sigma_part1} to \ref{fig:expdeconvolution_sigma_part5}) and therefore only the latter have been included in the figure. Symbols denote the DFT-calculated orbital energies of bisanthene/Cu(110), using the following exchange-correlation functionals: PBE0 (\textit{open circles}), B3LYP (\textit{open squares}), HSE (\textit{open diamonds}). The bars and symbols appear at the $k_x$, $k_y$ position of the strongest emission lobe of the corresponding orbital in the dispersion direction of the respective band (\textit{dotted vertical lines}). Note that the $k_x$ or $k_y$ at which the data cube of graphene is sliced to generate these band maps are chosen such that the main lobes of the respective orbitals are located within the band map. These values are: (a) $k_y = 2.40$~\AA$^{-1}$, (b) $k_y = 2.18$~\AA$^{-1}$, (c) $k_y = 1.94$~\AA$^{-1}$, and (d) $k_y = 1.58$~\AA$^{-1}$. }
 \label{fig:expdispersion_sigma_part1} 
\end{figure*}	

\begin{figure*}[h]
\begin{center}
	\includegraphics[width=\textwidth]{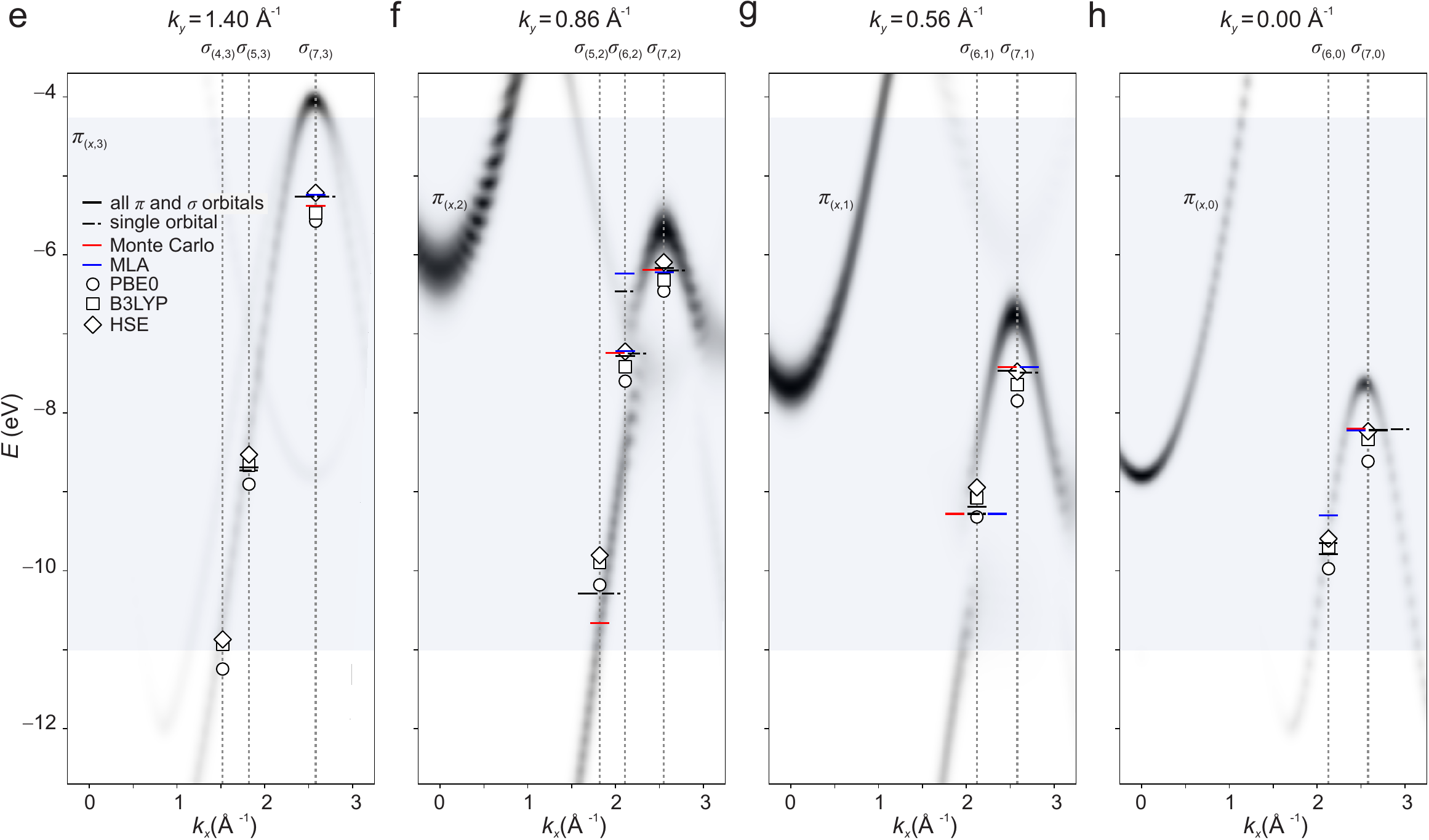}
\end{center}
	\caption{Intramolecular dispersion of $\sigma$  bands, part 2: panels (e) $\sigma_{(n,3)}$, (f) $\sigma_{(n,2)}$, (g) $\sigma_{(n,1)}$, (h) $\sigma_{(n,0)}$. Note that the $k_x$ or $k_y$ at which the data cube of graphene is sliced to generate these band maps are chosen such that the main lobes of the respective orbitals are located within the band map. These values are: (e) $k_y = 1.40$~\AA$^{-1}$, (f) $k_y = 0.86$~\AA$^{-1}$, (g) $k_y = 0.56$~\AA$^{-1}$, and (h) $k_y = 0.00$~\AA$^{-1}$. For details see caption of Fig.~\ref{fig:expdispersion_sigma_part1}.}
 \label{fig:expdispersion_sigma_part2} 
\end{figure*}

\begin{figure*}[h]
\begin{center}
	\includegraphics[width=\textwidth]{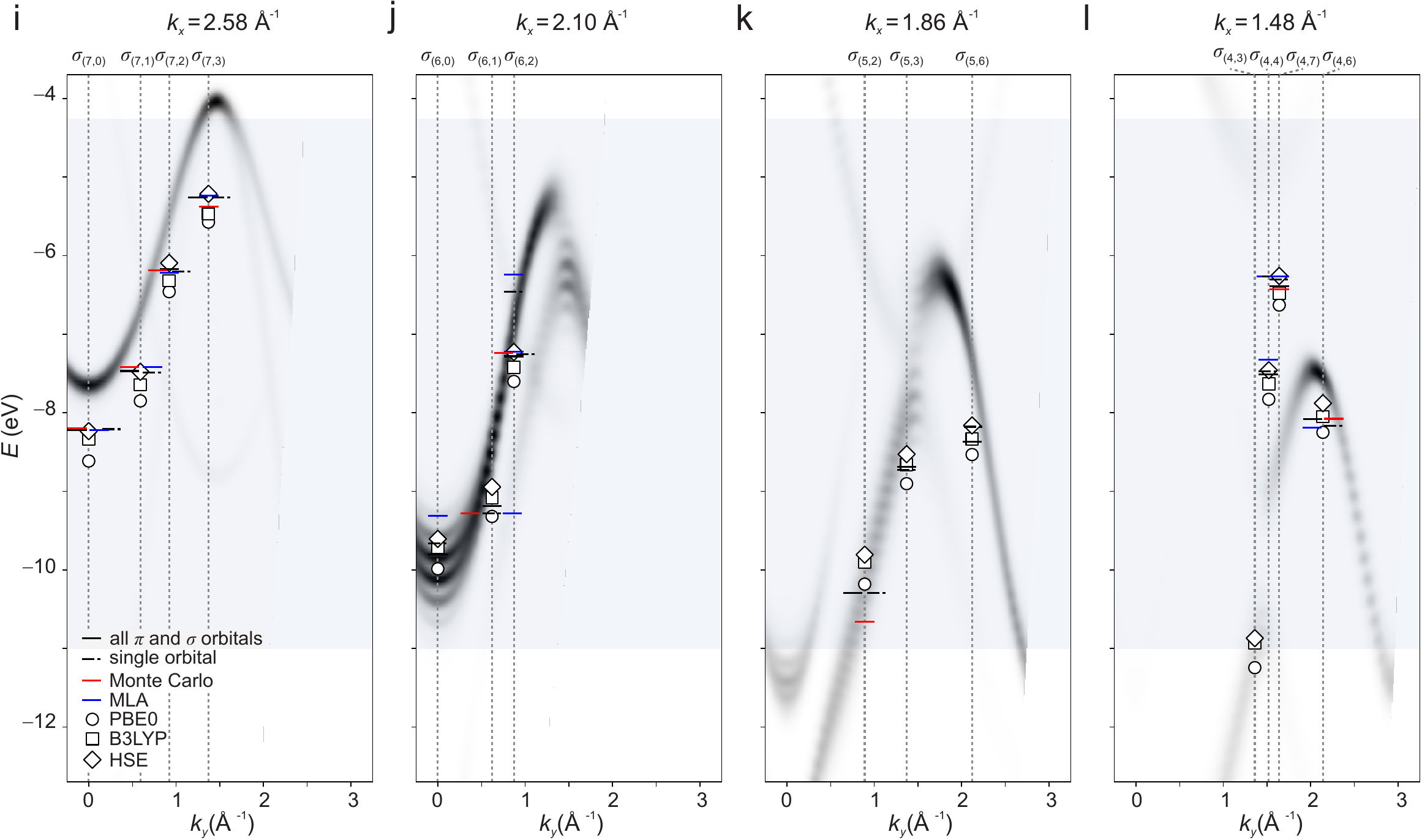}
\end{center}
	\caption{Intramolecular dispersion of  $\sigma$  bands, part 3: panels (i) $\sigma_{(7,m)}$, (j) $\sigma_{(6,m)}$, (k) $\sigma_{(5,m)}$, (l) $\sigma_{(4,m)}$. Note that the $k_x$ or $k_y$ at which the data cube of graphene is sliced to generate these band maps are chosen such that the main lobes of the respective orbitals are located within the band map. These values are: (i) $k_x = 2.58$~\AA$^{-1}$, (j) $k_x = 2.10$~\AA$^{-1}$, (k) $k_x = 1.86$~\AA$^{-1}$, and (l) $k_x = 1.48$~\AA$^{-1}$. For details see caption of Fig.~\ref{fig:expdispersion_sigma_part1}.}
 \label{fig:expdispersion_sigma_part3} 
\end{figure*}	

\begin{figure*}[h]
\begin{center}
	\includegraphics[width=\textwidth]{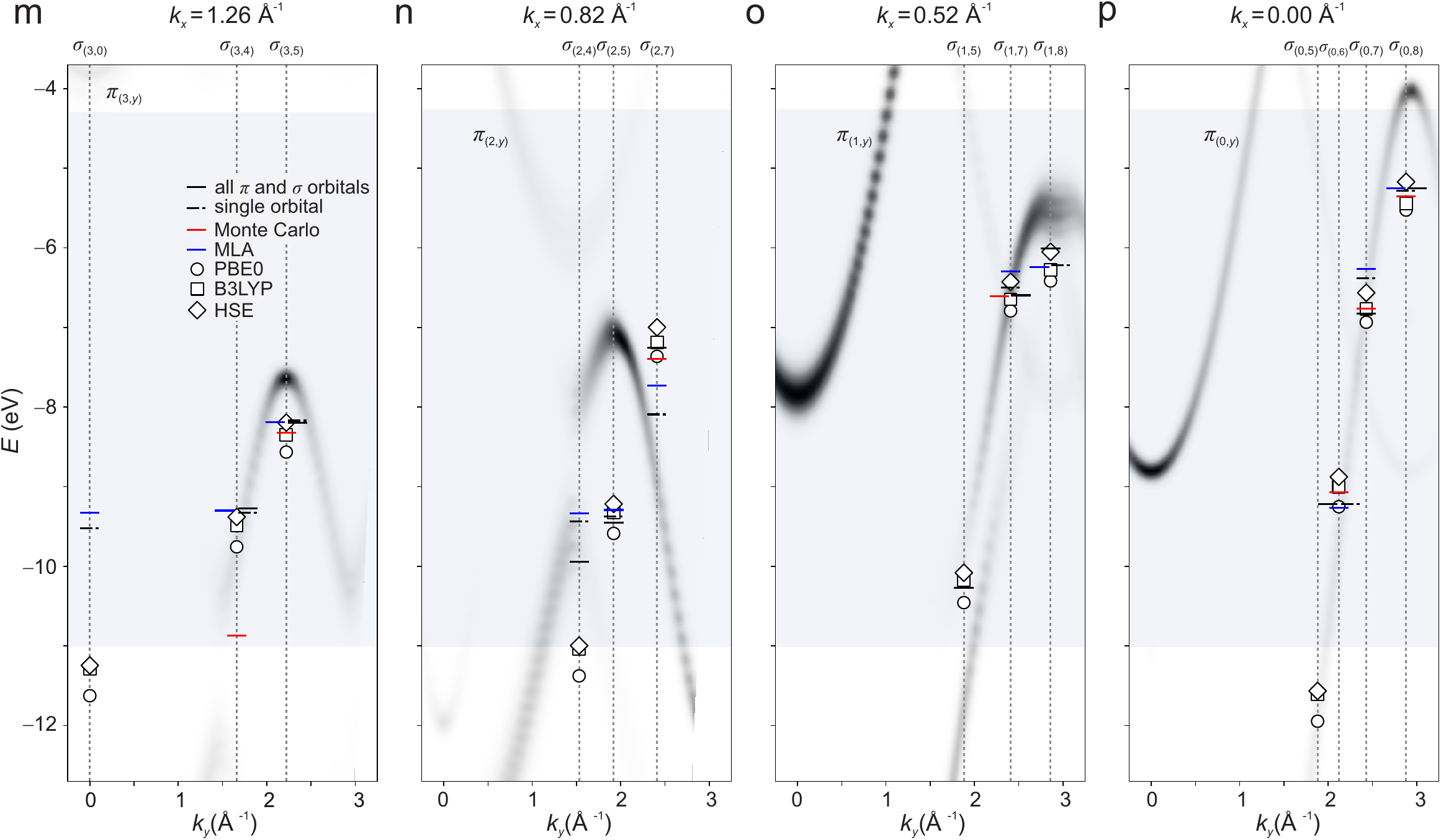}
\end{center}
	\caption{Intramolecular dispersion of  $\sigma$  bands, part 4: panels (m) $\sigma_{(3,m)}$, (n) $\sigma_{(2,m)}$, (o) $\sigma_{(1,m)}$, (p) $\sigma_{(0,m)}$. Note that the $k_x$ or $k_y$ at which the data cube of graphene is sliced to generate these band maps are chosen such that the main lobes of the respective orbitals are located within the band map. These values are: (m) $k_x = 1.26$~\AA$^{-1}$, (n) $k_x = 0.82$~\AA$^{-1}$, (o) $k_x = 0.52$~\AA$^{-1}$, and (p) $k_x = 0.00$~\AA$^{-1}$. For details see caption of Fig.~\ref{fig:expdispersion_sigma_part1}. }
 \label{fig:expdispersion_sigma_part4}
\end{figure*}

\clearpage
\twocolumngrid

\bibliography{bib} 

\begin{thebibliography}{89}%
\makeatletter
\providecommand \@ifxundefined [1]{%
 \@ifx{#1\undefined}
}%
\providecommand \@ifnum [1]{%
 \ifnum #1\expandafter \@firstoftwo
 \else \expandafter \@secondoftwo
 \fi
}%
\providecommand \@ifx [1]{%
 \ifx #1\expandafter \@firstoftwo
 \else \expandafter \@secondoftwo
 \fi
}%
\providecommand \natexlab [1]{#1}%
\providecommand \enquote  [1]{``#1''}%
\providecommand \bibnamefont  [1]{#1}%
\providecommand \bibfnamefont [1]{#1}%
\providecommand \citenamefont [1]{#1}%
\providecommand \href@noop [0]{\@secondoftwo}%
\providecommand \href [0]{\begingroup \@sanitize@url \@href}%
\providecommand \@href[1]{\@@startlink{#1}\@@href}%
\providecommand \@@href[1]{\endgroup#1\@@endlink}%
\providecommand \@sanitize@url [0]{\catcode `\\12\catcode `\$12\catcode
  `\&12\catcode `\#12\catcode `\^12\catcode `\_12\catcode `\%12\relax}%
\providecommand \@@startlink[1]{}%
\providecommand \@@endlink[0]{}%
\providecommand \url  [0]{\begingroup\@sanitize@url \@url }%
\providecommand \@url [1]{\endgroup\@href {#1}{\urlprefix }}%
\providecommand \urlprefix  [0]{URL }%
\providecommand \Eprint [0]{\href }%
\providecommand \doibase [0]{http://dx.doi.org/}%
\providecommand \selectlanguage [0]{\@gobble}%
\providecommand \bibinfo  [0]{\@secondoftwo}%
\providecommand \bibfield  [0]{\@secondoftwo}%
\providecommand \translation [1]{[#1]}%
\providecommand \BibitemOpen [0]{}%
\providecommand \bibitemStop [0]{}%
\providecommand \bibitemNoStop [0]{.\EOS\space}%
\providecommand \EOS [0]{\spacefactor3000\relax}%
\providecommand \BibitemShut  [1]{\csname bibitem#1\endcsname}%
\let\auto@bib@innerbib\@empty
\bibitem [{\citenamefont {Ishii}\ \emph {et~al.}(1999)\citenamefont {Ishii},
  \citenamefont {Sugiyama}, \citenamefont {Ito},\ and\ \citenamefont
  {Seki}}]{Ishii1999}%
  \BibitemOpen
  \bibfield  {author} {\bibinfo {author} {\bibfnamefont {H.}~\bibnamefont
  {Ishii}}, \bibinfo {author} {\bibfnamefont {K.}~\bibnamefont {Sugiyama}},
  \bibinfo {author} {\bibfnamefont {E.}~\bibnamefont {Ito}}, \ and\ \bibinfo
  {author} {\bibfnamefont {K.}~\bibnamefont {Seki}},\ }\href {\doibase
  https://doi.org/10.1002/(SICI)1521-4095(199906)11:8<605::AID-ADMA605>3.0.CO;2-Q}
  {\bibfield  {journal} {\bibinfo  {journal} {Advanced Materials}\ }\textbf
  {\bibinfo {volume} {11}},\ \bibinfo {pages} {605} (\bibinfo {year}
  {1999})}\BibitemShut {NoStop}%
\bibitem [{\citenamefont {Kahn}\ \emph {et~al.}(2003)\citenamefont {Kahn},
  \citenamefont {Koch},\ and\ \citenamefont {Gao}}]{Kahn2003}%
  \BibitemOpen
  \bibfield  {author} {\bibinfo {author} {\bibfnamefont {A.}~\bibnamefont
  {Kahn}}, \bibinfo {author} {\bibfnamefont {N.}~\bibnamefont {Koch}}, \ and\
  \bibinfo {author} {\bibfnamefont {W.}~\bibnamefont {Gao}},\ }\href {\doibase
  https://doi.org/10.1002/polb.10642} {\bibfield  {journal} {\bibinfo
  {journal} {Journal of Polymer Science Part B: Polymer Physics}\ }\textbf
  {\bibinfo {volume} {41}},\ \bibinfo {pages} {2529} (\bibinfo {year}
  {2003})}\BibitemShut {NoStop}%
\bibitem [{\citenamefont {Ueno}\ and\ \citenamefont {Kera}(2008)}]{Ueno2008}%
  \BibitemOpen
  \bibfield  {author} {\bibinfo {author} {\bibfnamefont {N.}~\bibnamefont
  {Ueno}}\ and\ \bibinfo {author} {\bibfnamefont {S.}~\bibnamefont {Kera}},\
  }\href {\doibase https://doi.org/10.1016/j.progsurf.2008.10.002} {\bibfield
  {journal} {\bibinfo  {journal} {Progress in Surface Science}\ }\textbf
  {\bibinfo {volume} {83}},\ \bibinfo {pages} {490} (\bibinfo {year}
  {2008})}\BibitemShut {NoStop}%
\bibitem [{\citenamefont {Braun}\ \emph {et~al.}(2009)\citenamefont {Braun},
  \citenamefont {Salaneck},\ and\ \citenamefont {Fahlman}}]{Braun2009}%
  \BibitemOpen
  \bibfield  {author} {\bibinfo {author} {\bibfnamefont {S.}~\bibnamefont
  {Braun}}, \bibinfo {author} {\bibfnamefont {W.~R.}\ \bibnamefont {Salaneck}},
  \ and\ \bibinfo {author} {\bibfnamefont {M.}~\bibnamefont {Fahlman}},\ }\href
  {\doibase https://doi.org/10.1002/adma.200802893} {\bibfield  {journal}
  {\bibinfo  {journal} {Advanced Materials}\ }\textbf {\bibinfo {volume}
  {21}},\ \bibinfo {pages} {1450} (\bibinfo {year} {2009})}\BibitemShut
  {NoStop}%
\bibitem [{\citenamefont {B\"urker}\ \emph {et~al.}(2013)\citenamefont
  {B\"urker}, \citenamefont {Chen}, \citenamefont {Chou}, \citenamefont
  {Flores}, \citenamefont {Gerlach}, \citenamefont {Hosokai}, \citenamefont
  {Huang}, \citenamefont {Kahn}, \citenamefont {Kanai}, \citenamefont {Kera},
  \citenamefont {Koch}, \citenamefont {Kronik}, \citenamefont {Morikawa},
  \citenamefont {Ortega}, \citenamefont {Qi}, \citenamefont {Schreiber},
  \citenamefont {Ueno}, \citenamefont {Wee},\ and\ \citenamefont
  {Wong}}]{MoleculeMetalInterface}%
  \BibitemOpen
  \bibfield  {author} {\bibinfo {author} {\bibfnamefont {C.}~\bibnamefont
  {B\"urker}}, \bibinfo {author} {\bibfnamefont {W.}~\bibnamefont {Chen}},
  \bibinfo {author} {\bibfnamefont {W.-Y.}\ \bibnamefont {Chou}}, \bibinfo
  {author} {\bibfnamefont {F.}~\bibnamefont {Flores}}, \bibinfo {author}
  {\bibfnamefont {A.}~\bibnamefont {Gerlach}}, \bibinfo {author} {\bibfnamefont
  {T.}~\bibnamefont {Hosokai}}, \bibinfo {author} {\bibfnamefont
  {H.}~\bibnamefont {Huang}}, \bibinfo {author} {\bibfnamefont
  {A.}~\bibnamefont {Kahn}}, \bibinfo {author} {\bibfnamefont {K.}~\bibnamefont
  {Kanai}}, \bibinfo {author} {\bibfnamefont {S.}~\bibnamefont {Kera}},
  \bibinfo {author} {\bibfnamefont {N.}~\bibnamefont {Koch}}, \bibinfo {author}
  {\bibfnamefont {L.}~\bibnamefont {Kronik}}, \bibinfo {author} {\bibfnamefont
  {Y.}~\bibnamefont {Morikawa}}, \bibinfo {author} {\bibfnamefont
  {J.}~\bibnamefont {Ortega}}, \bibinfo {author} {\bibfnamefont {D.-C.}\
  \bibnamefont {Qi}}, \bibinfo {author} {\bibfnamefont {F.}~\bibnamefont
  {Schreiber}}, \bibinfo {author} {\bibfnamefont {N.}~\bibnamefont {Ueno}},
  \bibinfo {author} {\bibfnamefont {A.~T.}\ \bibnamefont {Wee}}, \ and\
  \bibinfo {author} {\bibfnamefont {S.~L.}\ \bibnamefont {Wong}},\ }\href@noop
  {} {\emph {\bibinfo {title} {The Molecule-Metal Interface}}},\ edited by\
  \bibinfo {editor} {\bibfnamefont {N.}~\bibnamefont {Koch}}, \bibinfo {editor}
  {\bibfnamefont {N.}~\bibnamefont {Ueno}}, \ and\ \bibinfo {editor}
  {\bibfnamefont {A.~T.}\ \bibnamefont {Wee}}\ (\bibinfo  {publisher}
  {WILEY-VCH Verlag GmbH \& Co. KGaA},\ \bibinfo {year} {2013})\BibitemShut
  {NoStop}%
\bibitem [{\citenamefont {Willenbockel}\ \emph
  {et~al.}(2015{\natexlab{a}})\citenamefont {Willenbockel}, \citenamefont
  {Lüftner}, \citenamefont {Stadtmüller}, \citenamefont {Koller},
  \citenamefont {Kumpf}, \citenamefont {Soubatch}, \citenamefont {Puschnig},
  \citenamefont {Ramsey},\ and\ \citenamefont {Tautz}}]{Willenbockel2014}%
  \BibitemOpen
  \bibfield  {author} {\bibinfo {author} {\bibfnamefont {M.}~\bibnamefont
  {Willenbockel}}, \bibinfo {author} {\bibfnamefont {D.}~\bibnamefont
  {Lüftner}}, \bibinfo {author} {\bibfnamefont {B.}~\bibnamefont
  {Stadtmüller}}, \bibinfo {author} {\bibfnamefont {G.}~\bibnamefont
  {Koller}}, \bibinfo {author} {\bibfnamefont {C.}~\bibnamefont {Kumpf}},
  \bibinfo {author} {\bibfnamefont {S.}~\bibnamefont {Soubatch}}, \bibinfo
  {author} {\bibfnamefont {P.}~\bibnamefont {Puschnig}}, \bibinfo {author}
  {\bibfnamefont {M.~G.}\ \bibnamefont {Ramsey}}, \ and\ \bibinfo {author}
  {\bibfnamefont {F.~S.}\ \bibnamefont {Tautz}},\ }\href
  {https://pubs.rsc.org/en/content/articlelanding/2015/cp/c4cp04595e}
  {\bibfield  {journal} {\bibinfo  {journal} {Phys. Chem. Chem. Phys.}\
  }\textbf {\bibinfo {volume} {17}},\ \bibinfo {pages} {1530} (\bibinfo {year}
  {2015}{\natexlab{a}})}\BibitemShut {NoStop}%
\bibitem [{\citenamefont {Liu}\ \emph {et~al.}(2017)\citenamefont {Liu},
  \citenamefont {Egger}, \citenamefont {Refaely-Abramson}, \citenamefont
  {Kronik},\ and\ \citenamefont {Neaton}}]{Liu2017}%
  \BibitemOpen
  \bibfield  {author} {\bibinfo {author} {\bibfnamefont {Z.~F.}\ \bibnamefont
  {Liu}}, \bibinfo {author} {\bibfnamefont {D.~A.}\ \bibnamefont {Egger}},
  \bibinfo {author} {\bibfnamefont {S.}~\bibnamefont {Refaely-Abramson}},
  \bibinfo {author} {\bibfnamefont {L.}~\bibnamefont {Kronik}}, \ and\ \bibinfo
  {author} {\bibfnamefont {J.~B.}\ \bibnamefont {Neaton}},\ }\href
  {https://pubs.aip.org/aip/jcp/article/146/9/092326/76781} {\bibfield
  {journal} {\bibinfo  {journal} {J. Chem. Phys.}\ }\textbf {\bibinfo {volume}
  {146}},\ \bibinfo {pages} {092326} (\bibinfo {year} {2017})}\BibitemShut
  {NoStop}%
\bibitem [{\citenamefont {Ferri}\ \emph {et~al.}(2017)\citenamefont {Ferri},
  \citenamefont {Ambrosetti},\ and\ \citenamefont {Tkatchenko}}]{Ferri2017}%
  \BibitemOpen
  \bibfield  {author} {\bibinfo {author} {\bibfnamefont {N.}~\bibnamefont
  {Ferri}}, \bibinfo {author} {\bibfnamefont {A.}~\bibnamefont {Ambrosetti}}, \
  and\ \bibinfo {author} {\bibfnamefont {A.}~\bibnamefont {Tkatchenko}},\
  }\href {\doibase 10.1103/PhysRevMaterials.1.026003} {\bibfield  {journal}
  {\bibinfo  {journal} {Phys. Rev. Mater.}\ }\textbf {\bibinfo {volume} {1}},\
  \bibinfo {pages} {026003} (\bibinfo {year} {2017})}\BibitemShut {NoStop}%
\bibitem [{\citenamefont {Berland}\ \emph {et~al.}(2015)\citenamefont
  {Berland}, \citenamefont {Cooper}, \citenamefont {Lee}, \citenamefont
  {Schröder}, \citenamefont {Thonhauser}, \citenamefont {Hyldgaard},\ and\
  \citenamefont {Lundqvist}}]{Berland2015}%
  \BibitemOpen
  \bibfield  {author} {\bibinfo {author} {\bibfnamefont {K.}~\bibnamefont
  {Berland}}, \bibinfo {author} {\bibfnamefont {V.~R.}\ \bibnamefont {Cooper}},
  \bibinfo {author} {\bibfnamefont {K.}~\bibnamefont {Lee}}, \bibinfo {author}
  {\bibfnamefont {E.}~\bibnamefont {Schröder}}, \bibinfo {author}
  {\bibfnamefont {T.}~\bibnamefont {Thonhauser}}, \bibinfo {author}
  {\bibfnamefont {P.}~\bibnamefont {Hyldgaard}}, \ and\ \bibinfo {author}
  {\bibfnamefont {B.~I.}\ \bibnamefont {Lundqvist}},\ }\href {\doibase
  10.1088/0034-4885/78/6/066501} {\bibfield  {journal} {\bibinfo  {journal}
  {Reports on Progress in Physics}\ }\textbf {\bibinfo {volume} {78}},\
  \bibinfo {pages} {066501} (\bibinfo {year} {2015})}\BibitemShut {NoStop}%
\bibitem [{\citenamefont {Hermann}\ \emph {et~al.}(2017)\citenamefont
  {Hermann}, \citenamefont {DiStasio},\ and\ \citenamefont
  {Tkatchenko}}]{Hermann2017}%
  \BibitemOpen
  \bibfield  {author} {\bibinfo {author} {\bibfnamefont {J.}~\bibnamefont
  {Hermann}}, \bibinfo {author} {\bibfnamefont {R.~A.~J.}\ \bibnamefont
  {DiStasio}}, \ and\ \bibinfo {author} {\bibfnamefont {A.}~\bibnamefont
  {Tkatchenko}},\ }\href {\doibase 10.1021/acs.chemrev.6b00446} {\bibfield
  {journal} {\bibinfo  {journal} {Chemical Reviews}\ }\textbf {\bibinfo
  {volume} {117}},\ \bibinfo {pages} {4714} (\bibinfo {year} {2017})},\
  \bibinfo {note} {pMID: 28272886}\BibitemShut {NoStop}%
\bibitem [{\citenamefont {Neaton}\ \emph {et~al.}(2006)\citenamefont {Neaton},
  \citenamefont {Hybertsen},\ and\ \citenamefont {Louie}}]{Neaton2006}%
  \BibitemOpen
  \bibfield  {author} {\bibinfo {author} {\bibfnamefont {J.~B.}\ \bibnamefont
  {Neaton}}, \bibinfo {author} {\bibfnamefont {M.~S.}\ \bibnamefont
  {Hybertsen}}, \ and\ \bibinfo {author} {\bibfnamefont {S.~G.}\ \bibnamefont
  {Louie}},\ }\href
  {https://journals.aps.org/prl/abstract/10.1103/PhysRevLett.97.216405}
  {\bibfield  {journal} {\bibinfo  {journal} {Phys. Rev. Lett.}\ }\textbf
  {\bibinfo {volume} {97}},\ \bibinfo {pages} {216405} (\bibinfo {year}
  {2006})}\BibitemShut {NoStop}%
\bibitem [{\citenamefont {Garcia-Lastra}\ \emph {et~al.}(2009)\citenamefont
  {Garcia-Lastra}, \citenamefont {Rostgaard}, \citenamefont {Rubio},\ and\
  \citenamefont {Thygesen}}]{Garcia-Lastra2009}%
  \BibitemOpen
  \bibfield  {author} {\bibinfo {author} {\bibfnamefont {J.~M.}\ \bibnamefont
  {Garcia-Lastra}}, \bibinfo {author} {\bibfnamefont {C.}~\bibnamefont
  {Rostgaard}}, \bibinfo {author} {\bibfnamefont {A.}~\bibnamefont {Rubio}}, \
  and\ \bibinfo {author} {\bibfnamefont {K.~S.}\ \bibnamefont {Thygesen}},\
  }\href {\doibase 10.1103/PhysRevB.80.245427} {\bibfield  {journal} {\bibinfo
  {journal} {Phys. Rev. B}\ }\textbf {\bibinfo {volume} {80}},\ \bibinfo
  {pages} {245427} (\bibinfo {year} {2009})}\BibitemShut {NoStop}%
\bibitem [{\citenamefont {Thygesen}\ and\ \citenamefont
  {Rubio}(2009)}]{Thygesen2009}%
  \BibitemOpen
  \bibfield  {author} {\bibinfo {author} {\bibfnamefont {K.~S.}\ \bibnamefont
  {Thygesen}}\ and\ \bibinfo {author} {\bibfnamefont {A.}~\bibnamefont
  {Rubio}},\ }\href {\doibase 10.1103/PhysRevLett.102.046802} {\bibfield
  {journal} {\bibinfo  {journal} {Phys. Rev. Lett.}\ }\textbf {\bibinfo
  {volume} {102}},\ \bibinfo {pages} {046802} (\bibinfo {year}
  {2009})}\BibitemShut {NoStop}%
\bibitem [{\citenamefont {Soubatch}\ \emph {et~al.}(2009)\citenamefont
  {Soubatch}, \citenamefont {Weiss}, \citenamefont {Temirov},\ and\
  \citenamefont {Tautz}}]{Soubatch2009}%
  \BibitemOpen
  \bibfield  {author} {\bibinfo {author} {\bibfnamefont {S.}~\bibnamefont
  {Soubatch}}, \bibinfo {author} {\bibfnamefont {C.}~\bibnamefont {Weiss}},
  \bibinfo {author} {\bibfnamefont {R.}~\bibnamefont {Temirov}}, \ and\
  \bibinfo {author} {\bibfnamefont {F.~S.}\ \bibnamefont {Tautz}},\ }\href
  {\doibase 10.1103/PhysRevLett.102.177405} {\bibfield  {journal} {\bibinfo
  {journal} {Phys. Rev. Lett.}\ }\textbf {\bibinfo {volume} {102}},\ \bibinfo
  {pages} {177405} (\bibinfo {year} {2009})}\BibitemShut {NoStop}%
\bibitem [{\citenamefont {Puschnig}\ \emph {et~al.}(2012)\citenamefont
  {Puschnig}, \citenamefont {Amiri},\ and\ \citenamefont
  {Draxl}}]{Puschnig2012}%
  \BibitemOpen
  \bibfield  {author} {\bibinfo {author} {\bibfnamefont {P.}~\bibnamefont
  {Puschnig}}, \bibinfo {author} {\bibfnamefont {P.}~\bibnamefont {Amiri}}, \
  and\ \bibinfo {author} {\bibfnamefont {C.}~\bibnamefont {Draxl}},\ }\href
  {\doibase 10.1103/PhysRevB.86.085107} {\bibfield  {journal} {\bibinfo
  {journal} {Phys. Rev. B}\ }\textbf {\bibinfo {volume} {86}},\ \bibinfo
  {pages} {085107} (\bibinfo {year} {2012})}\BibitemShut {NoStop}%
\bibitem [{\citenamefont {Yamane}\ \emph {et~al.}(2007)\citenamefont {Yamane},
  \citenamefont {Yoshimura}, \citenamefont {Kawabe}, \citenamefont {Sumii},
  \citenamefont {Kanai}, \citenamefont {Ouchi}, \citenamefont {Ueno},\ and\
  \citenamefont {Seki}}]{Yamane2007}%
  \BibitemOpen
  \bibfield  {author} {\bibinfo {author} {\bibfnamefont {H.}~\bibnamefont
  {Yamane}}, \bibinfo {author} {\bibfnamefont {D.}~\bibnamefont {Yoshimura}},
  \bibinfo {author} {\bibfnamefont {E.}~\bibnamefont {Kawabe}}, \bibinfo
  {author} {\bibfnamefont {R.}~\bibnamefont {Sumii}}, \bibinfo {author}
  {\bibfnamefont {K.}~\bibnamefont {Kanai}}, \bibinfo {author} {\bibfnamefont
  {Y.}~\bibnamefont {Ouchi}}, \bibinfo {author} {\bibfnamefont
  {N.}~\bibnamefont {Ueno}}, \ and\ \bibinfo {author} {\bibfnamefont
  {K.}~\bibnamefont {Seki}},\ }\href {\doibase 10.1103/PhysRevB.76.165436}
  {\bibfield  {journal} {\bibinfo  {journal} {Phys. Rev. B}\ }\textbf {\bibinfo
  {volume} {76}},\ \bibinfo {pages} {165436} (\bibinfo {year}
  {2007})}\BibitemShut {NoStop}%
\bibitem [{\citenamefont {Ziroff}\ \emph {et~al.}(2010)\citenamefont {Ziroff},
  \citenamefont {Forster}, \citenamefont {Schöll}, \citenamefont {Puschnig},\
  and\ \citenamefont {Reinert}}]{Ziroff2010}%
  \BibitemOpen
  \bibfield  {author} {\bibinfo {author} {\bibfnamefont {J.}~\bibnamefont
  {Ziroff}}, \bibinfo {author} {\bibfnamefont {F.}~\bibnamefont {Forster}},
  \bibinfo {author} {\bibfnamefont {A.}~\bibnamefont {Schöll}}, \bibinfo
  {author} {\bibfnamefont {P.}~\bibnamefont {Puschnig}}, \ and\ \bibinfo
  {author} {\bibfnamefont {F.}~\bibnamefont {Reinert}},\ }\href
  {https://journals.aps.org/prl/abstract/10.1103/PhysRevLett.104.233004}
  {\bibfield  {journal} {\bibinfo  {journal} {Phys. Rev. Lett.}\ }\textbf
  {\bibinfo {volume} {104}},\ \bibinfo {pages} {233004} (\bibinfo {year}
  {2010})}\BibitemShut {NoStop}%
\bibitem [{\citenamefont {Berkebile}\ \emph {et~al.}(2011)\citenamefont
  {Berkebile}, \citenamefont {Ules}, \citenamefont {Puschnig}, \citenamefont
  {Romaner}, \citenamefont {Koller}, \citenamefont {Fleming}, \citenamefont
  {Emtsev}, \citenamefont {Seyller}, \citenamefont {Ambrosch-Draxl},
  \citenamefont {Netzer},\ and\ \citenamefont {Ramsey}}]{Berkebile2011}%
  \BibitemOpen
  \bibfield  {author} {\bibinfo {author} {\bibfnamefont {S.}~\bibnamefont
  {Berkebile}}, \bibinfo {author} {\bibfnamefont {T.}~\bibnamefont {Ules}},
  \bibinfo {author} {\bibfnamefont {P.}~\bibnamefont {Puschnig}}, \bibinfo
  {author} {\bibfnamefont {L.}~\bibnamefont {Romaner}}, \bibinfo {author}
  {\bibfnamefont {G.}~\bibnamefont {Koller}}, \bibinfo {author} {\bibfnamefont
  {A.~J.}\ \bibnamefont {Fleming}}, \bibinfo {author} {\bibfnamefont
  {K.}~\bibnamefont {Emtsev}}, \bibinfo {author} {\bibfnamefont
  {T.}~\bibnamefont {Seyller}}, \bibinfo {author} {\bibfnamefont
  {C.}~\bibnamefont {Ambrosch-Draxl}}, \bibinfo {author} {\bibfnamefont
  {F.~P.}\ \bibnamefont {Netzer}}, \ and\ \bibinfo {author} {\bibfnamefont
  {M.~G.}\ \bibnamefont {Ramsey}},\ }\href
  {https://pubs.rsc.org/en/content/articlelanding/2011/cp/c0cp01458c}
  {\bibfield  {journal} {\bibinfo  {journal} {Phys. Chem. Chem. Phys.}\
  }\textbf {\bibinfo {volume} {13}},\ \bibinfo {pages} {3604} (\bibinfo {year}
  {2011})}\BibitemShut {NoStop}%
\bibitem [{\citenamefont {Wießner}\ \emph {et~al.}(2013)\citenamefont
  {Wießner}, \citenamefont {Ziroff}, \citenamefont {Forster}, \citenamefont
  {Arita}, \citenamefont {Shimada}, \citenamefont {Puschnig}, \citenamefont
  {Schöll},\ and\ \citenamefont {Reinert}}]{Wiessner2013}%
  \BibitemOpen
  \bibfield  {author} {\bibinfo {author} {\bibfnamefont {M.}~\bibnamefont
  {Wießner}}, \bibinfo {author} {\bibfnamefont {J.}~\bibnamefont {Ziroff}},
  \bibinfo {author} {\bibfnamefont {F.}~\bibnamefont {Forster}}, \bibinfo
  {author} {\bibfnamefont {M.}~\bibnamefont {Arita}}, \bibinfo {author}
  {\bibfnamefont {K.}~\bibnamefont {Shimada}}, \bibinfo {author} {\bibfnamefont
  {P.}~\bibnamefont {Puschnig}}, \bibinfo {author} {\bibfnamefont
  {A.}~\bibnamefont {Schöll}}, \ and\ \bibinfo {author} {\bibfnamefont
  {F.}~\bibnamefont {Reinert}},\ }\href
  {https://www.nature.com/articles/ncomms2522} {\bibfield  {journal} {\bibinfo
  {journal} {Nat. Commun.}\ }\textbf {\bibinfo {volume} {4}},\ \bibinfo {pages}
  {1514} (\bibinfo {year} {2013})}\BibitemShut {NoStop}%
\bibitem [{\citenamefont {Ules}\ \emph {et~al.}(2014)\citenamefont {Ules},
  \citenamefont {Lüftner}, \citenamefont {Reinisch}, \citenamefont {Koller},
  \citenamefont {Puschnig},\ and\ \citenamefont {Ramsey}}]{Ules2014}%
  \BibitemOpen
  \bibfield  {author} {\bibinfo {author} {\bibfnamefont {T.}~\bibnamefont
  {Ules}}, \bibinfo {author} {\bibfnamefont {D.}~\bibnamefont {Lüftner}},
  \bibinfo {author} {\bibfnamefont {E.~M.}\ \bibnamefont {Reinisch}}, \bibinfo
  {author} {\bibfnamefont {G.}~\bibnamefont {Koller}}, \bibinfo {author}
  {\bibfnamefont {P.}~\bibnamefont {Puschnig}}, \ and\ \bibinfo {author}
  {\bibfnamefont {M.~G.}\ \bibnamefont {Ramsey}},\ }\href
  {https://journals.aps.org/prb/abstract/10.1103/PhysRevB.90.155430} {\bibfield
   {journal} {\bibinfo  {journal} {Phys. Rev. B}\ }\textbf {\bibinfo {volume}
  {90}},\ \bibinfo {pages} {155430} (\bibinfo {year} {2014})}\BibitemShut
  {NoStop}%
\bibitem [{\citenamefont {Yang}\ \emph {et~al.}(2022)\citenamefont {Yang},
  \citenamefont {Jugovac}, \citenamefont {Zamborlini}, \citenamefont {Feyer},
  \citenamefont {Koller}, \citenamefont {Puschnig}, \citenamefont {Soubatch},
  \citenamefont {Ramsey},\ and\ \citenamefont {Tautz}}]{Yang2022}%
  \BibitemOpen
  \bibfield  {author} {\bibinfo {author} {\bibfnamefont {X.}~\bibnamefont
  {Yang}}, \bibinfo {author} {\bibfnamefont {M.}~\bibnamefont {Jugovac}},
  \bibinfo {author} {\bibfnamefont {G.}~\bibnamefont {Zamborlini}}, \bibinfo
  {author} {\bibfnamefont {V.}~\bibnamefont {Feyer}}, \bibinfo {author}
  {\bibfnamefont {G.}~\bibnamefont {Koller}}, \bibinfo {author} {\bibfnamefont
  {P.}~\bibnamefont {Puschnig}}, \bibinfo {author} {\bibfnamefont
  {S.}~\bibnamefont {Soubatch}}, \bibinfo {author} {\bibfnamefont {M.~G.}\
  \bibnamefont {Ramsey}}, \ and\ \bibinfo {author} {\bibfnamefont {F.~S.}\
  \bibnamefont {Tautz}},\ }\href
  {https://www.nature.com/articles/s41467-022-32643-z#citeas} {\bibfield
  {journal} {\bibinfo  {journal} {Nat. Commun.}\ }\textbf {\bibinfo {volume}
  {13}} (\bibinfo {year} {2022})}\BibitemShut {NoStop}%
\bibitem [{\citenamefont {Duhm}\ \emph {et~al.}(2008)\citenamefont {Duhm},
  \citenamefont {Gerlach}, \citenamefont {Salzmann}, \citenamefont {Bröker},
  \citenamefont {Johnson}, \citenamefont {Schreiber},\ and\ \citenamefont
  {Koch}}]{Duhm2008}%
  \BibitemOpen
  \bibfield  {author} {\bibinfo {author} {\bibfnamefont {S.}~\bibnamefont
  {Duhm}}, \bibinfo {author} {\bibfnamefont {A.}~\bibnamefont {Gerlach}},
  \bibinfo {author} {\bibfnamefont {I.}~\bibnamefont {Salzmann}}, \bibinfo
  {author} {\bibfnamefont {B.}~\bibnamefont {Bröker}}, \bibinfo {author}
  {\bibfnamefont {R.}~\bibnamefont {Johnson}}, \bibinfo {author} {\bibfnamefont
  {F.}~\bibnamefont {Schreiber}}, \ and\ \bibinfo {author} {\bibfnamefont
  {N.}~\bibnamefont {Koch}},\ }\href {\doibase
  https://doi.org/10.1016/j.orgel.2007.10.004} {\bibfield  {journal} {\bibinfo
  {journal} {Organic Electronics}\ }\textbf {\bibinfo {volume} {9}},\ \bibinfo
  {pages} {111} (\bibinfo {year} {2008})}\BibitemShut {NoStop}%
\bibitem [{\citenamefont {Rangger}\ \emph {et~al.}(2009)\citenamefont
  {Rangger}, \citenamefont {Hofmann}, \citenamefont {Romaner}, \citenamefont
  {Heimel}, \citenamefont {Br\"oker}, \citenamefont {Blum}, \citenamefont
  {Johnson}, \citenamefont {Koch},\ and\ \citenamefont {Zojer}}]{Rangger2009}%
  \BibitemOpen
  \bibfield  {author} {\bibinfo {author} {\bibfnamefont {G.~M.}\ \bibnamefont
  {Rangger}}, \bibinfo {author} {\bibfnamefont {O.~T.}\ \bibnamefont
  {Hofmann}}, \bibinfo {author} {\bibfnamefont {L.}~\bibnamefont {Romaner}},
  \bibinfo {author} {\bibfnamefont {G.}~\bibnamefont {Heimel}}, \bibinfo
  {author} {\bibfnamefont {B.}~\bibnamefont {Br\"oker}}, \bibinfo {author}
  {\bibfnamefont {R.-P.}\ \bibnamefont {Blum}}, \bibinfo {author}
  {\bibfnamefont {R.~L.}\ \bibnamefont {Johnson}}, \bibinfo {author}
  {\bibfnamefont {N.}~\bibnamefont {Koch}}, \ and\ \bibinfo {author}
  {\bibfnamefont {E.}~\bibnamefont {Zojer}},\ }\href {\doibase
  10.1103/PhysRevB.79.165306} {\bibfield  {journal} {\bibinfo  {journal} {Phys.
  Rev. B}\ }\textbf {\bibinfo {volume} {79}},\ \bibinfo {pages} {165306}
  (\bibinfo {year} {2009})}\BibitemShut {NoStop}%
\bibitem [{\citenamefont {Puschnig}\ \emph {et~al.}(2009)\citenamefont
  {Puschnig}, \citenamefont {Berkebile}, \citenamefont {Fleming}, \citenamefont
  {Koller}, \citenamefont {Emtsev}, \citenamefont {Seyller}, \citenamefont
  {Riley}, \citenamefont {Ambrosch-Draxl}, \citenamefont {Netzer},\ and\
  \citenamefont {Ramsey}}]{Puschnig2009a}%
  \BibitemOpen
  \bibfield  {author} {\bibinfo {author} {\bibfnamefont {P.}~\bibnamefont
  {Puschnig}}, \bibinfo {author} {\bibfnamefont {S.}~\bibnamefont {Berkebile}},
  \bibinfo {author} {\bibfnamefont {A.~J.}\ \bibnamefont {Fleming}}, \bibinfo
  {author} {\bibfnamefont {G.}~\bibnamefont {Koller}}, \bibinfo {author}
  {\bibfnamefont {K.}~\bibnamefont {Emtsev}}, \bibinfo {author} {\bibfnamefont
  {T.}~\bibnamefont {Seyller}}, \bibinfo {author} {\bibfnamefont {J.~D.}\
  \bibnamefont {Riley}}, \bibinfo {author} {\bibfnamefont {C.}~\bibnamefont
  {Ambrosch-Draxl}}, \bibinfo {author} {\bibfnamefont {F.~P.}\ \bibnamefont
  {Netzer}}, \ and\ \bibinfo {author} {\bibfnamefont {M.~G.}\ \bibnamefont
  {Ramsey}},\ }\href {https://www.science.org/doi/10.1126/science.1176105}
  {\bibfield  {journal} {\bibinfo  {journal} {Science}\ }\textbf {\bibinfo
  {volume} {326}},\ \bibinfo {pages} {702} (\bibinfo {year}
  {2009})}\BibitemShut {NoStop}%
\bibitem [{\citenamefont {Heimel}\ \emph {et~al.}(2013)\citenamefont {Heimel},
  \citenamefont {Duhm}, \citenamefont {Salzmann}, \citenamefont {Gerlach},
  \citenamefont {Strozecka}, \citenamefont {Niederhausen}, \citenamefont
  {Bürker}, \citenamefont {Hosokai}, \citenamefont {Fernandez-Torrente},
  \citenamefont {Schulze}, \citenamefont {Winkler}, \citenamefont {Wilke},
  \citenamefont {Schlesinger}, \citenamefont {Frisch}, \citenamefont {Bröker},
  \citenamefont {Vollmer}, \citenamefont {Detlefs}, \citenamefont {Pflaum},
  \citenamefont {Kera}, \citenamefont {Franke}, \citenamefont {Ueno},
  \citenamefont {Pascual}, \citenamefont {Schreiber},\ and\ \citenamefont
  {Koch}}]{Heimel2013}%
  \BibitemOpen
  \bibfield  {author} {\bibinfo {author} {\bibfnamefont {G.}~\bibnamefont
  {Heimel}}, \bibinfo {author} {\bibfnamefont {S.}~\bibnamefont {Duhm}},
  \bibinfo {author} {\bibfnamefont {I.}~\bibnamefont {Salzmann}}, \bibinfo
  {author} {\bibfnamefont {A.}~\bibnamefont {Gerlach}}, \bibinfo {author}
  {\bibfnamefont {A.}~\bibnamefont {Strozecka}}, \bibinfo {author}
  {\bibfnamefont {J.}~\bibnamefont {Niederhausen}}, \bibinfo {author}
  {\bibfnamefont {C.}~\bibnamefont {Bürker}}, \bibinfo {author} {\bibfnamefont
  {T.}~\bibnamefont {Hosokai}}, \bibinfo {author} {\bibfnamefont
  {I.}~\bibnamefont {Fernandez-Torrente}}, \bibinfo {author} {\bibfnamefont
  {G.}~\bibnamefont {Schulze}}, \bibinfo {author} {\bibfnamefont
  {S.}~\bibnamefont {Winkler}}, \bibinfo {author} {\bibfnamefont
  {A.}~\bibnamefont {Wilke}}, \bibinfo {author} {\bibfnamefont
  {R.}~\bibnamefont {Schlesinger}}, \bibinfo {author} {\bibfnamefont
  {J.}~\bibnamefont {Frisch}}, \bibinfo {author} {\bibfnamefont
  {B.}~\bibnamefont {Bröker}}, \bibinfo {author} {\bibfnamefont
  {A.}~\bibnamefont {Vollmer}}, \bibinfo {author} {\bibfnamefont
  {B.}~\bibnamefont {Detlefs}}, \bibinfo {author} {\bibfnamefont
  {J.}~\bibnamefont {Pflaum}}, \bibinfo {author} {\bibfnamefont
  {S.}~\bibnamefont {Kera}}, \bibinfo {author} {\bibfnamefont {K.~J.}\
  \bibnamefont {Franke}}, \bibinfo {author} {\bibfnamefont {N.}~\bibnamefont
  {Ueno}}, \bibinfo {author} {\bibfnamefont {J.~I.}\ \bibnamefont {Pascual}},
  \bibinfo {author} {\bibfnamefont {F.}~\bibnamefont {Schreiber}}, \ and\
  \bibinfo {author} {\bibfnamefont {N.}~\bibnamefont {Koch}},\ }\href
  {https://doi.org/10.1038/nchem.1572} {\bibfield  {journal} {\bibinfo
  {journal} {Nature Chem}\ }\textbf {\bibinfo {volume} {5}},\ \bibinfo {pages}
  {187–194} (\bibinfo {year} {2013})}\BibitemShut {NoStop}%
\bibitem [{\citenamefont {Hofmann}\ \emph {et~al.}(2015)\citenamefont
  {Hofmann}, \citenamefont {Rinke}, \citenamefont {Scheffler},\ and\
  \citenamefont {Heimel}}]{Hofmann2015}%
  \BibitemOpen
  \bibfield  {author} {\bibinfo {author} {\bibfnamefont {O.~T.}\ \bibnamefont
  {Hofmann}}, \bibinfo {author} {\bibfnamefont {P.}~\bibnamefont {Rinke}},
  \bibinfo {author} {\bibfnamefont {M.}~\bibnamefont {Scheffler}}, \ and\
  \bibinfo {author} {\bibfnamefont {G.}~\bibnamefont {Heimel}},\ }\href
  {\doibase 10.1021/acsnano.5b01164} {\bibfield  {journal} {\bibinfo  {journal}
  {ACS Nano}\ }\textbf {\bibinfo {volume} {9}},\ \bibinfo {pages} {5391}
  (\bibinfo {year} {2015})},\ \bibinfo {note} {pMID: 25905769}\BibitemShut
  {NoStop}%
\bibitem [{\citenamefont {Schönauer}\ \emph {et~al.}(2016)\citenamefont
  {Schönauer}, \citenamefont {Weiss}, \citenamefont {Feyer}, \citenamefont
  {Lüftner}, \citenamefont {Stadtmüller}, \citenamefont {Schwarz},
  \citenamefont {Sueyoshi}, \citenamefont {Kumpf}, \citenamefont {Puschnig},
  \citenamefont {Ramsey}, \citenamefont {Tautz},\ and\ \citenamefont
  {Soubatch}}]{Schonauer2016}%
  \BibitemOpen
  \bibfield  {author} {\bibinfo {author} {\bibfnamefont {K.}~\bibnamefont
  {Schönauer}}, \bibinfo {author} {\bibfnamefont {S.}~\bibnamefont {Weiss}},
  \bibinfo {author} {\bibfnamefont {V.}~\bibnamefont {Feyer}}, \bibinfo
  {author} {\bibfnamefont {D.}~\bibnamefont {Lüftner}}, \bibinfo {author}
  {\bibfnamefont {B.}~\bibnamefont {Stadtmüller}}, \bibinfo {author}
  {\bibfnamefont {D.}~\bibnamefont {Schwarz}}, \bibinfo {author} {\bibfnamefont
  {T.}~\bibnamefont {Sueyoshi}}, \bibinfo {author} {\bibfnamefont
  {C.}~\bibnamefont {Kumpf}}, \bibinfo {author} {\bibfnamefont
  {P.}~\bibnamefont {Puschnig}}, \bibinfo {author} {\bibfnamefont {M.~G.}\
  \bibnamefont {Ramsey}}, \bibinfo {author} {\bibfnamefont {F.~S.}\
  \bibnamefont {Tautz}}, \ and\ \bibinfo {author} {\bibfnamefont
  {S.}~\bibnamefont {Soubatch}},\ }\href
  {https://journals.aps.org/prb/abstract/10.1103/PhysRevB.94.205144} {\bibfield
   {journal} {\bibinfo  {journal} {Phys. Rev. B}\ }\textbf {\bibinfo {volume}
  {94}},\ \bibinfo {pages} {205144} (\bibinfo {year} {2016})}\BibitemShut
  {NoStop}%
\bibitem [{\citenamefont {Hollerer}\ \emph {et~al.}(2017)\citenamefont
  {Hollerer}, \citenamefont {L\"{u}ftner}, \citenamefont {Hurdax},
  \citenamefont {Ules}, \citenamefont {Soubatch}, \citenamefont {Tautz},
  \citenamefont {Koller}, \citenamefont {Puschnig}, \citenamefont {Sterrer},\
  and\ \citenamefont {Ramsey}}]{Hollerer2017}%
  \BibitemOpen
  \bibfield  {author} {\bibinfo {author} {\bibfnamefont {M.}~\bibnamefont
  {Hollerer}}, \bibinfo {author} {\bibfnamefont {D.}~\bibnamefont
  {L\"{u}ftner}}, \bibinfo {author} {\bibfnamefont {P.}~\bibnamefont {Hurdax}},
  \bibinfo {author} {\bibfnamefont {T.}~\bibnamefont {Ules}}, \bibinfo {author}
  {\bibfnamefont {S.}~\bibnamefont {Soubatch}}, \bibinfo {author}
  {\bibfnamefont {F.~S.}\ \bibnamefont {Tautz}}, \bibinfo {author}
  {\bibfnamefont {G.}~\bibnamefont {Koller}}, \bibinfo {author} {\bibfnamefont
  {P.}~\bibnamefont {Puschnig}}, \bibinfo {author} {\bibfnamefont
  {M.}~\bibnamefont {Sterrer}}, \ and\ \bibinfo {author} {\bibfnamefont
  {M.~G.}\ \bibnamefont {Ramsey}},\ }\href
  {https://doi.org/10.1021/acsnano.7b02449} {\bibfield  {journal} {\bibinfo
  {journal} {ACS Nano}\ }\textbf {\bibinfo {volume} {11}},\ \bibinfo {pages}
  {6252} (\bibinfo {year} {2017})}\BibitemShut {NoStop}%
\bibitem [{\citenamefont {Seidl}\ \emph {et~al.}(1996)\citenamefont {Seidl},
  \citenamefont {G\"orling}, \citenamefont {Vogl}, \citenamefont {Majewski},\
  and\ \citenamefont {Levy}}]{Seidl1996}%
  \BibitemOpen
  \bibfield  {author} {\bibinfo {author} {\bibfnamefont {A.}~\bibnamefont
  {Seidl}}, \bibinfo {author} {\bibfnamefont {A.}~\bibnamefont {G\"orling}},
  \bibinfo {author} {\bibfnamefont {P.}~\bibnamefont {Vogl}}, \bibinfo {author}
  {\bibfnamefont {J.~A.}\ \bibnamefont {Majewski}}, \ and\ \bibinfo {author}
  {\bibfnamefont {M.}~\bibnamefont {Levy}},\ }\href {\doibase
  10.1103/PhysRevB.53.3764} {\bibfield  {journal} {\bibinfo  {journal} {Phys.
  Rev. B}\ }\textbf {\bibinfo {volume} {53}},\ \bibinfo {pages} {3764}
  (\bibinfo {year} {1996})}\BibitemShut {NoStop}%
\bibitem [{\citenamefont {Kronik}\ \emph {et~al.}(2012)\citenamefont {Kronik},
  \citenamefont {Stein}, \citenamefont {Refaely-Abramson},\ and\ \citenamefont
  {Baer}}]{Kronik2012}%
  \BibitemOpen
  \bibfield  {author} {\bibinfo {author} {\bibfnamefont {L.}~\bibnamefont
  {Kronik}}, \bibinfo {author} {\bibfnamefont {T.}~\bibnamefont {Stein}},
  \bibinfo {author} {\bibfnamefont {S.}~\bibnamefont {Refaely-Abramson}}, \
  and\ \bibinfo {author} {\bibfnamefont {R.}~\bibnamefont {Baer}},\ }\href
  {\doibase 10.1021/ct2009363} {\bibfield  {journal} {\bibinfo  {journal}
  {Journal of Chemical Theory and Computation}\ }\textbf {\bibinfo {volume}
  {8}},\ \bibinfo {pages} {1515} (\bibinfo {year} {2012})},\ \bibinfo {note}
  {pMID: 26593646}\BibitemShut {NoStop}%
\bibitem [{\citenamefont {Perdew}\ \emph
  {et~al.}(1996{\natexlab{a}})\citenamefont {Perdew}, \citenamefont {Burke},\
  and\ \citenamefont {Ernzerhof}}]{Perdew1996}%
  \BibitemOpen
  \bibfield  {author} {\bibinfo {author} {\bibfnamefont {J.~P.}\ \bibnamefont
  {Perdew}}, \bibinfo {author} {\bibfnamefont {K.}~\bibnamefont {Burke}}, \
  and\ \bibinfo {author} {\bibfnamefont {M.}~\bibnamefont {Ernzerhof}},\ }\href
  {https://journals.aps.org/prl/abstract/10.1103/PhysRevLett.77.3865}
  {\bibfield  {journal} {\bibinfo  {journal} {Phys. Rev. Lett.}\ }\textbf
  {\bibinfo {volume} {77}},\ \bibinfo {pages} {3865} (\bibinfo {year}
  {1996}{\natexlab{a}})}\BibitemShut {NoStop}%
\bibitem [{\citenamefont {Heyd}\ and\ \citenamefont
  {Scuseria}(2004)}]{Heyd2004}%
  \BibitemOpen
  \bibfield  {author} {\bibinfo {author} {\bibfnamefont {J.}~\bibnamefont
  {Heyd}}\ and\ \bibinfo {author} {\bibfnamefont {G.~E.}\ \bibnamefont
  {Scuseria}},\ }\href
  {https://pubs.aip.org/aip/jcp/article-abstract/121/3/1187/186757/Efficient-hybrid-density-functional-calculations?redirectedFrom=fulltext}
  {\bibfield  {journal} {\bibinfo  {journal} {J. Chem. Phys.}\ }\textbf
  {\bibinfo {volume} {121}},\ \bibinfo {pages} {1187} (\bibinfo {year}
  {2004})}\BibitemShut {NoStop}%
\bibitem [{\citenamefont {Heyd}\ \emph {et~al.}(2006)\citenamefont {Heyd},
  \citenamefont {Scuseria},\ and\ \citenamefont {Ernzerhof}}]{Heyd2006}%
  \BibitemOpen
  \bibfield  {author} {\bibinfo {author} {\bibfnamefont {J.}~\bibnamefont
  {Heyd}}, \bibinfo {author} {\bibfnamefont {G.~E.}\ \bibnamefont {Scuseria}},
  \ and\ \bibinfo {author} {\bibfnamefont {M.}~\bibnamefont {Ernzerhof}},\
  }\href
  {https://pubs.aip.org/aip/jcp/article/124/21/219906/895871/Erratum-Hybrid-functionals-based-on-a-screened}
  {\bibfield  {journal} {\bibinfo  {journal} {J. Chem. Phys.}\ }\textbf
  {\bibinfo {volume} {124}},\ \bibinfo {pages} {219906} (\bibinfo {year}
  {2006})}\BibitemShut {NoStop}%
\bibitem [{\citenamefont {Blase}\ \emph {et~al.}(2011)\citenamefont {Blase},
  \citenamefont {Attaccalite},\ and\ \citenamefont {Olevano}}]{Blase2011}%
  \BibitemOpen
  \bibfield  {author} {\bibinfo {author} {\bibfnamefont {X.}~\bibnamefont
  {Blase}}, \bibinfo {author} {\bibfnamefont {C.}~\bibnamefont {Attaccalite}},
  \ and\ \bibinfo {author} {\bibfnamefont {V.}~\bibnamefont {Olevano}},\ }\href
  {\doibase 10.1103/PhysRevB.83.115103} {\bibfield  {journal} {\bibinfo
  {journal} {Phys. Rev. B}\ }\textbf {\bibinfo {volume} {83}},\ \bibinfo
  {pages} {115103} (\bibinfo {year} {2011})}\BibitemShut {NoStop}%
\bibitem [{\citenamefont {Faber}\ \emph {et~al.}(2014)\citenamefont {Faber},
  \citenamefont {Boulanger}, \citenamefont {Attaccalite}, \citenamefont
  {Duchemin},\ and\ \citenamefont {Blase}}]{Faber2014}%
  \BibitemOpen
  \bibfield  {author} {\bibinfo {author} {\bibfnamefont {C.}~\bibnamefont
  {Faber}}, \bibinfo {author} {\bibfnamefont {P.}~\bibnamefont {Boulanger}},
  \bibinfo {author} {\bibfnamefont {C.}~\bibnamefont {Attaccalite}}, \bibinfo
  {author} {\bibfnamefont {I.}~\bibnamefont {Duchemin}}, \ and\ \bibinfo
  {author} {\bibfnamefont {X.}~\bibnamefont {Blase}},\ }\href {\doibase
  10.1098/rsta.2013.0271} {\bibfield  {journal} {\bibinfo  {journal}
  {Philosophical Transactions of the Royal Society A: Mathematical, Physical
  and Engineering Sciences}\ }\textbf {\bibinfo {volume} {372}},\ \bibinfo
  {pages} {20130271} (\bibinfo {year} {2014})}\BibitemShut {NoStop}%
\bibitem [{\citenamefont {Draxl}\ \emph {et~al.}(2014)\citenamefont {Draxl},
  \citenamefont {Nabok},\ and\ \citenamefont {Hannewald}}]{Draxl2014}%
  \BibitemOpen
  \bibfield  {author} {\bibinfo {author} {\bibfnamefont {C.}~\bibnamefont
  {Draxl}}, \bibinfo {author} {\bibfnamefont {D.}~\bibnamefont {Nabok}}, \ and\
  \bibinfo {author} {\bibfnamefont {K.}~\bibnamefont {Hannewald}},\ }\href
  {\doibase 10.1021/ar500096q} {\bibfield  {journal} {\bibinfo  {journal}
  {Accounts of Chemical Research}\ }\textbf {\bibinfo {volume} {47}},\ \bibinfo
  {pages} {3225} (\bibinfo {year} {2014})},\ \bibinfo {note} {pMID:
  25171272}\BibitemShut {NoStop}%
\bibitem [{\citenamefont {Marom}(2017)}]{Marom2017}%
  \BibitemOpen
  \bibfield  {author} {\bibinfo {author} {\bibfnamefont {N.}~\bibnamefont
  {Marom}},\ }\href {\doibase 10.1088/1361-648X/29/10/103003} {\bibfield
  {journal} {\bibinfo  {journal} {Journal of Physics: Condensed Matter}\
  }\textbf {\bibinfo {volume} {29}},\ \bibinfo {pages} {103003} (\bibinfo
  {year} {2017})}\BibitemShut {NoStop}%
\bibitem [{\citenamefont {Golze}\ \emph {et~al.}(2019)\citenamefont {Golze},
  \citenamefont {Dvorak},\ and\ \citenamefont {Rinke}}]{Golze2019}%
  \BibitemOpen
  \bibfield  {author} {\bibinfo {author} {\bibfnamefont {D.}~\bibnamefont
  {Golze}}, \bibinfo {author} {\bibfnamefont {M.}~\bibnamefont {Dvorak}}, \
  and\ \bibinfo {author} {\bibfnamefont {P.}~\bibnamefont {Rinke}},\ }\href
  {\doibase 10.3389/fchem.2019.00377} {\bibfield  {journal} {\bibinfo
  {journal} {Frontiers in Chemistry}\ }\textbf {\bibinfo {volume} {7}}
  (\bibinfo {year} {2019}),\ 10.3389/fchem.2019.00377}\BibitemShut {NoStop}%
\bibitem [{\citenamefont {Truhlar}\ \emph {et~al.}(2019)\citenamefont
  {Truhlar}, \citenamefont {Hiberty}, \citenamefont {Shaik}, \citenamefont
  {Gordon},\ and\ \citenamefont {Danovich}}]{Truhlar2019}%
  \BibitemOpen
  \bibfield  {author} {\bibinfo {author} {\bibfnamefont {D.~G.}\ \bibnamefont
  {Truhlar}}, \bibinfo {author} {\bibfnamefont {P.~C.}\ \bibnamefont
  {Hiberty}}, \bibinfo {author} {\bibfnamefont {S.}~\bibnamefont {Shaik}},
  \bibinfo {author} {\bibfnamefont {M.~S.}\ \bibnamefont {Gordon}}, \ and\
  \bibinfo {author} {\bibfnamefont {D.}~\bibnamefont {Danovich}},\ }\href
  {\doibase 10.1002/ange.201904609} {\bibfield  {journal} {\bibinfo  {journal}
  {Angewandte Chemie}\ }\textbf {\bibinfo {volume} {131}},\ \bibinfo {pages}
  {12460} (\bibinfo {year} {2019})}\BibitemShut {NoStop}%
\bibitem [{\citenamefont {Krylov}(2020)}]{Krylov2020}%
  \BibitemOpen
  \bibfield  {author} {\bibinfo {author} {\bibfnamefont {A.~I.}\ \bibnamefont
  {Krylov}},\ }\href {\doibase 10.1063/5.0018597} {\bibfield  {journal}
  {\bibinfo  {journal} {J. Chem. Phys.}\ }\textbf {\bibinfo {volume} {153}},\
  \bibinfo {pages} {080901} (\bibinfo {year} {2020})}\BibitemShut {NoStop}%
\bibitem [{\citenamefont {Stowasser}\ and\ \citenamefont
  {Hoffmann}(1999)}]{Stowasser1999}%
  \BibitemOpen
  \bibfield  {author} {\bibinfo {author} {\bibfnamefont {R.}~\bibnamefont
  {Stowasser}}\ and\ \bibinfo {author} {\bibfnamefont {R.}~\bibnamefont
  {Hoffmann}},\ }\href {\doibase 10.1021/ja9826892} {\bibfield  {journal}
  {\bibinfo  {journal} {Journal of the American Chemical Society}\ }\textbf
  {\bibinfo {volume} {121}},\ \bibinfo {pages} {3414} (\bibinfo {year}
  {1999})}\BibitemShut {NoStop}%
\bibitem [{\citenamefont {Chong}\ \emph {et~al.}(2002)\citenamefont {Chong},
  \citenamefont {Gritsenko},\ and\ \citenamefont {Baerends}}]{Chong2002}%
  \BibitemOpen
  \bibfield  {author} {\bibinfo {author} {\bibfnamefont {D.~P.}\ \bibnamefont
  {Chong}}, \bibinfo {author} {\bibfnamefont {O.~V.}\ \bibnamefont
  {Gritsenko}}, \ and\ \bibinfo {author} {\bibfnamefont {E.~J.}\ \bibnamefont
  {Baerends}},\ }\href {\doibase 10.1063/1.1430255} {\bibfield  {journal}
  {\bibinfo  {journal} {J. Chem. Phys.}\ }\textbf {\bibinfo {volume} {116}},\
  \bibinfo {pages} {1760} (\bibinfo {year} {2002})}\BibitemShut {NoStop}%
\bibitem [{\citenamefont {Dori}\ \emph {et~al.}(2006)\citenamefont {Dori},
  \citenamefont {Menon}, \citenamefont {Kilian}, \citenamefont {Sokolowski},
  \citenamefont {Kronik},\ and\ \citenamefont {Umbach}}]{Dori2006}%
  \BibitemOpen
  \bibfield  {author} {\bibinfo {author} {\bibfnamefont {N.}~\bibnamefont
  {Dori}}, \bibinfo {author} {\bibfnamefont {M.}~\bibnamefont {Menon}},
  \bibinfo {author} {\bibfnamefont {L.}~\bibnamefont {Kilian}}, \bibinfo
  {author} {\bibfnamefont {M.}~\bibnamefont {Sokolowski}}, \bibinfo {author}
  {\bibfnamefont {L.}~\bibnamefont {Kronik}}, \ and\ \bibinfo {author}
  {\bibfnamefont {E.}~\bibnamefont {Umbach}},\ }\href {\doibase
  10.1103/PhysRevB.73.195208} {\bibfield  {journal} {\bibinfo  {journal} {Phys.
  Rev. B}\ }\textbf {\bibinfo {volume} {73}},\ \bibinfo {pages} {195208}
  (\bibinfo {year} {2006})}\BibitemShut {NoStop}%
\bibitem [{\citenamefont {Hwang}\ \emph {et~al.}(2007)\citenamefont {Hwang},
  \citenamefont {Kim}, \citenamefont {Liu}, \citenamefont {Brédas},
  \citenamefont {Duggal},\ and\ \citenamefont {Kahn}}]{Hwang2007}%
  \BibitemOpen
  \bibfield  {author} {\bibinfo {author} {\bibfnamefont {J.}~\bibnamefont
  {Hwang}}, \bibinfo {author} {\bibfnamefont {E.-G.}\ \bibnamefont {Kim}},
  \bibinfo {author} {\bibfnamefont {J.}~\bibnamefont {Liu}}, \bibinfo {author}
  {\bibfnamefont {J.-L.}\ \bibnamefont {Brédas}}, \bibinfo {author}
  {\bibfnamefont {A.}~\bibnamefont {Duggal}}, \ and\ \bibinfo {author}
  {\bibfnamefont {A.}~\bibnamefont {Kahn}},\ }\href {\doibase
  10.1021/jp067004w} {\bibfield  {journal} {\bibinfo  {journal} {The Journal of
  Physical Chemistry C}\ }\textbf {\bibinfo {volume} {111}},\ \bibinfo {pages}
  {1378} (\bibinfo {year} {2007})}\BibitemShut {NoStop}%
\bibitem [{\citenamefont {Heimel}\ \emph {et~al.}(2008)\citenamefont {Heimel},
  \citenamefont {Romaner}, \citenamefont {Zojer},\ and\ \citenamefont
  {Bredas}}]{Heimel2008}%
  \BibitemOpen
  \bibfield  {author} {\bibinfo {author} {\bibfnamefont {G.}~\bibnamefont
  {Heimel}}, \bibinfo {author} {\bibfnamefont {L.}~\bibnamefont {Romaner}},
  \bibinfo {author} {\bibfnamefont {E.}~\bibnamefont {Zojer}}, \ and\ \bibinfo
  {author} {\bibfnamefont {J.-L.}\ \bibnamefont {Bredas}},\ }\href {\doibase
  10.1021/ar700284q} {\bibfield  {journal} {\bibinfo  {journal} {Accounts of
  Chemical Research}\ }\textbf {\bibinfo {volume} {41}},\ \bibinfo {pages}
  {721} (\bibinfo {year} {2008})}\BibitemShut {NoStop}%
\bibitem [{\citenamefont {K\"orzd\"orfer}\ \emph {et~al.}(2009)\citenamefont
  {K\"orzd\"orfer}, \citenamefont {K\"ummel}, \citenamefont {Marom},\ and\
  \citenamefont {Kronik}}]{Korzdorfer2009}%
  \BibitemOpen
  \bibfield  {author} {\bibinfo {author} {\bibfnamefont {T.}~\bibnamefont
  {K\"orzd\"orfer}}, \bibinfo {author} {\bibfnamefont {S.}~\bibnamefont
  {K\"ummel}}, \bibinfo {author} {\bibfnamefont {N.}~\bibnamefont {Marom}}, \
  and\ \bibinfo {author} {\bibfnamefont {L.}~\bibnamefont {Kronik}},\ }\href
  {\doibase 10.1103/PhysRevB.79.201205} {\bibfield  {journal} {\bibinfo
  {journal} {Phys. Rev. B}\ }\textbf {\bibinfo {volume} {79}},\ \bibinfo
  {pages} {201205} (\bibinfo {year} {2009})}\BibitemShut {NoStop}%
\bibitem [{\citenamefont {Dauth}\ \emph {et~al.}(2014)\citenamefont {Dauth},
  \citenamefont {Wiessner}, \citenamefont {Feyer}, \citenamefont {Sch\"oll},
  \citenamefont {Puschnig}, \citenamefont {Reinert},\ and\ \citenamefont
  {K\"ummel}}]{Dauth2014}%
  \BibitemOpen
  \bibfield  {author} {\bibinfo {author} {\bibfnamefont {M.}~\bibnamefont
  {Dauth}}, \bibinfo {author} {\bibfnamefont {M.}~\bibnamefont {Wiessner}},
  \bibinfo {author} {\bibfnamefont {V.}~\bibnamefont {Feyer}}, \bibinfo
  {author} {\bibfnamefont {A.}~\bibnamefont {Sch\"oll}}, \bibinfo {author}
  {\bibfnamefont {P.}~\bibnamefont {Puschnig}}, \bibinfo {author}
  {\bibfnamefont {F.}~\bibnamefont {Reinert}}, \ and\ \bibinfo {author}
  {\bibfnamefont {S.}~\bibnamefont {K\"ummel}},\ }\href {\doibase
  10.1088/1367-2630/16/10/103005} {\bibfield  {journal} {\bibinfo  {journal}
  {New J. Phys.}\ }\textbf {\bibinfo {volume} {16}},\ \bibinfo {pages} {103005}
  (\bibinfo {year} {2014})}\BibitemShut {NoStop}%
\bibitem [{\citenamefont {Woodruff}(2016)}]{Woodruff2016}%
  \BibitemOpen
  \bibfield  {author} {\bibinfo {author} {\bibfnamefont {D.~P.}\ \bibnamefont
  {Woodruff}},\ }\href
  {https://books.google.de/books?hl=de&lr=&id=8KUODQAAQBAJ&oi=fnd&pg=PR7&dq=Modern+techniques+of+surface+science+woodruff&ots=WPswCR7sLE&sig=RHlXAdxofrv_o0j_cgItPZt3y4o#v=onepage&q=Modern%20techniques%20of%20surface%20science%20woodruff&f=false}
  {\emph {\bibinfo {title} {Modern techniques of surface science}}},\ \bibinfo
  {edition} {3rd}\ ed.\ (\bibinfo  {publisher} {Cambridge University Press},\
  \bibinfo {year} {2016})\BibitemShut {NoStop}%
\bibitem [{\citenamefont {Puschnig}\ \emph {et~al.}(2017)\citenamefont
  {Puschnig}, \citenamefont {Boese}, \citenamefont {Willenbockel},
  \citenamefont {Meyer}, \citenamefont {Lüftner}, \citenamefont {Reinisch},
  \citenamefont {Ules}, \citenamefont {Koller}, \citenamefont {Soubatch},
  \citenamefont {Ramsey},\ and\ \citenamefont {Tautz}}]{Puschnig2017}%
  \BibitemOpen
  \bibfield  {author} {\bibinfo {author} {\bibfnamefont {P.}~\bibnamefont
  {Puschnig}}, \bibinfo {author} {\bibfnamefont {A.~D.}\ \bibnamefont {Boese}},
  \bibinfo {author} {\bibfnamefont {M.}~\bibnamefont {Willenbockel}}, \bibinfo
  {author} {\bibfnamefont {M.}~\bibnamefont {Meyer}}, \bibinfo {author}
  {\bibfnamefont {D.}~\bibnamefont {Lüftner}}, \bibinfo {author}
  {\bibfnamefont {E.~M.}\ \bibnamefont {Reinisch}}, \bibinfo {author}
  {\bibfnamefont {T.}~\bibnamefont {Ules}}, \bibinfo {author} {\bibfnamefont
  {G.}~\bibnamefont {Koller}}, \bibinfo {author} {\bibfnamefont
  {S.}~\bibnamefont {Soubatch}}, \bibinfo {author} {\bibfnamefont {M.~G.}\
  \bibnamefont {Ramsey}}, \ and\ \bibinfo {author} {\bibfnamefont {F.~S.}\
  \bibnamefont {Tautz}},\ }\href
  {https://pubs.acs.org/doi/10.1021/acs.jpclett.6b02517} {\bibfield  {journal}
  {\bibinfo  {journal} {J. Phys. Chem. Lett.}\ }\textbf {\bibinfo {volume}
  {8}},\ \bibinfo {pages} {208} (\bibinfo {year} {2017})}\BibitemShut {NoStop}%
\bibitem [{\citenamefont {Dauth}\ \emph {et~al.}(2011)\citenamefont {Dauth},
  \citenamefont {Körzdörfer}, \citenamefont {Kümmel}, \citenamefont
  {Reinert}, \citenamefont {Ziroff}, \citenamefont {Wiessner}, \citenamefont
  {Schöll}, \citenamefont {Arita},\ and\ \citenamefont {Shimada}}]{Dauth2011}%
  \BibitemOpen
  \bibfield  {author} {\bibinfo {author} {\bibfnamefont {M.}~\bibnamefont
  {Dauth}}, \bibinfo {author} {\bibfnamefont {T.}~\bibnamefont {Körzdörfer}},
  \bibinfo {author} {\bibfnamefont {S.}~\bibnamefont {Kümmel}}, \bibinfo
  {author} {\bibfnamefont {F.}~\bibnamefont {Reinert}}, \bibinfo {author}
  {\bibfnamefont {J.}~\bibnamefont {Ziroff}}, \bibinfo {author} {\bibfnamefont
  {M.}~\bibnamefont {Wiessner}}, \bibinfo {author} {\bibfnamefont
  {A.}~\bibnamefont {Schöll}}, \bibinfo {author} {\bibfnamefont
  {M.}~\bibnamefont {Arita}}, \ and\ \bibinfo {author} {\bibfnamefont
  {K.}~\bibnamefont {Shimada}},\ }\href
  {https://journals.aps.org/prl/abstract/10.1103/PhysRevLett.107.193002}
  {\bibfield  {journal} {\bibinfo  {journal} {Phys. Rev. Lett.}\ }\textbf
  {\bibinfo {volume} {107}},\ \bibinfo {pages} {193002} (\bibinfo {year}
  {2011})}\BibitemShut {NoStop}%
\bibitem [{\citenamefont {Puschnig}\ \emph {et~al.}(2011)\citenamefont
  {Puschnig}, \citenamefont {Reinisch}, \citenamefont {Ules}, \citenamefont
  {Koller}, \citenamefont {Soubatch}, \citenamefont {Ostler}, \citenamefont
  {Romaner}, \citenamefont {Tautz}, \citenamefont {Ambrosch-Draxl},\ and\
  \citenamefont {Ramsey}}]{Puschnig2011}%
  \BibitemOpen
  \bibfield  {author} {\bibinfo {author} {\bibfnamefont {P.}~\bibnamefont
  {Puschnig}}, \bibinfo {author} {\bibfnamefont {E.-M.}\ \bibnamefont
  {Reinisch}}, \bibinfo {author} {\bibfnamefont {T.}~\bibnamefont {Ules}},
  \bibinfo {author} {\bibfnamefont {G.}~\bibnamefont {Koller}}, \bibinfo
  {author} {\bibfnamefont {S.}~\bibnamefont {Soubatch}}, \bibinfo {author}
  {\bibfnamefont {M.}~\bibnamefont {Ostler}}, \bibinfo {author} {\bibfnamefont
  {L.}~\bibnamefont {Romaner}}, \bibinfo {author} {\bibfnamefont {F.~S.}\
  \bibnamefont {Tautz}}, \bibinfo {author} {\bibfnamefont {C.}~\bibnamefont
  {Ambrosch-Draxl}}, \ and\ \bibinfo {author} {\bibfnamefont {M.~G.}\
  \bibnamefont {Ramsey}},\ }\href {\doibase 10.1103/PhysRevB.84.235427}
  {\bibfield  {journal} {\bibinfo  {journal} {Phys. Rev. B}\ }\textbf {\bibinfo
  {volume} {84}},\ \bibinfo {pages} {235427} (\bibinfo {year}
  {2011})}\BibitemShut {NoStop}%
\bibitem [{\citenamefont {Zamborlini}\ \emph {et~al.}(2017)\citenamefont
  {Zamborlini}, \citenamefont {Lüftner}, \citenamefont {Feng}, \citenamefont
  {Kollmann}, \citenamefont {Puschnig}, \citenamefont {Dri}, \citenamefont
  {Panighel}, \citenamefont {Santo}, \citenamefont {Goldoni}, \citenamefont
  {Comelli}, \citenamefont {Jugovac}, \citenamefont {Feyer},\ and\
  \citenamefont {Schneider}}]{Zamborlini2017}%
  \BibitemOpen
  \bibfield  {author} {\bibinfo {author} {\bibfnamefont {G.}~\bibnamefont
  {Zamborlini}}, \bibinfo {author} {\bibfnamefont {D.}~\bibnamefont
  {Lüftner}}, \bibinfo {author} {\bibfnamefont {Z.}~\bibnamefont {Feng}},
  \bibinfo {author} {\bibfnamefont {B.}~\bibnamefont {Kollmann}}, \bibinfo
  {author} {\bibfnamefont {P.}~\bibnamefont {Puschnig}}, \bibinfo {author}
  {\bibfnamefont {C.}~\bibnamefont {Dri}}, \bibinfo {author} {\bibfnamefont
  {M.}~\bibnamefont {Panighel}}, \bibinfo {author} {\bibfnamefont {G.~D.}\
  \bibnamefont {Santo}}, \bibinfo {author} {\bibfnamefont {A.}~\bibnamefont
  {Goldoni}}, \bibinfo {author} {\bibfnamefont {G.}~\bibnamefont {Comelli}},
  \bibinfo {author} {\bibfnamefont {M.}~\bibnamefont {Jugovac}}, \bibinfo
  {author} {\bibfnamefont {V.}~\bibnamefont {Feyer}}, \ and\ \bibinfo {author}
  {\bibfnamefont {C.~M.}\ \bibnamefont {Schneider}},\ }\href
  {https://www.nature.com/articles/s41467-017-00402-0} {\bibfield  {journal}
  {\bibinfo  {journal} {Nat. Commun.}\ }\textbf {\bibinfo {volume} {8}},\
  \bibinfo {pages} {335} (\bibinfo {year} {2017})}\BibitemShut {NoStop}%
\bibitem [{\citenamefont {Yang}\ \emph {et~al.}(2019)\citenamefont {Yang},
  \citenamefont {Egger}, \citenamefont {Hurdax}, \citenamefont {Kaser},
  \citenamefont {Lüftner}, \citenamefont {Bocquet}, \citenamefont {Koller},
  \citenamefont {Gottwald}, \citenamefont {Tegeder}, \citenamefont {Richter},
  \citenamefont {Ramsey}, \citenamefont {Puschnig}, \citenamefont {Soubatch},\
  and\ \citenamefont {Tautz}}]{Yang2019}%
  \BibitemOpen
  \bibfield  {author} {\bibinfo {author} {\bibfnamefont {X.}~\bibnamefont
  {Yang}}, \bibinfo {author} {\bibfnamefont {L.}~\bibnamefont {Egger}},
  \bibinfo {author} {\bibfnamefont {P.}~\bibnamefont {Hurdax}}, \bibinfo
  {author} {\bibfnamefont {H.}~\bibnamefont {Kaser}}, \bibinfo {author}
  {\bibfnamefont {D.}~\bibnamefont {Lüftner}}, \bibinfo {author}
  {\bibfnamefont {F.~C.}\ \bibnamefont {Bocquet}}, \bibinfo {author}
  {\bibfnamefont {G.}~\bibnamefont {Koller}}, \bibinfo {author} {\bibfnamefont
  {A.}~\bibnamefont {Gottwald}}, \bibinfo {author} {\bibfnamefont
  {P.}~\bibnamefont {Tegeder}}, \bibinfo {author} {\bibfnamefont
  {M.}~\bibnamefont {Richter}}, \bibinfo {author} {\bibfnamefont {M.~G.}\
  \bibnamefont {Ramsey}}, \bibinfo {author} {\bibfnamefont {P.}~\bibnamefont
  {Puschnig}}, \bibinfo {author} {\bibfnamefont {S.}~\bibnamefont {Soubatch}},
  \ and\ \bibinfo {author} {\bibfnamefont {F.~S.}\ \bibnamefont {Tautz}},\
  }\href {https://www.nature.com/articles/s41467-019-11133-9} {\bibfield
  {journal} {\bibinfo  {journal} {Nat. Commun.}\ }\textbf {\bibinfo {volume}
  {10}},\ \bibinfo {pages} {3189} (\bibinfo {year} {2019})}\BibitemShut
  {NoStop}%
\bibitem [{\citenamefont {Kliuiev}\ \emph {et~al.}(2019)\citenamefont
  {Kliuiev}, \citenamefont {Zamborlini}, \citenamefont {Jugovac}, \citenamefont
  {Gurdal}, \citenamefont {Arx}, \citenamefont {Waltar}, \citenamefont
  {Schnidrig}, \citenamefont {Alberto}, \citenamefont {Iannuzzi}, \citenamefont
  {Feyer}, \citenamefont {Hengsberger}, \citenamefont {Osterwalder},\ and\
  \citenamefont {Castiglioni}}]{Kliuiev2019}%
  \BibitemOpen
  \bibfield  {author} {\bibinfo {author} {\bibfnamefont {P.}~\bibnamefont
  {Kliuiev}}, \bibinfo {author} {\bibfnamefont {G.}~\bibnamefont {Zamborlini}},
  \bibinfo {author} {\bibfnamefont {M.}~\bibnamefont {Jugovac}}, \bibinfo
  {author} {\bibfnamefont {Y.}~\bibnamefont {Gurdal}}, \bibinfo {author}
  {\bibfnamefont {K.~V.}\ \bibnamefont {Arx}}, \bibinfo {author} {\bibfnamefont
  {K.}~\bibnamefont {Waltar}}, \bibinfo {author} {\bibfnamefont
  {S.}~\bibnamefont {Schnidrig}}, \bibinfo {author} {\bibfnamefont
  {R.}~\bibnamefont {Alberto}}, \bibinfo {author} {\bibfnamefont
  {M.}~\bibnamefont {Iannuzzi}}, \bibinfo {author} {\bibfnamefont
  {V.}~\bibnamefont {Feyer}}, \bibinfo {author} {\bibfnamefont
  {M.}~\bibnamefont {Hengsberger}}, \bibinfo {author} {\bibfnamefont
  {J.}~\bibnamefont {Osterwalder}}, \ and\ \bibinfo {author} {\bibfnamefont
  {L.}~\bibnamefont {Castiglioni}},\ }\href
  {https://www.nature.com/articles/s41467-019-13254-7} {\bibfield  {journal}
  {\bibinfo  {journal} {Nat. Commun.}\ }\textbf {\bibinfo {volume} {10}},\
  \bibinfo {pages} {5255} (\bibinfo {year} {2019})}\BibitemShut {NoStop}%
\bibitem [{\citenamefont {Brandstetter}\ \emph {et~al.}(2021)\citenamefont
  {Brandstetter}, \citenamefont {Yang}, \citenamefont {Lüftner}, \citenamefont
  {Tautz},\ and\ \citenamefont {Puschnig}}]{Brandstetter2021}%
  \BibitemOpen
  \bibfield  {author} {\bibinfo {author} {\bibfnamefont {D.}~\bibnamefont
  {Brandstetter}}, \bibinfo {author} {\bibfnamefont {X.}~\bibnamefont {Yang}},
  \bibinfo {author} {\bibfnamefont {D.}~\bibnamefont {Lüftner}}, \bibinfo
  {author} {\bibfnamefont {F.~S.}\ \bibnamefont {Tautz}}, \ and\ \bibinfo
  {author} {\bibfnamefont {P.}~\bibnamefont {Puschnig}},\ }\href {\doibase
  https://doi.org/10.1016/j.cpc.2021.107905} {\bibfield  {journal} {\bibinfo
  {journal} {Comput. Phys. Commun.}\ }\textbf {\bibinfo {volume} {263}},\
  \bibinfo {pages} {107905} (\bibinfo {year} {2021})}\BibitemShut {NoStop}%
\bibitem [{\citenamefont {Sättele}\ \emph {et~al.}(2021)\citenamefont
  {Sättele}, \citenamefont {Windischbacher}, \citenamefont {Egger},
  \citenamefont {Haags}, \citenamefont {Hurdax}, \citenamefont {Kirschner},
  \citenamefont {Gottwald}, \citenamefont {Richter}, \citenamefont {Bocquet},
  \citenamefont {Soubatch}, \citenamefont {Tautz}, \citenamefont {Bettinger},
  \citenamefont {Peisert}, \citenamefont {Chassé}, \citenamefont {Ramsey},
  \citenamefont {Puschnig},\ and\ \citenamefont {Koller}}]{Saettele2021}%
  \BibitemOpen
  \bibfield  {author} {\bibinfo {author} {\bibfnamefont {M.~S.}\ \bibnamefont
  {Sättele}}, \bibinfo {author} {\bibfnamefont {A.}~\bibnamefont
  {Windischbacher}}, \bibinfo {author} {\bibfnamefont {L.}~\bibnamefont
  {Egger}}, \bibinfo {author} {\bibfnamefont {A.}~\bibnamefont {Haags}},
  \bibinfo {author} {\bibfnamefont {P.}~\bibnamefont {Hurdax}}, \bibinfo
  {author} {\bibfnamefont {H.}~\bibnamefont {Kirschner}}, \bibinfo {author}
  {\bibfnamefont {A.}~\bibnamefont {Gottwald}}, \bibinfo {author}
  {\bibfnamefont {M.}~\bibnamefont {Richter}}, \bibinfo {author} {\bibfnamefont
  {F.~C.}\ \bibnamefont {Bocquet}}, \bibinfo {author} {\bibfnamefont
  {S.}~\bibnamefont {Soubatch}}, \bibinfo {author} {\bibfnamefont {F.~S.}\
  \bibnamefont {Tautz}}, \bibinfo {author} {\bibfnamefont {H.~F.}\ \bibnamefont
  {Bettinger}}, \bibinfo {author} {\bibfnamefont {H.}~\bibnamefont {Peisert}},
  \bibinfo {author} {\bibfnamefont {T.}~\bibnamefont {Chassé}}, \bibinfo
  {author} {\bibfnamefont {M.~G.}\ \bibnamefont {Ramsey}}, \bibinfo {author}
  {\bibfnamefont {P.}~\bibnamefont {Puschnig}}, \ and\ \bibinfo {author}
  {\bibfnamefont {G.}~\bibnamefont {Koller}},\ }\href
  {https://pubs.acs.org/doi/10.1021/acs.jpcc.0c09062} {\bibfield  {journal}
  {\bibinfo  {journal} {J. Phys. Chem. C}\ }\textbf {\bibinfo {volume} {125}},\
  \bibinfo {pages} {2918} (\bibinfo {year} {2021})}\BibitemShut {NoStop}%
\bibitem [{\citenamefont {Haags}\ \emph {et~al.}(2022)\citenamefont {Haags},
  \citenamefont {Yang}, \citenamefont {Egger}, \citenamefont {Brandstetter},
  \citenamefont {Kirschner}, \citenamefont {Bocquet}, \citenamefont {Koller},
  \citenamefont {Gottwald}, \citenamefont {Richter}, \citenamefont {Gottfried},
  \citenamefont {Ramsey}, \citenamefont {Puschnig}, \citenamefont {Soubatch},\
  and\ \citenamefont {Tautz}}]{Haags2022}%
  \BibitemOpen
  \bibfield  {author} {\bibinfo {author} {\bibfnamefont {A.}~\bibnamefont
  {Haags}}, \bibinfo {author} {\bibfnamefont {X.}~\bibnamefont {Yang}},
  \bibinfo {author} {\bibfnamefont {L.}~\bibnamefont {Egger}}, \bibinfo
  {author} {\bibfnamefont {D.}~\bibnamefont {Brandstetter}}, \bibinfo {author}
  {\bibfnamefont {H.}~\bibnamefont {Kirschner}}, \bibinfo {author}
  {\bibfnamefont {F.~C.}\ \bibnamefont {Bocquet}}, \bibinfo {author}
  {\bibfnamefont {G.}~\bibnamefont {Koller}}, \bibinfo {author} {\bibfnamefont
  {A.}~\bibnamefont {Gottwald}}, \bibinfo {author} {\bibfnamefont
  {M.}~\bibnamefont {Richter}}, \bibinfo {author} {\bibfnamefont {J.~M.}\
  \bibnamefont {Gottfried}}, \bibinfo {author} {\bibfnamefont {M.~G.}\
  \bibnamefont {Ramsey}}, \bibinfo {author} {\bibfnamefont {P.}~\bibnamefont
  {Puschnig}}, \bibinfo {author} {\bibfnamefont {S.}~\bibnamefont {Soubatch}},
  \ and\ \bibinfo {author} {\bibfnamefont {F.~S.}\ \bibnamefont {Tautz}},\
  }\href {https://www.science.org/doi/full/10.1126/sciadv.abn0819} {\bibfield
  {journal} {\bibinfo  {journal} {Sci. Adv.}\ }\textbf {\bibinfo {volume}
  {8}},\ \bibinfo {pages} {819} (\bibinfo {year} {2022})}\BibitemShut {NoStop}%
\bibitem [{\citenamefont {Valiev}\ \emph {et~al.}(2010)\citenamefont {Valiev},
  \citenamefont {Bylaska}, \citenamefont {Govind}, \citenamefont {Kowalski},
  \citenamefont {Straatsma}, \citenamefont {{Van Dam}}, \citenamefont {Wang},
  \citenamefont {Nieplocha}, \citenamefont {Apra}, \citenamefont {Windus},\
  and\ \citenamefont {{de Jong}}}]{Valiev2010}%
  \BibitemOpen
  \bibfield  {author} {\bibinfo {author} {\bibfnamefont {M.}~\bibnamefont
  {Valiev}}, \bibinfo {author} {\bibfnamefont {E.}~\bibnamefont {Bylaska}},
  \bibinfo {author} {\bibfnamefont {N.}~\bibnamefont {Govind}}, \bibinfo
  {author} {\bibfnamefont {K.}~\bibnamefont {Kowalski}}, \bibinfo {author}
  {\bibfnamefont {T.}~\bibnamefont {Straatsma}}, \bibinfo {author}
  {\bibfnamefont {H.}~\bibnamefont {{Van Dam}}}, \bibinfo {author}
  {\bibfnamefont {D.}~\bibnamefont {Wang}}, \bibinfo {author} {\bibfnamefont
  {J.}~\bibnamefont {Nieplocha}}, \bibinfo {author} {\bibfnamefont
  {E.}~\bibnamefont {Apra}}, \bibinfo {author} {\bibfnamefont {T.}~\bibnamefont
  {Windus}}, \ and\ \bibinfo {author} {\bibfnamefont {W.}~\bibnamefont {{de
  Jong}}},\ }\href {\doibase https://doi.org/10.1016/j.cpc.2010.04.018}
  {\bibfield  {journal} {\bibinfo  {journal} {Computer Physics Communications}\
  }\textbf {\bibinfo {volume} {181}},\ \bibinfo {pages} {1477} (\bibinfo {year}
  {2010})}\BibitemShut {NoStop}%
\bibitem [{\citenamefont {Perdew}\ \emph
  {et~al.}(1996{\natexlab{b}})\citenamefont {Perdew}, \citenamefont
  {Ernzerhof},\ and\ \citenamefont {Burke}}]{Perdew1996a}%
  \BibitemOpen
  \bibfield  {author} {\bibinfo {author} {\bibfnamefont {J.~P.}\ \bibnamefont
  {Perdew}}, \bibinfo {author} {\bibfnamefont {M.}~\bibnamefont {Ernzerhof}}, \
  and\ \bibinfo {author} {\bibfnamefont {K.}~\bibnamefont {Burke}},\ }\href
  {https://pubs.aip.org/aip/jcp/article-abstract/105/22/9982/478038/Rationale-for-mixing-exact-exchange-with-density?redirectedFrom=fulltext}
  {\bibfield  {journal} {\bibinfo  {journal} {J. Chem. Phys.}\ }\textbf
  {\bibinfo {volume} {105}},\ \bibinfo {pages} {9982} (\bibinfo {year}
  {1996}{\natexlab{b}})}\BibitemShut {NoStop}%
\bibitem [{\citenamefont {Becke}(1993)}]{Becke1993}%
  \BibitemOpen
  \bibfield  {author} {\bibinfo {author} {\bibfnamefont {A.~D.}\ \bibnamefont
  {Becke}},\ }\href
  {https://pubs.aip.org/aip/jcp/article-abstract/98/7/5648/842114/Density-functional-thermochemistry-III-The-role-of?redirectedFrom=fulltext}
  {\bibfield  {journal} {\bibinfo  {journal} {J. Chem. Phys.}\ }\textbf
  {\bibinfo {volume} {98}},\ \bibinfo {pages} {5648} (\bibinfo {year}
  {1993})}\BibitemShut {NoStop}%
\bibitem [{\citenamefont {Puschnig}(2020)}]{Puschnig2020}%
  \BibitemOpen
  \bibfield  {author} {\bibinfo {author} {\bibfnamefont {P.}~\bibnamefont
  {Puschnig}},\ }\href {http://physikmdb.uni-graz.at:5001} {\enquote {\bibinfo
  {title} {Organic molecule database: a database for molecular orbitals of
  $\pi$-conjugated organic molecules based on the atomic simulation environment
  (ase) and nwchem as the dft calculator
  (http://physikmdb.uni-graz.at:5001)},}\ } (\bibinfo {year}
  {2020})\BibitemShut {NoStop}%
\bibitem [{\citenamefont {Kresse}\ and\ \citenamefont
  {Furthmüller}(1996)}]{Kresse1996}%
  \BibitemOpen
  \bibfield  {author} {\bibinfo {author} {\bibfnamefont {G.}~\bibnamefont
  {Kresse}}\ and\ \bibinfo {author} {\bibfnamefont {J.}~\bibnamefont
  {Furthmüller}},\ }\href {\doibase
  https://doi.org/10.1016/0927-0256(96)00008-0} {\bibfield  {journal} {\bibinfo
   {journal} {Computational Materials Science}\ }\textbf {\bibinfo {volume}
  {6}},\ \bibinfo {pages} {15} (\bibinfo {year} {1996})}\BibitemShut {NoStop}%
\bibitem [{\citenamefont {Kresse}\ and\ \citenamefont
  {Furthm\"uller}(1996)}]{Kresse1996b}%
  \BibitemOpen
  \bibfield  {author} {\bibinfo {author} {\bibfnamefont {G.}~\bibnamefont
  {Kresse}}\ and\ \bibinfo {author} {\bibfnamefont {J.}~\bibnamefont
  {Furthm\"uller}},\ }\href {\doibase 10.1103/PhysRevB.54.11169} {\bibfield
  {journal} {\bibinfo  {journal} {Phys. Rev. B}\ }\textbf {\bibinfo {volume}
  {54}},\ \bibinfo {pages} {11169} (\bibinfo {year} {1996})}\BibitemShut
  {NoStop}%
\bibitem [{\citenamefont {Kresse}\ and\ \citenamefont
  {Joubert}(1999)}]{Kresse1999}%
  \BibitemOpen
  \bibfield  {author} {\bibinfo {author} {\bibfnamefont {G.}~\bibnamefont
  {Kresse}}\ and\ \bibinfo {author} {\bibfnamefont {D.}~\bibnamefont
  {Joubert}},\ }\href {\doibase 10.1103/PhysRevB.59.1758} {\bibfield  {journal}
  {\bibinfo  {journal} {Phys. Rev. B}\ }\textbf {\bibinfo {volume} {59}},\
  \bibinfo {pages} {1758} (\bibinfo {year} {1999})}\BibitemShut {NoStop}%
\bibitem [{\citenamefont {Neugebauer}\ and\ \citenamefont
  {Scheffler}(1992)}]{Neugebauer1992}%
  \BibitemOpen
  \bibfield  {author} {\bibinfo {author} {\bibfnamefont {J.}~\bibnamefont
  {Neugebauer}}\ and\ \bibinfo {author} {\bibfnamefont {M.}~\bibnamefont
  {Scheffler}},\ }\href {\doibase 10.1103/PhysRevB.46.16067} {\bibfield
  {journal} {\bibinfo  {journal} {Phys. Rev. B}\ }\textbf {\bibinfo {volume}
  {46}},\ \bibinfo {pages} {16067} (\bibinfo {year} {1992})}\BibitemShut
  {NoStop}%
\bibitem [{\citenamefont {Grimme}\ \emph {et~al.}(2010)\citenamefont {Grimme},
  \citenamefont {Antony}, \citenamefont {Ehrlich},\ and\ \citenamefont
  {Krieg}}]{Grimme2010}%
  \BibitemOpen
  \bibfield  {author} {\bibinfo {author} {\bibfnamefont {S.}~\bibnamefont
  {Grimme}}, \bibinfo {author} {\bibfnamefont {J.}~\bibnamefont {Antony}},
  \bibinfo {author} {\bibfnamefont {S.}~\bibnamefont {Ehrlich}}, \ and\
  \bibinfo {author} {\bibfnamefont {H.}~\bibnamefont {Krieg}},\ }\href
  {\doibase 10.1063/1.3382344} {\bibfield  {journal} {\bibinfo  {journal} {The
  Journal of Chemical Physics}\ }\textbf {\bibinfo {volume} {132}},\ \bibinfo
  {pages} {154104} (\bibinfo {year} {2010})}\BibitemShut {NoStop}%
\bibitem [{\citenamefont {Bl\"ochl}(1994)}]{Bloechl1994}%
  \BibitemOpen
  \bibfield  {author} {\bibinfo {author} {\bibfnamefont {P.~E.}\ \bibnamefont
  {Bl\"ochl}},\ }\href {\doibase 10.1103/PhysRevB.50.17953} {\bibfield
  {journal} {\bibinfo  {journal} {Phys. Rev. B}\ }\textbf {\bibinfo {volume}
  {50}},\ \bibinfo {pages} {17953} (\bibinfo {year} {1994})}\BibitemShut
  {NoStop}%
\bibitem [{\citenamefont {Lüftner}\ \emph {et~al.}(2017)\citenamefont
  {Lüftner}, \citenamefont {Weiß}, \citenamefont {Yang}, \citenamefont
  {Hurdax}, \citenamefont {Feyer}, \citenamefont {Gottwald}, \citenamefont
  {Koller}, \citenamefont {Soubatch}, \citenamefont {Puschnig}, \citenamefont
  {Ramsey},\ and\ \citenamefont {Tautz}}]{Lueftner2017}%
  \BibitemOpen
  \bibfield  {author} {\bibinfo {author} {\bibfnamefont {D.}~\bibnamefont
  {Lüftner}}, \bibinfo {author} {\bibfnamefont {S.}~\bibnamefont {Weiß}},
  \bibinfo {author} {\bibfnamefont {X.}~\bibnamefont {Yang}}, \bibinfo {author}
  {\bibfnamefont {P.}~\bibnamefont {Hurdax}}, \bibinfo {author} {\bibfnamefont
  {V.}~\bibnamefont {Feyer}}, \bibinfo {author} {\bibfnamefont
  {A.}~\bibnamefont {Gottwald}}, \bibinfo {author} {\bibfnamefont
  {G.}~\bibnamefont {Koller}}, \bibinfo {author} {\bibfnamefont
  {S.}~\bibnamefont {Soubatch}}, \bibinfo {author} {\bibfnamefont
  {P.}~\bibnamefont {Puschnig}}, \bibinfo {author} {\bibfnamefont {M.~G.}\
  \bibnamefont {Ramsey}}, \ and\ \bibinfo {author} {\bibfnamefont {F.~S.}\
  \bibnamefont {Tautz}},\ }\href
  {https://journals.aps.org/prb/abstract/10.1103/PhysRevB.96.125402} {\bibfield
   {journal} {\bibinfo  {journal} {Phys. Rev. B}\ }\textbf {\bibinfo {volume}
  {96}},\ \bibinfo {pages} {125402} (\bibinfo {year} {2017})}\BibitemShut
  {NoStop}%
\bibitem [{\citenamefont {Feibelman}\ and\ \citenamefont
  {Eastman}(1974)}]{Feibelman1974}%
  \BibitemOpen
  \bibfield  {author} {\bibinfo {author} {\bibfnamefont {P.~J.}\ \bibnamefont
  {Feibelman}}\ and\ \bibinfo {author} {\bibfnamefont {D.~E.}\ \bibnamefont
  {Eastman}},\ }\href
  {https://journals.aps.org/prb/abstract/10.1103/PhysRevB.10.4932} {\bibfield
  {journal} {\bibinfo  {journal} {Phys. Rev. B}\ }\textbf {\bibinfo {volume}
  {10}},\ \bibinfo {pages} {4932} (\bibinfo {year} {1974})}\BibitemShut
  {NoStop}%
\bibitem [{\citenamefont {Gottwald}\ \emph {et~al.}(2019)\citenamefont
  {Gottwald}, \citenamefont {Kaser},\ and\ \citenamefont
  {Kolbe}}]{Gottwald2019}%
  \BibitemOpen
  \bibfield  {author} {\bibinfo {author} {\bibfnamefont {A.}~\bibnamefont
  {Gottwald}}, \bibinfo {author} {\bibfnamefont {H.}~\bibnamefont {Kaser}}, \
  and\ \bibinfo {author} {\bibfnamefont {M.}~\bibnamefont {Kolbe}},\ }\href
  {https://scripts.iucr.org/cgi-bin/paper?S1600577518018428} {\bibfield
  {journal} {\bibinfo  {journal} {J. Synchrotron Rad.}\ }\textbf {\bibinfo
  {volume} {26}},\ \bibinfo {pages} {535} (\bibinfo {year} {2019})}\BibitemShut
  {NoStop}%
\bibitem [{\citenamefont {Broekman}\ \emph {et~al.}(2005)\citenamefont
  {Broekman}, \citenamefont {Tadich}, \citenamefont {Huwald}, \citenamefont
  {Riley}, \citenamefont {Leckey}, \citenamefont {Seyller}, \citenamefont
  {Emtsev},\ and\ \citenamefont {Ley}}]{Broekman2005}%
  \BibitemOpen
  \bibfield  {author} {\bibinfo {author} {\bibfnamefont {L.}~\bibnamefont
  {Broekman}}, \bibinfo {author} {\bibfnamefont {A.}~\bibnamefont {Tadich}},
  \bibinfo {author} {\bibfnamefont {E.}~\bibnamefont {Huwald}}, \bibinfo
  {author} {\bibfnamefont {J.}~\bibnamefont {Riley}}, \bibinfo {author}
  {\bibfnamefont {R.}~\bibnamefont {Leckey}}, \bibinfo {author} {\bibfnamefont
  {T.}~\bibnamefont {Seyller}}, \bibinfo {author} {\bibfnamefont
  {K.}~\bibnamefont {Emtsev}}, \ and\ \bibinfo {author} {\bibfnamefont
  {L.}~\bibnamefont {Ley}},\ }\href
  {https://www.sciencedirect.com/science/article/pii/S0368204805000605}
  {\bibfield  {journal} {\bibinfo  {journal} {J. Electron Spectrosc. Relat.
  Phenom.}\ }\textbf {\bibinfo {volume} {144-147}},\ \bibinfo {pages} {1001}
  (\bibinfo {year} {2005})}\BibitemShut {NoStop}%
\bibitem [{\citenamefont {Seah}\ and\ \citenamefont {Dench}(1979)}]{Seah1979}%
  \BibitemOpen
  \bibfield  {author} {\bibinfo {author} {\bibfnamefont {M.~P.}\ \bibnamefont
  {Seah}}\ and\ \bibinfo {author} {\bibfnamefont {W.~A.}\ \bibnamefont
  {Dench}},\ }\href {\doibase https://doi.org/10.1002/sia.740010103} {\bibfield
   {journal} {\bibinfo  {journal} {Surf. Interface Anal.}\ }\textbf {\bibinfo
  {volume} {1}},\ \bibinfo {pages} {2} (\bibinfo {year} {1979})}\BibitemShut
  {NoStop}%
\bibitem [{\citenamefont {Weiß}\ \emph {et~al.}(2015)\citenamefont {Weiß},
  \citenamefont {Lüftner}, \citenamefont {Ules}, \citenamefont {Reinisch},
  \citenamefont {Kaser}, \citenamefont {Gottwald}, \citenamefont {Richter},
  \citenamefont {Soubatch}, \citenamefont {Koller}, \citenamefont {Ramsey},
  \citenamefont {Tautz},\ and\ \citenamefont {Puschnig}}]{Weiss2015}%
  \BibitemOpen
  \bibfield  {author} {\bibinfo {author} {\bibfnamefont {S.}~\bibnamefont
  {Weiß}}, \bibinfo {author} {\bibfnamefont {D.}~\bibnamefont {Lüftner}},
  \bibinfo {author} {\bibfnamefont {T.}~\bibnamefont {Ules}}, \bibinfo {author}
  {\bibfnamefont {E.~M.}\ \bibnamefont {Reinisch}}, \bibinfo {author}
  {\bibfnamefont {H.}~\bibnamefont {Kaser}}, \bibinfo {author} {\bibfnamefont
  {A.}~\bibnamefont {Gottwald}}, \bibinfo {author} {\bibfnamefont
  {M.}~\bibnamefont {Richter}}, \bibinfo {author} {\bibfnamefont
  {S.}~\bibnamefont {Soubatch}}, \bibinfo {author} {\bibfnamefont
  {G.}~\bibnamefont {Koller}}, \bibinfo {author} {\bibfnamefont {M.~G.}\
  \bibnamefont {Ramsey}}, \bibinfo {author} {\bibfnamefont {F.~S.}\
  \bibnamefont {Tautz}}, \ and\ \bibinfo {author} {\bibfnamefont
  {P.}~\bibnamefont {Puschnig}},\ }\href
  {https://www.nature.com/articles/ncomms9287} {\bibfield  {journal} {\bibinfo
  {journal} {Nat. Commun.}\ }\textbf {\bibinfo {volume} {6}},\ \bibinfo {pages}
  {8287} (\bibinfo {year} {2015})}\BibitemShut {NoStop}%
\bibitem [{\citenamefont {Ruiz}\ \emph {et~al.}(2023)\citenamefont {Ruiz},
  \citenamefont {Wagner}, \citenamefont {Maaß}, \citenamefont {Arefi},
  \citenamefont {Stremlau}, \citenamefont {Tegeder}, \citenamefont {Tautz},\
  and\ \citenamefont {Tkatchenko}}]{Ruiz2023}%
  \BibitemOpen
  \bibfield  {author} {\bibinfo {author} {\bibfnamefont {V.~G.}\ \bibnamefont
  {Ruiz}}, \bibinfo {author} {\bibfnamefont {C.}~\bibnamefont {Wagner}},
  \bibinfo {author} {\bibfnamefont {F.}~\bibnamefont {Maaß}}, \bibinfo
  {author} {\bibfnamefont {H.~H.}\ \bibnamefont {Arefi}}, \bibinfo {author}
  {\bibfnamefont {S.}~\bibnamefont {Stremlau}}, \bibinfo {author}
  {\bibfnamefont {P.}~\bibnamefont {Tegeder}}, \bibinfo {author} {\bibfnamefont
  {F.~S.}\ \bibnamefont {Tautz}}, \ and\ \bibinfo {author} {\bibfnamefont
  {A.}~\bibnamefont {Tkatchenko}},\ }\href {\doibase
  https://doi.org/10.1038/s42004-023-00925-2} {\bibfield  {journal} {\bibinfo
  {journal} {Commun. Chem.}\ }\textbf {\bibinfo {volume} {6}} (\bibinfo {year}
  {2023}),\ https://doi.org/10.1038/s42004-023-00925-2}\BibitemShut {NoStop}%
\bibitem [{\citenamefont {Alemzadeh~M.}\ and\ \citenamefont
  {R}(2006)}]{Alemzadeh2006}%
  \BibitemOpen
  \bibfield  {author} {\bibinfo {author} {\bibfnamefont {B.~S.~S.}\
  \bibnamefont {Alemzadeh~M.}}\ and\ \bibinfo {author} {\bibfnamefont
  {H.}~\bibnamefont {R}},\ }\href {\doibase
  https://doi.org/10.5220/0001206001220126} {\bibfield  {journal} {\bibinfo
  {journal} {Proceedings of the Third International Conference on Informatics
  in Control, Automation and Robotics}\ ,\ \bibinfo {pages} {122}} (\bibinfo
  {year} {2006})}\BibitemShut {NoStop}%
\bibitem [{\citenamefont {Emtsev}\ \emph {et~al.}(2006)\citenamefont {Emtsev},
  \citenamefont {Seyller}, \citenamefont {Ley}, \citenamefont {Broekman},
  \citenamefont {Tadich}, \citenamefont {Riley}, \citenamefont {Leckey},\ and\
  \citenamefont {Preuss}}]{Emtsev2006}%
  \BibitemOpen
  \bibfield  {author} {\bibinfo {author} {\bibfnamefont {K.~V.}\ \bibnamefont
  {Emtsev}}, \bibinfo {author} {\bibfnamefont {T.}~\bibnamefont {Seyller}},
  \bibinfo {author} {\bibfnamefont {L.}~\bibnamefont {Ley}}, \bibinfo {author}
  {\bibfnamefont {L.}~\bibnamefont {Broekman}}, \bibinfo {author}
  {\bibfnamefont {A.}~\bibnamefont {Tadich}}, \bibinfo {author} {\bibfnamefont
  {J.~D.}\ \bibnamefont {Riley}}, \bibinfo {author} {\bibfnamefont {R.~G.~C.}\
  \bibnamefont {Leckey}}, \ and\ \bibinfo {author} {\bibfnamefont
  {M.}~\bibnamefont {Preuss}},\ }\href {\doibase 10.1103/PhysRevB.73.075412}
  {\bibfield  {journal} {\bibinfo  {journal} {Phys. Rev. B}\ }\textbf {\bibinfo
  {volume} {73}},\ \bibinfo {pages} {075412} (\bibinfo {year}
  {2006})}\BibitemShut {NoStop}%
\bibitem [{\citenamefont {Willenbockel}\ \emph {et~al.}(2013)\citenamefont
  {Willenbockel}, \citenamefont {Stadtmüller}, \citenamefont {Schönauer},
  \citenamefont {Bocquet}, \citenamefont {Lüftner}, \citenamefont {Reinisch},
  \citenamefont {Ules}, \citenamefont {Koller}, \citenamefont {Kumpf},
  \citenamefont {Soubatch}, \citenamefont {Puschnig}, \citenamefont {Ramsey},\
  and\ \citenamefont {Tautz}}]{Willenbockel2013}%
  \BibitemOpen
  \bibfield  {author} {\bibinfo {author} {\bibfnamefont {M.}~\bibnamefont
  {Willenbockel}}, \bibinfo {author} {\bibfnamefont {B.}~\bibnamefont
  {Stadtmüller}}, \bibinfo {author} {\bibfnamefont {K.}~\bibnamefont
  {Schönauer}}, \bibinfo {author} {\bibfnamefont {F.~C.}\ \bibnamefont
  {Bocquet}}, \bibinfo {author} {\bibfnamefont {D.}~\bibnamefont {Lüftner}},
  \bibinfo {author} {\bibfnamefont {E.~M.}\ \bibnamefont {Reinisch}}, \bibinfo
  {author} {\bibfnamefont {T.}~\bibnamefont {Ules}}, \bibinfo {author}
  {\bibfnamefont {G.}~\bibnamefont {Koller}}, \bibinfo {author} {\bibfnamefont
  {C.}~\bibnamefont {Kumpf}}, \bibinfo {author} {\bibfnamefont
  {S.}~\bibnamefont {Soubatch}}, \bibinfo {author} {\bibfnamefont
  {P.}~\bibnamefont {Puschnig}}, \bibinfo {author} {\bibfnamefont {M.~G.}\
  \bibnamefont {Ramsey}}, \ and\ \bibinfo {author} {\bibfnamefont {F.~S.}\
  \bibnamefont {Tautz}},\ }\href
  {https://iopscience.iop.org/article/10.1088/1367-2630/15/3/033017} {\bibfield
   {journal} {\bibinfo  {journal} {New J. Phys.}\ }\textbf {\bibinfo {volume}
  {15}},\ \bibinfo {pages} {033017} (\bibinfo {year} {2013})}\BibitemShut
  {NoStop}%
\bibitem [{\citenamefont {Willenbockel}\ \emph
  {et~al.}(2015{\natexlab{b}})\citenamefont {Willenbockel}, \citenamefont
  {Lüftner}, \citenamefont {Stadtmüller}, \citenamefont {Koller},
  \citenamefont {Kumpf}, \citenamefont {Soubatch}, \citenamefont {Puschnig},
  \citenamefont {Ramsey},\ and\ \citenamefont {Tautz}}]{Willenbockel2015}%
  \BibitemOpen
  \bibfield  {author} {\bibinfo {author} {\bibfnamefont {M.}~\bibnamefont
  {Willenbockel}}, \bibinfo {author} {\bibfnamefont {D.}~\bibnamefont
  {Lüftner}}, \bibinfo {author} {\bibfnamefont {B.}~\bibnamefont
  {Stadtmüller}}, \bibinfo {author} {\bibfnamefont {G.}~\bibnamefont
  {Koller}}, \bibinfo {author} {\bibfnamefont {C.}~\bibnamefont {Kumpf}},
  \bibinfo {author} {\bibfnamefont {S.}~\bibnamefont {Soubatch}}, \bibinfo
  {author} {\bibfnamefont {P.}~\bibnamefont {Puschnig}}, \bibinfo {author}
  {\bibfnamefont {M.~G.}\ \bibnamefont {Ramsey}}, \ and\ \bibinfo {author}
  {\bibfnamefont {F.~S.}\ \bibnamefont {Tautz}},\ }\href {\doibase
  10.1039/C4CP04595E} {\bibfield  {journal} {\bibinfo  {journal} {Phys. Chem.
  Chem. Phys.}\ }\textbf {\bibinfo {volume} {17}},\ \bibinfo {pages} {1530}
  (\bibinfo {year} {2015}{\natexlab{b}})}\BibitemShut {NoStop}%
\bibitem [{\citenamefont {Stadtmüller}\ \emph {et~al.}(2012)\citenamefont
  {Stadtmüller}, \citenamefont {Willenbockel}, \citenamefont {Reinisch},
  \citenamefont {Ules}, \citenamefont {Bocquet}, \citenamefont {Soubatch},
  \citenamefont {Puschnig}, \citenamefont {Koller}, \citenamefont {Ramsey},
  \citenamefont {Tautz},\ and\ \citenamefont {Kumpf}}]{Stadtmueller2012}%
  \BibitemOpen
  \bibfield  {author} {\bibinfo {author} {\bibfnamefont {B.}~\bibnamefont
  {Stadtmüller}}, \bibinfo {author} {\bibfnamefont {M.}~\bibnamefont
  {Willenbockel}}, \bibinfo {author} {\bibfnamefont {E.~M.}\ \bibnamefont
  {Reinisch}}, \bibinfo {author} {\bibfnamefont {T.}~\bibnamefont {Ules}},
  \bibinfo {author} {\bibfnamefont {F.~C.}\ \bibnamefont {Bocquet}}, \bibinfo
  {author} {\bibfnamefont {S.}~\bibnamefont {Soubatch}}, \bibinfo {author}
  {\bibfnamefont {P.}~\bibnamefont {Puschnig}}, \bibinfo {author}
  {\bibfnamefont {G.}~\bibnamefont {Koller}}, \bibinfo {author} {\bibfnamefont
  {M.~G.}\ \bibnamefont {Ramsey}}, \bibinfo {author} {\bibfnamefont {F.~S.}\
  \bibnamefont {Tautz}}, \ and\ \bibinfo {author} {\bibfnamefont
  {C.}~\bibnamefont {Kumpf}},\ }\href {\doibase 10.1209/0295-5075/100/26008}
  {\bibfield  {journal} {\bibinfo  {journal} {Europhysics Letters}\ }\textbf
  {\bibinfo {volume} {100}},\ \bibinfo {pages} {26008} (\bibinfo {year}
  {2012})}\BibitemShut {NoStop}%
\bibitem [{\citenamefont {Puschnig}\ and\ \citenamefont
  {L\"uftner}(2015)}]{Puschnig2015}%
  \BibitemOpen
  \bibfield  {author} {\bibinfo {author} {\bibfnamefont {P.}~\bibnamefont
  {Puschnig}}\ and\ \bibinfo {author} {\bibfnamefont {D.}~\bibnamefont
  {L\"uftner}},\ }\href {\doibase 10.1016/j.elspec.2015.06.003} {\bibfield
  {journal} {\bibinfo  {journal} {J. Elec. Spec. Rel. Phen.}\ }\textbf
  {\bibinfo {volume} {200}},\ \bibinfo {pages} {193} (\bibinfo {year}
  {2015})}\BibitemShut {NoStop}%
\bibitem [{\citenamefont {Refaely-Abramson}\ \emph {et~al.}(2013)\citenamefont
  {Refaely-Abramson}, \citenamefont {Sharifzadeh}, \citenamefont {Jain},
  \citenamefont {Baer}, \citenamefont {Neaton},\ and\ \citenamefont
  {Kronik}}]{Refaely-Abramson2013}%
  \BibitemOpen
  \bibfield  {author} {\bibinfo {author} {\bibfnamefont {S.}~\bibnamefont
  {Refaely-Abramson}}, \bibinfo {author} {\bibfnamefont {S.}~\bibnamefont
  {Sharifzadeh}}, \bibinfo {author} {\bibfnamefont {M.}~\bibnamefont {Jain}},
  \bibinfo {author} {\bibfnamefont {R.}~\bibnamefont {Baer}}, \bibinfo {author}
  {\bibfnamefont {J.~B.}\ \bibnamefont {Neaton}}, \ and\ \bibinfo {author}
  {\bibfnamefont {L.}~\bibnamefont {Kronik}},\ }\href {\doibase
  10.1103/PhysRevB.88.081204} {\bibfield  {journal} {\bibinfo  {journal} {Phys.
  Rev. B}\ }\textbf {\bibinfo {volume} {88}},\ \bibinfo {pages} {081204}
  (\bibinfo {year} {2013})}\BibitemShut {NoStop}%
\bibitem [{\citenamefont {Lüftner}\ \emph {et~al.}(2014)\citenamefont
  {Lüftner}, \citenamefont {Milko}, \citenamefont {Huppmann}, \citenamefont
  {Scholz}, \citenamefont {Ngyuen}, \citenamefont {Wießner}, \citenamefont
  {Schöll}, \citenamefont {Reinert},\ and\ \citenamefont
  {Puschnig}}]{Lueftner2014}%
  \BibitemOpen
  \bibfield  {author} {\bibinfo {author} {\bibfnamefont {D.}~\bibnamefont
  {Lüftner}}, \bibinfo {author} {\bibfnamefont {M.}~\bibnamefont {Milko}},
  \bibinfo {author} {\bibfnamefont {S.}~\bibnamefont {Huppmann}}, \bibinfo
  {author} {\bibfnamefont {M.}~\bibnamefont {Scholz}}, \bibinfo {author}
  {\bibfnamefont {N.}~\bibnamefont {Ngyuen}}, \bibinfo {author} {\bibfnamefont
  {M.}~\bibnamefont {Wießner}}, \bibinfo {author} {\bibfnamefont
  {A.}~\bibnamefont {Schöll}}, \bibinfo {author} {\bibfnamefont
  {F.}~\bibnamefont {Reinert}}, \ and\ \bibinfo {author} {\bibfnamefont
  {P.}~\bibnamefont {Puschnig}},\ }\href
  {https://www.sciencedirect.com/science/article/pii/S0368204814001376}
  {\bibfield  {journal} {\bibinfo  {journal} {J. Electron Spectrosc. Relat.
  Phenom.}\ }\textbf {\bibinfo {volume} {195}},\ \bibinfo {pages} {293}
  (\bibinfo {year} {2014})}\BibitemShut {NoStop}%
\bibitem [{\citenamefont {Kronik}\ and\ \citenamefont
  {Kümmel}(2018)}]{Kronik2018}%
  \BibitemOpen
  \bibfield  {author} {\bibinfo {author} {\bibfnamefont {L.}~\bibnamefont
  {Kronik}}\ and\ \bibinfo {author} {\bibfnamefont {S.}~\bibnamefont
  {Kümmel}},\ }\href {\doibase https://doi.org/10.1002/adma.201706560}
  {\bibfield  {journal} {\bibinfo  {journal} {Advanced Materials}\ }\textbf
  {\bibinfo {volume} {30}},\ \bibinfo {pages} {1706560} (\bibinfo {year}
  {2018})}\BibitemShut {NoStop}%
\bibitem [{\citenamefont {Gerlach}\ \emph {et~al.}(2001)\citenamefont
  {Gerlach}, \citenamefont {Berge}, \citenamefont {Goldmann}, \citenamefont
  {Campillo}, \citenamefont {Rubio}, \citenamefont {Pitarke},\ and\
  \citenamefont {Echenique}}]{Gerlach2001}%
  \BibitemOpen
  \bibfield  {author} {\bibinfo {author} {\bibfnamefont {A.}~\bibnamefont
  {Gerlach}}, \bibinfo {author} {\bibfnamefont {K.}~\bibnamefont {Berge}},
  \bibinfo {author} {\bibfnamefont {A.}~\bibnamefont {Goldmann}}, \bibinfo
  {author} {\bibfnamefont {I.}~\bibnamefont {Campillo}}, \bibinfo {author}
  {\bibfnamefont {A.}~\bibnamefont {Rubio}}, \bibinfo {author} {\bibfnamefont
  {J.~M.}\ \bibnamefont {Pitarke}}, \ and\ \bibinfo {author} {\bibfnamefont
  {P.~M.}\ \bibnamefont {Echenique}},\ }\href {\doibase
  10.1103/PhysRevB.64.085423} {\bibfield  {journal} {\bibinfo  {journal} {Phys.
  Rev. B}\ }\textbf {\bibinfo {volume} {64}},\ \bibinfo {pages} {085423}
  (\bibinfo {year} {2001})}\BibitemShut {NoStop}%
\bibitem [{\citenamefont {Marini}\ \emph {et~al.}(2002)\citenamefont {Marini},
  \citenamefont {Del~Sole}, \citenamefont {Rubio},\ and\ \citenamefont
  {Onida}}]{Marini2002}%
  \BibitemOpen
  \bibfield  {author} {\bibinfo {author} {\bibfnamefont {A.}~\bibnamefont
  {Marini}}, \bibinfo {author} {\bibfnamefont {R.}~\bibnamefont {Del~Sole}},
  \bibinfo {author} {\bibfnamefont {A.}~\bibnamefont {Rubio}}, \ and\ \bibinfo
  {author} {\bibfnamefont {G.}~\bibnamefont {Onida}},\ }\href {\doibase
  10.1103/PhysRevB.66.161104} {\bibfield  {journal} {\bibinfo  {journal} {Phys.
  Rev. B}\ }\textbf {\bibinfo {volume} {66}},\ \bibinfo {pages} {161104}
  (\bibinfo {year} {2002})}\BibitemShut {NoStop}%
\bibitem [{\citenamefont {Yi}\ \emph {et~al.}(2010)\citenamefont {Yi},
  \citenamefont {Ma}, \citenamefont {Rohlfing}, \citenamefont {Silkin},\ and\
  \citenamefont {Chulkov}}]{Yi2010}%
  \BibitemOpen
  \bibfield  {author} {\bibinfo {author} {\bibfnamefont {Z.}~\bibnamefont
  {Yi}}, \bibinfo {author} {\bibfnamefont {Y.}~\bibnamefont {Ma}}, \bibinfo
  {author} {\bibfnamefont {M.}~\bibnamefont {Rohlfing}}, \bibinfo {author}
  {\bibfnamefont {V.~M.}\ \bibnamefont {Silkin}}, \ and\ \bibinfo {author}
  {\bibfnamefont {E.~V.}\ \bibnamefont {Chulkov}},\ }\href {\doibase
  10.1103/PhysRevB.81.125125} {\bibfield  {journal} {\bibinfo  {journal} {Phys.
  Rev. B}\ }\textbf {\bibinfo {volume} {81}},\ \bibinfo {pages} {125125}
  (\bibinfo {year} {2010})}\BibitemShut {NoStop}%
\bibitem [{\citenamefont {Wallauer}\ \emph {et~al.}(2021)\citenamefont
  {Wallauer}, \citenamefont {Raths}, \citenamefont {Stallberg}, \citenamefont
  {Münster}, \citenamefont {Brandstetter}, \citenamefont {Yang}, \citenamefont
  {Güdde}, \citenamefont {Puschnig}, \citenamefont {Soubatch}, \citenamefont
  {Kumpf}, \citenamefont {Bocquet}, \citenamefont {Tautz},\ and\ \citenamefont
  {Höfer}}]{Wallauer2021}%
  \BibitemOpen
  \bibfield  {author} {\bibinfo {author} {\bibfnamefont {R.}~\bibnamefont
  {Wallauer}}, \bibinfo {author} {\bibfnamefont {M.}~\bibnamefont {Raths}},
  \bibinfo {author} {\bibfnamefont {K.}~\bibnamefont {Stallberg}}, \bibinfo
  {author} {\bibfnamefont {L.}~\bibnamefont {Münster}}, \bibinfo {author}
  {\bibfnamefont {D.}~\bibnamefont {Brandstetter}}, \bibinfo {author}
  {\bibfnamefont {X.}~\bibnamefont {Yang}}, \bibinfo {author} {\bibfnamefont
  {J.}~\bibnamefont {Güdde}}, \bibinfo {author} {\bibfnamefont
  {P.}~\bibnamefont {Puschnig}}, \bibinfo {author} {\bibfnamefont
  {S.}~\bibnamefont {Soubatch}}, \bibinfo {author} {\bibfnamefont
  {C.}~\bibnamefont {Kumpf}}, \bibinfo {author} {\bibfnamefont {F.~C.}\
  \bibnamefont {Bocquet}}, \bibinfo {author} {\bibfnamefont {F.~S.}\
  \bibnamefont {Tautz}}, \ and\ \bibinfo {author} {\bibfnamefont
  {U.}~\bibnamefont {Höfer}},\ }\href
  {https://www.science.org/doi/full/10.1126/science.abf3286} {\bibfield
  {journal} {\bibinfo  {journal} {Science}\ }\textbf {\bibinfo {volume}
  {371}},\ \bibinfo {pages} {1056} (\bibinfo {year} {2021})}\BibitemShut
  {NoStop}%
\bibitem [{\citenamefont {Baumgärtner}\ \emph {et~al.}(2022)\citenamefont
  {Baumgärtner}, \citenamefont {Reuner}, \citenamefont {Metzger},
  \citenamefont {Kutnyakhov}, \citenamefont {Heber}, \citenamefont {Pressacco},
  \citenamefont {Min}, \citenamefont {Peixoto}, \citenamefont {Reiser},
  \citenamefont {Kim}, \citenamefont {Lu}, \citenamefont {Shayduk},
  \citenamefont {Izquierdo}, \citenamefont {Brenner}, \citenamefont {Roth},
  \citenamefont {Schöll}, \citenamefont {Molodtsov}, \citenamefont {Wurth},
  \citenamefont {Reinert}, \citenamefont {Madsen}, \citenamefont
  {Popova-Gorelova},\ and\ \citenamefont {Scholz}}]{Baumgartner2022}%
  \BibitemOpen
  \bibfield  {author} {\bibinfo {author} {\bibfnamefont {K.}~\bibnamefont
  {Baumgärtner}}, \bibinfo {author} {\bibfnamefont {M.}~\bibnamefont
  {Reuner}}, \bibinfo {author} {\bibfnamefont {C.}~\bibnamefont {Metzger}},
  \bibinfo {author} {\bibfnamefont {D.}~\bibnamefont {Kutnyakhov}}, \bibinfo
  {author} {\bibfnamefont {M.}~\bibnamefont {Heber}}, \bibinfo {author}
  {\bibfnamefont {F.}~\bibnamefont {Pressacco}}, \bibinfo {author}
  {\bibfnamefont {C.-H.}\ \bibnamefont {Min}}, \bibinfo {author} {\bibfnamefont
  {T.~R.~F.}\ \bibnamefont {Peixoto}}, \bibinfo {author} {\bibfnamefont
  {M.}~\bibnamefont {Reiser}}, \bibinfo {author} {\bibfnamefont
  {C.}~\bibnamefont {Kim}}, \bibinfo {author} {\bibfnamefont {W.}~\bibnamefont
  {Lu}}, \bibinfo {author} {\bibfnamefont {R.}~\bibnamefont {Shayduk}},
  \bibinfo {author} {\bibfnamefont {M.}~\bibnamefont {Izquierdo}}, \bibinfo
  {author} {\bibfnamefont {G.}~\bibnamefont {Brenner}}, \bibinfo {author}
  {\bibfnamefont {F.}~\bibnamefont {Roth}}, \bibinfo {author} {\bibfnamefont
  {A.}~\bibnamefont {Schöll}}, \bibinfo {author} {\bibfnamefont
  {S.}~\bibnamefont {Molodtsov}}, \bibinfo {author} {\bibfnamefont
  {W.}~\bibnamefont {Wurth}}, \bibinfo {author} {\bibfnamefont
  {F.}~\bibnamefont {Reinert}}, \bibinfo {author} {\bibfnamefont
  {A.}~\bibnamefont {Madsen}}, \bibinfo {author} {\bibfnamefont
  {D.}~\bibnamefont {Popova-Gorelova}}, \ and\ \bibinfo {author} {\bibfnamefont
  {M.}~\bibnamefont {Scholz}},\ }\href
  {https://doi.org/10.1038/s41467-022-30404-6} {\bibfield  {journal} {\bibinfo
  {journal} {Nat. Commun}\ }\textbf {\bibinfo {volume} {13}} (\bibinfo {year}
  {2022})}\BibitemShut {NoStop}%
\bibitem [{\citenamefont {Neef}\ \emph {et~al.}(2023)\citenamefont {Neef},
  \citenamefont {Beaulieu}, \citenamefont {Hammer}, \citenamefont {Dong},
  \citenamefont {Maklar}, \citenamefont {Pincelli}, \citenamefont {Xian},
  \citenamefont {Wolf}, \citenamefont {Rettig}, \citenamefont {Pflaum},\ and\
  \citenamefont {Ernstorfer}}]{Neef2023}%
  \BibitemOpen
  \bibfield  {author} {\bibinfo {author} {\bibfnamefont {A.}~\bibnamefont
  {Neef}}, \bibinfo {author} {\bibfnamefont {S.}~\bibnamefont {Beaulieu}},
  \bibinfo {author} {\bibfnamefont {S.}~\bibnamefont {Hammer}}, \bibinfo
  {author} {\bibfnamefont {S.}~\bibnamefont {Dong}}, \bibinfo {author}
  {\bibfnamefont {J.}~\bibnamefont {Maklar}}, \bibinfo {author} {\bibfnamefont
  {T.}~\bibnamefont {Pincelli}}, \bibinfo {author} {\bibfnamefont {R.~P.}\
  \bibnamefont {Xian}}, \bibinfo {author} {\bibfnamefont {M.}~\bibnamefont
  {Wolf}}, \bibinfo {author} {\bibfnamefont {L.}~\bibnamefont {Rettig}},
  \bibinfo {author} {\bibfnamefont {J.}~\bibnamefont {Pflaum}}, \ and\ \bibinfo
  {author} {\bibfnamefont {R.}~\bibnamefont {Ernstorfer}},\ }\href
  {https://doi.org/10.1038/s41586-023-05814-1} {\bibfield  {journal} {\bibinfo
  {journal} {Nature}\ }\textbf {\bibinfo {volume} {616}},\ \bibinfo {pages}
  {275–279} (\bibinfo {year} {2023})}\BibitemShut {NoStop}%
\end{thebibliography}%

\end{document}